\documentclass[prx, twocolumn, nofootinbib, showpacs, amsmath, amssymb, floatfix, eqsecnum]{revtex4-1}
\usepackage{amsmath}
\usepackage{amssymb}
\usepackage{amsthm}
\usepackage{amsfonts}

\usepackage{graphicx,color,framed}
\usepackage{hyperref}
\usepackage{times}
\usepackage{enumerate}
\usepackage{lipsum}
\usepackage{slashed}
\usepackage{url}
\usepackage{bbm}
\hypersetup{
    colorlinks=true, 
    linktoc=all,     
    linkcolor=blue,  
}

\def \beq {\begin{equation}}
\def \eeq {\end{equation}}
\def \beqa {\begin{eqnarray}}
\def \eeqa {\end{eqnarray}}
\def \bseq {\begin{subequations}}
\def \eseq {\end{subequations}}

\newcommand{\<}{\langle}
\renewcommand{\>}{\rangle}

\renewcommand{\v}[1]{\boldsymbol{#1}}
\newtheorem{theorem}{Theorem}

\newtheorem{lemma}{Lemma}
\newtheorem{corollary}{Corollary}
\newtheorem{definition}{Definition}
\usepackage[normalem]{ulem}

\begin{document}

\title{Bulk-boundary correspondence for interacting Floquet systems in two dimensions}




\author{Carolyn Zhang}
\author{Michael Levin}
\affiliation{Department of Physics, Kadanoff Center for Theoretical Physics, University of Chicago, Chicago, Illinois 60637,  USA}
\date{\today}

\begin{abstract}
We present a method for deriving bulk and edge invariants for interacting, many-body localized Floquet systems in two spatial dimensions. This method is based on a general mathematical object which we call a \emph{flow}. As an application of our method, we derive bulk invariants for Floquet systems without symmetry, as well as for systems with $U(1)$ symmetry. We also derive new formulations of previously known single-particle and many-body invariants. For bosonic systems without symmetry, our invariant gives a bulk counterpart of the rational-valued GNVW index $\frac{p}{q}$ quantifying transport of quantum information along the edge.
\end{abstract}

\maketitle

\section{Introduction}\label{sintroduction}
Periodically driven systems, also known as Floquet systems, can realize interesting topological phases that have no stationary analogue\cite{floquetreview,rudnerband}. One illustrative example of such a system was introduced in Refs.~\onlinecite{kitagawa2010, anomalousedge}. In these works, the authors constructed a single-particle Floquet system in two spatial dimensions with the property that (i) there are chiral edge modes propagating in each Floquet band gap and (ii) all of the Floquet bands have vanishing Chern number. 

This example leads to a puzzle, since it is not obvious how the information about the number of chiral edge modes is encoded in the bulk dynamics. This puzzle was resolved in Ref.~\onlinecite{anomalousedge}, which showed that the number of chiral edge modes is determined by a particular winding number that characterizes the time evolution of the bulk bands during a single period. Note that this winding number characterizes the bulk ``micromotion,'' or motion within a period, as opposed to the stroboscopic dynamics\cite{magnetization}. This bulk-boundary correspondence was further explored in Refs.~\onlinecite{nathan2015, afai, graf2018, shapiro2019, vu2022}.

In this paper, we consider an analogous problem involving \emph{many-body} Floquet systems in two spatial dimensions. A prototypical example of such a system is the ``SWAP circuit'', a many-body Floquet system constructed out of either bosonic or fermionic degrees of freedom living on the sites of the square lattice\cite{chiralbosons, harperorder}. Like the single-particle example mentioned above, the SWAP circuit displays interesting stroboscopic dynamics at its edge. In particular, when the SWAP circuit is defined on a lattice with a boundary, one finds that the lattice sites near the edge undergo a unit translation during each driving period. This behavior is significant because translations cannot be generated by a local, 1D Hamiltonian\cite{GNVW}. In this sense, the SWAP circuit has ``anomalous'' edge dynamics, just like the single-particle example discussed above. More quantitatively, the anomalous edge dynamics of the SWAP circuit or its relatives can be characterized by an edge invariant -- known as the GNVW index -- which takes values in the rational numbers\cite{chiralbosons,mpu,tracking,ranard2022}.

Again, we are faced with a puzzle: we have an edge invariant for these systems (i.e. the GNVW index) but we lack a corresponding bulk topological invariant analogous to the above single-particle winding number. A similar puzzle exists for $U(1)$ symmetric generalizations of the SWAP circuit\cite{u1floquet,nathanafi,eft}: there too, we have an edge invariant that quantifies the anomalous edge dynamics in these systems but the corresponding bulk invariant is missing. The goal of this paper is to construct these missing bulk invariants.

We investigate this problem in the context of two dimensional ``many-body localized'' Floquet systems. The reason we focus on many-body localized (MBL) Floquet systems is that these systems either do not thermalize, or take a long time to thermalize. As a result, they can display a rich array of long-lived dynamics\cite{floquetreview}, unlike generic interacting many-body Floquet systems, which heat up by absorbing energy from the drive\cite{rigollongtime,lazaridesgeneric,ponteergodic, fate, ponte, abanintheory}. 

\begin{figure}[tb]
   \centering
   \includegraphics[width=.85\columnwidth]{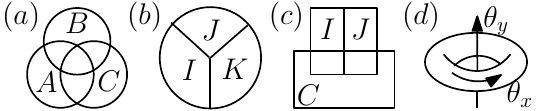} 
   \caption{Schematic geometries of the four types of bulk invariants that we discuss. (a) The most general bulk invariant (\ref{mag}), which applies to all the systems studied in this work, involves three overlapping disklike regions $A, B, C$. (b) Our invariant (\ref{nonoverlap}), which applies to single-particle systems and $U(1)$ symmetric many-body systems, involves three non-overlapping adjacent regions $I, J, K$. (c) We also obtain bulk invariants (\ref{bulkcurrbd}) for these systems involving regions $I, J, C$ as well as (d) invariants via flux threading on a torus (\ref{wintkspace}).}
   \label{fig:summarypic}
\end{figure}

Our central result is a method for constructing both bulk and edge invariants for 2D MBL Floquet systems with different symmetry groups $G$. We also show that our bulk and edge invariants are equal to one another, thereby establishing a bulk-boundary correspondence for these systems. Our results are summarized in Table~\ref{table:table1} and Fig.~\ref{fig:summarypic}. Notably, we find a bulk invariant for general 2D MBL Floquet systems without symmetry, as well as for systems with $U(1)$ symmetry. The first invariant gives a bulk formulation of the GNVW index, while the second invariant gives a bulk counterpart of the edge invariants in Ref.~\onlinecite{u1floquet}. We also derive different formulations of previously known edge invariants and single-particle invariants. 

Our method for constructing invariants involves a mathematical object which we call a ``flow.'' A flow $\Omega_{A,B}(U)$ is a real-valued function of a unitary $U$ and two subsets of lattice sites $A, B$ that obeys certain properties. We show that if one can find a flow for some symmetry group $G$, then one can immediately construct corresponding bulk and edge invariants for general 2D MBL Floquet systems. 

\begin{table}[tb]
\centering
\begin{tabular}{ |c|c|c|c| } 
\hline
 & Single-particle & \begin{tabular}{@{}c@{}}Many-body, \\ $U(1)$ symmetry \end{tabular} & \begin{tabular}{@{}c@{}}Many-body, \\ no symmetry\end{tabular} \\
\hline
Edge & \begin{tabular}{@{}c@{}} Eq.~(\ref{flownonintedge}) \end{tabular} & \begin{tabular}{@{}c@{}} Eq.~(\ref{flowu1}) \end{tabular} & \begin{tabular}{@{}c@{}} Eq.~(\ref{flownosymm1}) \end{tabular}\\
\hline
Bulk & \begin{tabular}{@{}c@{}} Eq.~(\ref{magnonint}) \end{tabular} & Eq.~(\ref{mformu1int})  & Eq.~(\ref{magswap}) \\ 
\hline
\end{tabular}
\caption{A summary of the bulk and edge invariants presented in this work.}
\label{table:table1}
\end{table}

The paper is structured as follows. For simplicity, we first present our results for a special kind of MBL Floquet system called a ``unitary loop''; later, we explain how to extend our results to general 2D MBL Floquet systems. In Sec.~\ref{ssetup}, we review the definitions of MBL Floquet systems and unitary loops and we give a precise statement of the problem we wish to solve. Sec.~\ref{sflows} presents the main results of this paper: we introduce the concept of a flow, and we show how to construct bulk and edge invariants from flows. In Sec.~\ref{sadditive}, we discuss a special kind of flow, called a ``spatially additive flow'', and we derive additional formulas for bulk and edge invariants for spatially additive flows. We then study the general results of the preceding two sections with three illustrative examples: single-particle systems (Sec.~\ref{snoninteracting}), interacting systems with $U(1)$ symmetry (Sec.~\ref{su1}), and interacting systems without symmetry (Sec.~\ref{snosymm}). In Sec.~\ref{smblgen}, we discuss the extension of our results from unitary loops to general MBL Floquet systems. We conclude with some open questions in Sec.~\ref{sdiscussion}. Additional details and technical arguments can be found in the appendices.

\section{Setup and definitions}\label{ssetup}
In this section we explain the basic setup of our problem and the objects that we study, namely MBL Floquet systems and unitary loops. We also explain the connection between $d$ dimensional unitary loops and $(d-1)$ dimensional locality preserving unitaries describing their stroboscopic edge dynamics\cite{alldimensions,chiralbosons}.  


\subsection{MBL Floquet systems}\label{mblfloqdef}
We begin by recalling the definition of an MBL Floquet system. Consider a bosonic\footnote{We discuss the generalization of our results to fermionic systems in Sec.~\ref{sdiscussion}.} many-body system built out of $k$-state spins living on an infinite $d$ dimensional lattice. We assume that Hamiltonian is periodic in time:
\begin{align}
H(t+T) = H(t),
\end{align}
where $T$ is the period. We also assume that $H(t)$ is local in the sense that it can be written as a sum of terms of the form
\begin{align}
H(t) = \sum_r H_r(t),
\label{Hsum}
\end{align}
where $H_r(t)$ is supported near site $r$. Let $U_F$ denote the Floquet unitary that describes the stroboscopic dynamics:
\begin{align}
U_F = \mathcal{T} e^{-i\int_0^T dt H(t)dt}.
\end{align}
An ``MBL Floquet system'' is a system of this type with the property that $U_F$ is many-body localized, i.e. $U_F$ can be written as a product of mutually commuting quasi-local unitaries\cite{chiralbosons}:
\begin{equation}\label{MBLcond1}
U_F=\prod_{r} U_r\qquad [U_r,U_{r'}]=0,
\end{equation}
where each $U_r$ is a unitary supported within a finite distance $\xi$ of site $r$, (possibly with exponentially decaying tails). 
The significance of the above condition (\ref{MBLcond1}) is that it guarantees that $U_F$ does not spread operators beyond the distance scale $\xi$, no matter how many times it is applied; consequently the stroboscopic dynamics described by $U_F$ does not result in thermalization.

In this paper, we will mostly focus on a special class of MBL Floquet systems, namely those with \emph{trivial} stroboscopic dynamics: 
\begin{align}
U_F = \mathbbm{1}.
\label{MBLstrong0}
\end{align}

It turns out that this special case contains all of the relevant physics of MBL Floquet systems, but in a simpler setting. Later, in Sec.~\ref{smblgen}, we will show that our results can be straightforwardly extended to general MBL Floquet systems obeying (\ref{MBLcond1}), but for now we will focus on systems obeying (\ref{MBLstrong0}). Our task is thus to find bulk and edge invaraints for MBL Floquet systems obeying (\ref{MBLstrong0}).

\subsection{Unitary loops}\label{sloopdef}

An equivalent way to think about MBL Floquet systems with $U_F = \mathbbm{1}$ is as ``unitary loops.'' Here, a ``unitary loop'' is a one-parameter family of unitaries $\{U(t)  : t\in[0,T] \}$, generated by a local Hamiltonian (\ref{Hsum}), with the property that
\begin{equation}
\label{MBLstrong}
U(T)= U(0) =\mathbbm{1},
\end{equation}
In this language, our problem is to find bulk and edge invariants for unitary loops.

But what does it mean to construct an invariant for a unitary loop? To answer this question, we need to define a notion of equivalence similar to the notion of adiabatic equivalence in equilibrium systems. We say that two unitary loops $\{U(t)\}$ and $\{U'(t)\}$ are ``equivalent'', denoted $\{U(t)\}\sim \{U'(t)\}$, if they can be smoothly deformed into one another. That is, $\{U(t)\}\sim \{U'(t)\}$ if there exists a one-parameter family of unitary loops, $\{U_s(t) : s \in[0,1]\}$, depending smoothly on $s$, such that 
\begin{align}
U_0(t) = U(t), \quad \quad U_1(t)=U'(t). \label{homotopy}
\end{align}
Importantly, this interpolation must maintain the loop condition (\ref{MBLstrong}) for all $s$. That is,
\begin{equation}\label{homotopymbl}
U_s(T)=\mathbbm{1}
\end{equation}
for all $s\in[0,1]$. We note that a similar notion of equivalence can be defined for more general MBL Floquet systems: in that case, we say that two MBL Floquet systems are equivalent if they can be smoothly deformed into one another while maintaining the MBL property (\ref{MBLcond1}).

\subsection{Locality preserving unitaries}

Another concept that we will need below is a ``locality preserving unitary'' (LPU). Roughly speaking, a locality preserving unitary $U$ is a unitary that transforms local operators to nearby local operators. More precisely, if $O_r$ is an operator supported on site $r$, then $U^\dagger O_rU$ is supported within a finite distance $\xi$ of the site $r$ (up to exponential tails). We will refer to the length scale $\xi$ as the ``operator spreading length'' of $U$.

There is a natural way to define equivalence classes of LPUs. We say that two LPUs $U$ and $U'$ are equivalent, denoted $U \simeq U'$, if they differ by a ``locally generated unitary'' (LGU) $W$:
\begin{equation}\label{simW}
U =W\cdot U'.
\end{equation}
Here, a locally generated unitary $W$ is a unitary that can be generated by the time evolution of a local Hamiltonian over a finite period of time: 
\begin{align}
W=\mathcal{T}e^{-i\int_0^1H(s)ds}. 
\end{align}

For some of our arguments, we will find it useful to consider LPUs with strict locality properties. We say that a unitary $U$ is a \emph{strict} LPU with operator spreading length $\xi$ if, for any operator $O_r$ supported on site $r$, the operator $U^\dagger O_r U$ is \emph{strictly} supported within a finite distance $\xi$ of $r$, without any exponential tails. 

We will also find it useful to consider a special class of LGUs with strict locality properties which we will call ``finite depth local unitaries'' (or FDLUs). An FDLU is a unitary that can be written as a finite depth quantum circuit. More specifically, we say that $W$ is an FDLU of depth $n$ and radius $\lambda$, if $W$ can be written as a finite depth quantum circuit of depth $n$, where each layer is a product of local unitary gates supported in (non-overlapping) balls of radius $\lambda$. Note that every LGU can be approximated to arbitrarily small error by an FDLU using a Trotter expansion.

\subsection{Mapping between $d$-dimensional unitary loops and $(d-1)$-dimensional LPUs}
We now explain an important mapping between $d$-dimensional unitary loops and $(d-1)$-dimensional LPUs\cite{chiralbosons, alldimensions}. The basic idea is that given any $d$-dimensional unitary loop $\{U(t)\}$, we can construct a corresponding $(d-1)$-dimensional LPU by considering the dynamics of $U(t)$ near a physical boundary or ``edge'' (Here, we use the term ``edge'' because we will be primarily interested in the case $d=2$, where the boundary is one dimensional). 

The precise construction is as follows. Given a $d$-dimensional unitary loop with Hamiltonian $H(t)$, we \emph{restrict} the Hamiltonian to a large, but finite, ball $C$ by discarding all terms that have support outside of $C$. We denote the restricted Hamiltonian by $H_C(t)$. We then define a boundary or ``edge'' unitary by 
\begin{align}
U_{\mathrm{edge}} = \mathcal{T}e^{-i\int_0^T dt H_C(t)}
\label{Uedgedef}
\end{align}
Comparing this definition with (\ref{MBLstrong}) it is clear that $U_{\mathrm{edge}}$ acts trivially deep in the interior of $C$ -- that is, $U_{\mathrm{edge}}$ is supported within a finite distance of the boundary of $C$ (up to exponential tails). Thus, $U_{\mathrm{edge}}$ can be thought of as a $(d-1)$-dimensional unitary acting on the boundary of $B$. It is also clear that $U_{\mathrm{edge}}$ is locality preserving, by Lieb-Robinson bounds.\footnote{See Sec.~\ref{smblgen} for the definition of $U_{\mathrm{edge}}$ for general MBL Floquet circuits.} Note that, in the context of Floquet systems, $U_{\mathrm{edge}}$ has a simple physical meaning: it describes the stroboscopic edge dynamics of the Floquet system corresponding to $\{U(t)\}$. 

Importantly, one can show that the above mapping is consistent with the two equivalence relations in the sense that
\begin{equation}\label{phase}
\{U(t)\}\sim\{U'(t)\}  \implies U_{\mathrm{edge}}\simeq U_{\mathrm{edge}}'.
\end{equation}
(see Appendix~\ref{shomotopy} for a proof). One implication of this result is that one can classify (or at least partially classify) unitary loops/Floquet systems by studying their corresponding edge unitaries.

\subsection{Incorporating symmetries}\label{symmetries}
We now discuss how to incorporate symmetries into these definitions. Consider a symmetry group $G$ and a corresponding collection of onsite unitary symmetry transformations $\{U_g : g \in G\}$. We say that a  unitary loop $\{U(t)\}$ is ``$G$-symmetric'' if it is generated by a $G$-symmetric Hamiltonian $H(t)$, i.e. $U_g H(t) U_g^{-1} = H(t)$ for all $t\in[0,T]$. Likewise, we say that two $G$-symmetric unitary loops are equivalent if they can be smoothly deformed into one another while preserving the symmetry, i.e. $\{U_s(t)\}$ should be generated by a local \emph{$G$-symmetric} Hamiltonian $H_s(t)$ for all $s \in [0,1]$.

We can also incorporate symmetry into the definition of an LPU in a natural way. We say that an LPU $U$ is ``$G$-symmetric'' if $U$ commutes with the symmetry transformation $U_g$ for all $g \in G$. Likewise we say that two $G$-symmetric LPUs are equivalent if they differ by a locally generated unitary $W$ whose generating Hamiltonian $H(s)$ is $G$-symmetric for all $s\in [0,1]$.

\subsection{Bulk and edge invariants}\label{sinvariants}
One of the main goals of this paper is to construct bulk and edge invariants for unitary loops. Here, a ``bulk invariant'' is a real-valued function $M(\{U(t)\}$ defined on unitary loops, with the property that it is invariant under the equivalence relation (\ref{homotopy}) in the sense that 
\begin{equation}
M(\{U(t)\})=M(\{U'(t)\}) \ \ \ \text{ if }\{U(t)\}\sim\{U'(t)\}
\end{equation}

Likewise, an ``edge invariant'' is a real valued function defined on the edge unitaries, $F(U_{\mathrm{edge}})$ that is invariant under the equivalence relation defined in (\ref{simW}) in the sense that
\begin{equation}
F(U_{\mathrm{edge}})=F(U_{\mathrm{edge}}') \ \ \ \text{ if }U_{\mathrm{edge}}\simeq U_{\mathrm{edge}}'
\label{edgeinvdef}
\end{equation}

In this paper we will construct bulk and edge invariants for \emph{two-dimensional} unitary loops (or equivalently, two-dimensional Floquet systems). That is, we will construct bulk invariants $M(\{U(t)\}$ for 2D unitary loops, and edge invariants $F(U_{\mathrm{edge}})$ for their 1D edge unitaries. Our invariants have the additional feature of obeying a bulk-boundary correspondence:
\begin{equation}\label{bulkboundarycorr}
M(\{U(t)\})=F(U_{\mathrm{edge}})
\end{equation}

\section{General theory of flows}\label{sflows}

In this section, we define a general mathematical object called a ``flow.'' This mathematical object will be our main tool for constructing bulk and edge invariants for unitary loops. 

\subsection{Prologue: a single-particle example}\label{sprologue}
To motivate our definition, we begin with an example of a flow in single-particle systems. Consider a single-particle system defined on a $d$-dimensional lattice $\Lambda$. Let $U$ be a single-particle unitary transformation, i.e.~a $|\Lambda| \times |\Lambda|$ unitary matrix $U_{ab} = \<a| U | b\>$ where $a,b \in \Lambda$. Given any two subsets of lattice sites $A, B \subset \Lambda$, we can define a real number $\omega_{A,B}(U)$ by
\begin{equation}\label{nonintflowU}
\omega_{A,B}(U)=\sum_{a\in A}\sum_{b\in B} (|U_{ab}|^2-\delta_{ab}).
\end{equation}
We can think of $\omega_{A,B}(U)$ as providing a quantitative measure of how much the unitary $U$ transports particles from $B$ to $A$. The first term $\sum_{a\in A}\sum_{b\in B} |U_{ab}|^2$ measures the magnitude of the matrix elements of $U$ between $B$ and $A$, while the second term, $-\sum_{a\in A}\sum_{b\in B} \delta_{ab}$ is a constant offset that guarantees that $\omega_{A,B}(U) = 0$ if $U = \mathbbm{1}$. 

The quantity $\omega_{A,B}(U)$ has two important properties. First, for any unitary $V_A$ that is supported entirely in $A$ or its complement $\overline{A}$, and for any unitary $U$,
\begin{align}
\omega_{A,B}(V_AU) = \omega_{A,B}(U),
\label{omab1}
\end{align}
Likewise, for any unitary $V_B$ that is supported entirely in $B$ or its complement $\overline{B}$,
\begin{align}
\omega_{A,B}(UV_B) = \omega_{A,B}(U).
\label{omab2}
\end{align}
where again $U$ is a general unitary. To derive the first property (\ref{omab1}), notice that any $V_A$ of this kind does not mix the sites within $A$ with those outside of $A$; therefore $\sum_{a\in A} |(V_AU)_{ab}|^2 = \sum_{a\in A} |U_{ab}|^2$. The second property (\ref{omab2}) follows from similar reasoning.

The above two properties (\ref{omab1} -\ref{omab2}) are important because they guarantee that $\omega_{A,B}(U)$ only depends on $U$ in a very limited way. As a result we can construct bulk and edge invariants out of $\omega_{A,B}(U)$.

The idea is as follows: consider the case where $\Lambda$ is a one dimensional lattice, and suppose that $U$ is a 1D unitary transformation that is locality preserving in the sense that it only mixes nearby lattice sites $a,a' \in \Lambda$. Choose $A$ and $B$ to be two large overlapping intervals. In this case, we can use (\ref{omab1} - \ref{omab2}) to prove that
\begin{align*}
\omega_{A,B}(VU) = \omega_{A,B}(U),
\end{align*}
for any unitary $V$ supported within an interval much smaller than the overlap of $A$ and $B$. The reason is that any such $V$ is either fully supported within $A$ or $\overline{A}$, in which case we can use (\ref{omab1}), or it is supported deep within $B$ or $\overline{B}$, in which case we can use (\ref{omab2}) after first commuting $V$ through $U$:
\begin{align*}
\omega_{A,B}(VU) &= \omega_{A,B} (U[U^{-1} VU]) \nonumber \\
&=\omega_{A,B}(U).
\end{align*}
Here, in the second equality, we are using the fact that $U$ is locality preserving and $V$ is supported \emph{deep} within $B$ or $\overline{B}$ and therefore $U^{-1} VU$ is supported within $B$ or $\overline{B}$.

By repeating the above argument multiple times, it follows that 
\begin{align*}
\omega_{A,B}(V_N V_{N-1} \cdots V_1U) = \omega_{A,B}(U),
\end{align*}
for any collection of unitaries $V_1,...,V_N$ that are supported within small intervals, as long as we take $A, B$ and $A \cap B$ sufficiently large. Next, consider any unitary $W$ that is generated by a local (1D) Hamiltonian over a finite period of time. Any such $W$ can be approximately arbitrarily closely by a product of the form $V_N V_{N-1} \cdots V_1$. Hence, we deduce that 
\begin{align*}
\omega_{A,B}(WU) = \omega_{A,B}(U),
\end{align*}
in the limit of large $A, B$ and large overlap $A \cap B$. The latter property means that $\omega_{A,B}(U)$ defines an edge invariant for 2D unitary loops, because it satisfies (\ref{edgeinvdef}).\footnote{This invariant is closely related to the ``flow'' of a unitary matrix, described in Appendix C.1 of Ref.~\onlinecite{kitaev}.} 

It turns out that one can also use $\omega_{A,B}(U)$ to construct \emph{bulk} invariants for 2D unitary loops (see Sec. ~\ref{sflownonint}). Thus, $\omega_{A,B}(U)$ provides a powerful tool for constructing both edge and bulk invariants for unitary loops in single-particle systems.

Motivated by this example, we now define the notion of a ``flow'' for \emph{many-body} systems.

\subsection{Definition of flow}\label{soverlap}

Consider a many-body system defined on a $d$-dimensional lattice $\Lambda$ with an on-site symmetry group $G$. In this context, we can define a general mathematical object that we call a ``flow'':
\\

\begin{definition} \label{flowdef}
A flow $\Omega_{A,B}(U)$ is a function that outputs a real number given a $G$-symmetric unitary $U$ and two subsets of lattice sites $A, B \subset \Lambda$, and that has the following properties:
\begin{enumerate}
\item{$\Omega_{A,B}(V_AU) = \Omega_{A,B}(U)$ if $\text{supp}(V_A) \subset A$ or $\overline{A}$.}
\item{$\Omega_{A,B}(UV_B) = \Omega_{A,B}(U)$ if $\text{supp}(V_B) \subset B$ or $\overline{B}$.}
\item{$\Omega_{A_1\cup A_2,B_1\cup B_2}(U_1 \otimes U_{2})=\Omega_{A_1,B_1}(U_{1})+\Omega_{A_2,B_2}(U_{2})$
for any $U_1, U_2$ defined on disjoint sets of lattice sites $\Lambda_1, \Lambda_2$ with $A_1, B_1 \subset \Lambda_1$ and $A_2, B_2 \subset \Lambda_2$.}
\item{$\Omega_{A,B}(\mathbbm{1})=0$.}
\end{enumerate}
\end{definition}

Each of these properties has a simple intuitive meaning. The first two properties tell us that $\Omega_{A,B}(U)$ is insensitive to $G$-symmetric unitaries that are supported entirely within $A$ or $B$ or their complements $\overline{A}, \overline{B}$. This is compatible with the idea that, roughly speaking, $\Omega_{A,B}(U)$ measures \emph{total} transport between $A$ and $B$. The third property tells us that the flow is additive under the tensor product (or ``stacking'') operation. The last property is simply a normalization convention.

Notice that the function $\omega_{A,B}(U)$ defined in Eq.~(\ref{nonintflowU}) obeys all of the above properties if we translate them to a single-particle framework -- i.e. replacing the tensor product $U_1 \otimes U_2$ with a direct sum $U_1 \oplus U_2$. Thus, $\omega_{A,B}(U)$ can be thought of as a single-particle analog of a flow. 

At this point, we should mention that there is a subtlety in the interpretation of the direction of transport: while a flow measures transport of \emph{states} from $B$ to $A$, it measures transport of \emph{operators} from $A$ to $B$.  While in Sec.~\ref{sprologue} we mentioned that $\omega_{A,B}(U)$ measures transport of particles from $B$ to $A$, in the many-body setting, it's often easiest to interpret the flow as transport of operators from $A$ to $B$.

\subsection{Examples of flows}\label{sexamples}
Here we briefly present two many-body examples of flows that will be discussed later in the paper.

\subsubsection{Example 1: $U(1)$ symmetry}
Our first example of a flow applies to lattice many-body systems with a global $U(1)$ symmetry. More specifically, consider lattice systems that conserve a total $U(1)$ charge $Q$ of the form $Q = \sum_r Q_r$ where $Q_r$ is a Hermitian operator supported on lattice site $r \in \Lambda$. Define
\begin{equation}\label{chargeomega}
\Omega_{A,B}(U)=\left\langle U^\dagger Q_AU Q_B\right\rangle_\rho-\left\langle Q_AQ_B\right\rangle_\rho,
\end{equation}
where 
\begin{align}
Q_A=\sum_{r\in A}Q_r ,\quad \quad Q_B=\sum_{r\in B}Q_r, 
\end{align}
and where the expectation value $\<\cdot\>_\rho$ is taken in the mixed state
\begin{equation}
\rho=\frac{1}{Z}e^{\mu Q}, \qquad Z=\mathrm{Tr}(e^{\mu Q}),
\end{equation}
for some real valued ``chemical potential'' $\mu$. 

It is easy to check that $\Omega_{A,B}(U)$ satisfies all the requirements for a flow. For example, to establish the first property in the above definition, we need to show that $\Omega_{A,B}(U)$ is invariant under replacing $U \rightarrow V_AU$ for any $U(1)$ symmetric $V_A$ supported in $A$ or $\overline{A}$. To prove this statement, notice that any such $V_A$ commutes with $Q_A$ and hence
\begin{align}
\left\langle \left(V_AU\right)^{\dagger}Q_A (V_AU) Q_B \right\rangle_\rho = \langle U^\dagger Q_A U Q_B\rangle_\rho.
\end{align}
It follows immediately that $\Omega_{A,B}(V_AU) = \Omega_{A,B}(U)$. 

Note that the parameter $\mu$ can take any real value, so we have constructed not just one flow but rather a continuous family of flows. We discuss this flow and its applications in more detail in Sec.~\ref{su1}.

\subsubsection{Example 2: No symmetry}
Our second example of a flow applies to interacting systems without any symmetry constraints. To explain this example, we first need to review the definition of $\eta(\mathcal{A}, \mathcal{B})$ -- a real-valued ``overlap'' between two operator algebras $\mathcal{A}, \mathcal{B}$, introduced in Ref.~\onlinecite{GNVW}. 

Let $\mathcal{A}, \mathcal{B}$ be any two operator algebras consisting of operators acting on some finite dimensional Hilbert space. Let $\{O_a\}$ be a complete orthonormal basis of operators in $\mathcal{A}$ -- that is, a collection of operators such that  (i) $\{O_a\}$ is a complete basis for $\mathcal{A}$ and (ii) $\{O_a\}$ satisfies $\mathrm{tr}(O_a^\dagger O_{a'}) = \delta_{aa'}$ where we use the lowercase symbol ``${\mathrm{tr}}$'' to denote a \emph{normalized} trace defined by ${\mathrm{tr}}(\mathbbm{1}) = 1$. Similarly, let $\{O_b\}$ be a complete orthonormal basis for $\mathcal{B}$. The ``overlap'' $\eta(\mathcal{A}, \mathcal{B})$ is defined by
\begin{align}
\eta(\mathcal{A}, \mathcal{B}) = 
\sqrt{\sum_{O_a\in\mathcal{A},O_b\in\mathcal{B}}| \mathrm{tr}(O_{a}^\dagger O_{b})|^2}.
\end{align}
One can check that $\eta(\mathcal{A}, \mathcal{B})$ only depends on the algebras $\mathcal{A}, \mathcal{B}$ and not on the choice of orthonormal bases $\{O_a\}, \{O_b\}$. Also, it is not hard to show that $\eta(\mathcal{A}, \mathcal{B}) \geq 1$ since the two algebras $\mathcal{A}, \mathcal{B}$ both contain the identity operator $\mathbbm{1}$.

With this notation, we are now ready to give an example of a flow for interacting systems without symmetries. Let $A, B$ be any two subsets of lattice sites, $A, B \subset \Lambda$, and let $\mathcal{A}, \mathcal{B}$ denote the corresponding operator algebras, consisting of operators supported on $A, B$, respectively. We can define a flow by
\begin{equation}
\Omega_{A,B}(U)=\log\left[\frac{\eta(U^{\dagger}\mathcal{A}U,\mathcal{B})}{\eta(\mathcal{A},\mathcal{B})}\right],
\label{flownosymm_ex}
\end{equation}
Again, it is easy to check that $\Omega_{A,B}(U)$ satisfies all the properties of a flow. For example, to prove the first property of a flow, namely that $\Omega_{A,B}(U)$ is invariant under replacing $U \rightarrow V_AU$ for any $V_A$ supported on $A$ or $\overline{A}$, notice that
\begin{align}
\eta\left(\left(V_AU\right)^{\dagger}\mathcal{A}(V_AU),\mathcal{B}\right) =  \eta(U^\dagger\mathcal{A}U,\mathcal{B}),
\end{align}
since $V_A$ can only shuffle operators in $\mathcal{A}$ and therefore $V_A^\dagger \mathcal{A} V_A = \mathcal{A}$.

This flow is closely related to the GNVW index for classifying 1D locality preserving unitaries\cite{GNVW}. We discuss this flow and its applications in more detail in Sec.~\ref{snosymm}. 

\subsection{Properties of flows}\label{sproperties}

We now state two important properties of flows that follow from Definition~\ref{flowdef}. First some notation: we define the $\ell$-\emph{boundary} of a set $A$, $\partial_{\ell}A$, as 
\begin{equation}
\partial_{\ell}A=\{r\in\Lambda:\mathrm{dist}(r,A)\leq \ell\text{ and }\mathrm{dist}(r,\overline{A})\leq \ell\}.
\end{equation}
One can think of $\partial_\ell A$ as a ``thickened boundary'' which consists of all lattice sites that are within distance $\ell$ from the boundary of $A$.
With this notation, we can now state the two properties of $\Omega_{A,B}(U)$:
\begin{theorem}\label{theorem1}
Let $U$ be a $G$-symmetric strict LPU with an operator spreading length $\xi$. Let $W$ be a $G$-symmetric FDLU of depth $n$ which is built out of unitary gates supported in balls of radius $\lambda$. Then:
\begin{enumerate}
\item{$\Omega_{A,B}(WU) = \Omega_{A,B}(W'U)$ where $W'$ is obtained by removing all gates from $W$ except for those fully supported in $(\partial_{2n \lambda} A)\cap(\partial_{2n \lambda+ \xi} B)$.}
\item{$\Omega_{A,B}(U) = \Omega_{A\setminus a,B}(U)$ for any $a \notin \partial_{4\xi}B$. $\Omega_{A,B}(U) = \Omega_{A,B\setminus b}(U)$ for any $b \notin \partial_{4\xi}A$.}
\end{enumerate}
\end{theorem}
Each of these properties tell us that $\Omega_{A,B}(U)$ is invariant under some kind of change in $A, B$ or $U$. The first property says that $\Omega_{A,B}(WU)$ doesn't change if we remove gates from $W$ that are far from the intersection of the two boundaries of $A, B$. The second property says that $\Omega_{A,B}(U)$ is invariant under adding or removing a lattice site $a \in A$ as long as $a$ is far from the boundary of $B$, and similarly $\Omega_{A,B}(U)$ is invariant under adding or removing a lattice site $b \in B$ as long as $b$ is far from the boundary of $A$. We prove Theorem~\ref{theorem1} in Appendix~\ref{slemmas}. 

We now state two useful corollaries of Theorem~\ref{theorem1}:
\begin{corollary}\label{corollary1}
Let $U$ be a $G$-symmetric strict LPU with an operator spreading length $\xi$. Let $W$ be a $G$-symmetric FDLU of depth $n$ which is built out of unitary gates supported in balls of radius $\lambda$. If
$(\partial_{2n \lambda} A)\cap(\partial_{2n \lambda+ \xi} B) = \emptyset$, then
\begin{align}
\Omega_{A,B}(W U) = \Omega_{A,B}(U).
\end{align}
\end{corollary}

\begin{corollary}\label{corollary2}
Let $W$ be a $G$-symmetric FDLU of depth $n$ which is built out of unitary gates supported in balls of radius $\lambda$. Then 
\begin{align}
\Omega_{A,B}(W) = \Omega_{A,B}(W'),
\end{align}
where $W'$ is obtained by removing all gates from $W$ except for those fully supported in $(\partial_{2n \lambda} A)\cap(\partial_{2n \lambda} B)$.
\end{corollary}

Both corollaries are immediate consequences of Theorem~\ref{theorem1}.1.

\subsection{Edge invariants from flows}\label{sflowedge}
\begin{figure}[tb]
   \centering
   \includegraphics[width=.67\columnwidth]{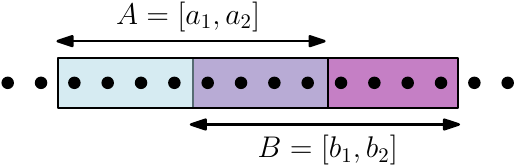} 
   \caption{We use overlapping intervals $A$ and $B$ to define our edge invariant $F(U_{\mathrm{edge}})=\Omega_{A,B}(U_{\mathrm{edge}})$.} 
   \label{fig:ABintervals}
\end{figure}

We now explain how to construct an edge invariant for 2D unitary loops given any flow $\Omega_{A,B}(U)$. As usual, our invariant $F(U_{\mathrm{edge}})$ is defined on 1D LPUs $U_{\mathrm{edge}}$. However, we will present the definition in the special case where $U_{\mathrm{edge}}$ is a 1D \emph{strict} LPU, because this allows for a simpler and more rigorous analysis.

Our invariant is defined as follows. Given a 1D strict locality preserving unitary $U_{\mathrm{edge}}$ with operator spreading length $\xi$, we choose two overlapping intervals, $A = [a_1, a_2]$, and $B = [b_1, b_2]$ with $a_1 < b_1 < a_2 < b_2$ such that $b_1 - a_1$, $a_2 - b_1$ and $b_2 - a_2$ are larger than $4\xi$ (see Fig.~\ref{fig:ABintervals}). We then define
\begin{equation}\label{flow}
F(U_{\mathrm{edge}})=\Omega_{A,B}(U_{\mathrm{edge}}).
\end{equation}

In order for this definition to be unambiguous, we need to check that $\Omega_{A,B}(U_{\mathrm{edge}})$ does not depend on the choice of $A, B$. Conveniently, this follows immediately from Theorem~\ref{theorem1}.2. Indeed, Theorem~\ref{theorem1}.2 guarantees that we can shift any of the endpoints $a_i \rightarrow a_i \pm 1$ or $b_i \rightarrow b_i \pm 1$, as long as $b_1 - a_1$, $a_2 - b_1$ and $b_2 - a_2$ are larger than $4\xi$. By shifting endpoints using Theorem~\ref{theorem1}.2, we can show that  
\begin{align}
\Omega_{A,B}(U_{\mathrm{edge}}) = \Omega_{A',B'}(U_{\mathrm{edge}}),
\end{align}
for any other choice of $A' = [a_1', a_2']$ and $B' = [b_1', b_2']$ obeying the constraint that $b_1' - a_1'$, $a_2' - b_1'$ and $b_2' - a_2'$ are larger than $4\xi$. 

To complete the discussion, we need to check that $F(U_{\mathrm{edge}})$ is a true edge invariant, i.e. $F(W U_{\mathrm{edge}}) = F( U_{\mathrm{edge}})$ for any $G$-symmetric locally generated unitary $W$. For simplicity, we will check this invariance in the case where $W$ is a $G$-symmetric FDLU. More specifically, suppose $W$ is an FDLU of depth $n$ built out of gates of radius $\lambda$. We wish to show that $F(W U_{\mathrm{edge}}) = F(U_{\mathrm{edge}})$. To prove this, we first choose $A, B$ so that $b_1 - a_1$, $a_2 - b_1$ and $b_2 - a_2$ are larger than $4(n\lambda+\xi)$, because the operator spreading length of $WU_{\mathrm{edge}}$ is $n\lambda+\xi$. The desired identity, $\Omega_{A,B}(WU_{\mathrm{edge}})=\Omega_{A,B}(U_{\mathrm{edge}})$, then follows from Corollary \ref{corollary1}.

A general propery of the above edge invariant (\ref{flow}) that is worth mentioning is that it is \emph{odd} under spatial reflections. That is,
\begin{equation}\label{antisymmetryab}
\Omega_{A,B}(U)=-\Omega_{B,A}(U)
\end{equation}
for any overlapping intervals $A, B$ with the geometry discussed above. In other words, switching the direction that we call ``positive" switches the sign of the edge invariant. We prove this result in Corollary~\ref{antisymmetric} in Appendix~\ref{slemmas} using general properties of flows. (Note that the above anti-symmetry property does not apply to general subsets $A, B \subset \Lambda$ -- only to the specific case of overlapping intervals in 1D).

\subsection{Bulk invariants from flows}\label{sflowbulk}

\begin{figure}[tb]
   \centering
   \includegraphics[width=.45\columnwidth]{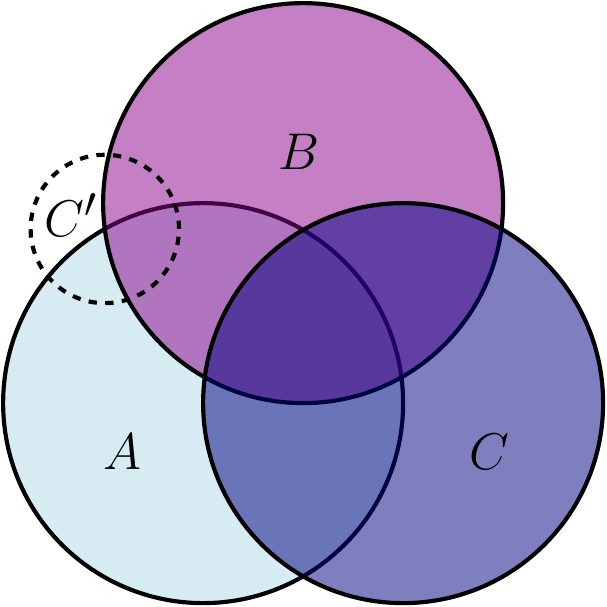} 
   \caption{We use three overlapping disk-like regions $A, B, C$ to define our bulk invariant $M(\{U(t)\})=\Omega_{A,B}^C(\{U(t)\})$. The boundaries of $A$ and $B$ intersect at two points: one in region $C$ and one in another region $C'$.}
   \label{fig:ABC}
\end{figure}

For any flow $\Omega_{A,B}(U)$, we can also construct a corresponding \emph{bulk} invariant for 2D unitary loops. This bulk invariant, denoted $M(\{U(t)\})$, is defined as follows. Let $U(t) = \mathcal{T} \exp\left[-i \int_0^t dt' H(t')\right]$ be a 2D unitary loop. We choose three overlapping disk-like regions $A, B, C$ as illustrated in Fig.~\ref{fig:ABC}. These disks must be large enough that all distances are much larger than the ``Lieb-Robinson length'' $\ell$ of $\{U(t)\}$ defined by $\ell = v_{LR} T$, where $v_{LR}$ is the Lieb-Robinson velocity associated with $H(t)$.

We define the bulk invariant $M(\{U(t)\})$ by
\begin{equation}\label{mag}
M(\{U(t)\})=\Omega_{A,B}^C(\{U(t)\}).
\end{equation}
where 
\begin{equation}\label{omegaabc}
\Omega_{A,B}^C(\{U(t)\})= \int_0^Tdt\frac{\partial}{\partial t_C}\Omega_{A,B}(U(t)),
\end{equation}
Here, we define the operation ``$\frac{\partial}{\partial t_C}$'' as follows. For any function $G[U(t)]$, 
\begin{align}
\frac{\partial}{\partial t_C} G[U(t)]  = \lim_{\epsilon \rightarrow 0} \frac{G\left[ e^{-i \epsilon H_C(t)} \cdot U(t) \right] - G[U(t)]}{\epsilon},
\end{align}
where $H_C(t)$ consists of all the terms in $H(t) = \sum_r H_r(t)$ (\ref{Hsum}) that are supported in region $C$:
\begin{align}
H_C(t) = \sum_{r \in C} H_r(t).
\end{align}

In explicit examples of $\Omega_{A,B}^C(\{U(t)\})$, we will see that the operation $\frac{\partial}{\partial t_C}$ can be implemented in a simple way. This is because the flow $\Omega_{A,B}(U(t))$ can often be expressed in terms of Heisenberg evolved operators $O(t) = U^\dagger(t) O U(t)$. Recall that the usual time derivative of a Heisenberg evolved operator $O(t)$ is given by 
\begin{align*}
\frac{\partial}{\partial t}O(t)=iU^\dagger (t)[H(t),O]U(t). 
\end{align*}
To instead compute $\frac{\partial}{\partial t_C}$, we simply replace $H(t) \rightarrow H_C(t)$ in the commutator, i.e. 
\begin{align*}
\frac{\partial}{\partial t_C}O(t)=iU^\dagger (t)[H_C(t),O]U(t).
\end{align*}

\subsection{Showing that $\Omega_{A,B}^C(\{U(t)\})$ does not depend on choice of $A, B, C$}\label{stopological}
To show that our bulk invariant is well-defined, we need to show that $\Omega_{A,B}^C(\{U(t)\})$ does not depend on the choice of $A, B, C$, as long as they are sufficiently large. We now prove this claim. 

To begin, consider another large disklike region $C'$ that surrounds the other intersection point of $\partial A$ and $\partial B$, which is not in $C$ (see Fig.~\ref{fig:ABC}). Let $U_C(t)$ and $U_{C'}(t)$ be the unitaries generated by $H_C(t)$ and $H_{C'}(t)$, respectively:
\begin{align}
U_C(t) &= \mathcal{T} \exp \left(\int_0^{t} H_C(s) ds \right) \nonumber \\
U_{C'}(t) &= \mathcal{T} \exp \left(\int_0^{t} H_{C'}(s) ds \right).
\end{align}
Below, we will prove the following two identities using the general properties of flows. First, we will show that
\begin{equation}\label{totflow2}
\Omega_{A,B}^{C}(\{U(t)\})+\Omega_{A,B}^{C'}(\{U(t)\}) = 0.
\end{equation}
Second, we will show that
\begin{equation}
\Omega_{A,B}^{C}(\{U(t)\}) = \int_0^T dt \frac{d}{dt} \Omega_{A \cap C, B \cap C}(U_C(t)).
\label{flowC}
\end{equation}
Using these two identities it is easy to see that $\Omega_{A,B}^C(\{U(t)\})$ is independent of the choice of $A, B, C$. Indeed, the fact that $\Omega_{A,B}^C(\{U(t)\})$ doesn't depend on $C$ follows from Eq.~(\ref{totflow2}) since the second term $\Omega_{A,B}^{C'}(\{U(t)\})$ is manifestly independent of $C$ and the two terms sum to zero. Likewise, to see that $\Omega_{A,B}^C(\{U(t)\})$ doesn't depend on $A, B$, notice that (\ref{flowC}) implies that $\Omega_{A,B}^C(\{U(t)\})$ doesn't change if we modify $A, B$ outside of $C$. By the same logic, (\ref{totflow2}) and (\ref{flowC}) together tell us that $\Omega_{A,B}^C(\{U(t)\})$ doesn't change if we modify $A, B$ outside of $C'$. Combining these two observations we see that $\Omega_{A,B}^C(\{U(t)\})$ doesn't change under any modification of $A, B$.

In addition Eq.~(\ref{totflow2}) tells us that $\Omega_{A,B}^{C}(\{U(t)\})$ must be invariant under any deformation of $U(t)$ that is far away from $C'$. It is also invariant under any deformation of $U(t)$ far away from $C$ by definition, so it is invariant under any local deformations of $U(t)$, as long as $C$ and $C'$ are sufficiently far separated.

We now derive the two identities (\ref{totflow2}) and (\ref{flowC}). To begin, we claim that
\begin{align}
\Omega_{A,B}(U(t)) &= \Omega_{A,B}(U_C(t) U_{C'}(t)),
\label{Uccprimeid}
\end{align}
as long as the regions $C, C'$ are sufficiently large. To see this, first suppose that $U(t)$ is an FDLU (rather than an LGU). In that case, Corollary~\ref{corollary2} implies that we can remove all the gates from $U(t)$ except for those near the intersection of the boundaries of $A$ and $B$. In particular this means we can remove all the gates from $U(t)$ except for those supported in $C$ and $C'$, implying Eq.~(\ref{Uccprimeid}) in this case. More generally, for any $U(t)$ that is generated by the time evolution of a local Hamiltonian $H(t)$, we can always \emph{approximate} $U(t)$ by an FDLU with arbitrarily small error. Hence (\ref{Uccprimeid}) must hold up to this error. We expect that this error vanishes exponentially in the separation between $C$ and $C'$, so (\ref{Uccprimeid}) becomes exact in the limit of large $A, B, C$.
 
Having established (\ref{Uccprimeid}), we next observe that property 3 in the definition of a flow (or more precisely, Lemma 1 in Appendix~\ref{slemmas}) guarantees that
\begin{align}
\Omega_{A,B}(U_C(t) U_{C'}(t)) &= \Omega_{A,B}(U_C(t)) + \Omega_{A,B}(U_{C'}(t)).
\end{align}
Combining this equation with (\ref{Uccprimeid}), we deduce that
\begin{align}
\Omega_{A,B}(U(t))  = \Omega_{A,B}(U_C(t)) + \Omega_{A,B}(U_{C'}(t)).
\label{idcc1}
\end{align}
Now consider the quantity $\frac{\partial}{\partial t_C} \Omega_{A,B}(U(t))$. By definition,
\begin{widetext}
\begin{align}
\frac{\partial}{\partial t_C}  \Omega_{A,B}(U(t)) &= 
 \lim_{\epsilon \rightarrow 0} \frac{\Omega_{A,B}( e^{-i \epsilon H_C(t)} \cdot U(t)) - \Omega_{A,B}(U(t))}{\epsilon}.
 \end{align}
Substituting the identity (\ref{idcc1}) for $\Omega_{A,B}(U(t))$ and using the analogous identity for $\Omega_{A,B}( e^{-i \epsilon H_C(t)} \cdot U(t))$, we derive
\begin{align}
\frac{\partial}{\partial t_C}  \Omega_{A,B}(U(t)) &=  \lim_{\epsilon \rightarrow 0} \frac{\Omega_{A,B}( e^{-i \epsilon H_C(t)} \cdot U_C(t)) + \Omega_{A,B}(U_{C'}(t))  - \Omega_{A,B}(U_C(t)) - \Omega_{A,B}(U_{C'}(t))}{\epsilon} \nonumber \\
&=  \lim_{\epsilon \rightarrow 0} \frac{\Omega_{A,B}( e^{-i \epsilon H_C(t)} \cdot U_C(t)) - \Omega_{A,B}(U_C(t))}{\epsilon} \nonumber \\
&= \frac{d}{dt} \Omega_{A,B}(U_C(t)).
\label{idc1}
\end{align}
\end{widetext}
Likewise
\begin{align}
\frac{\partial}{\partial t_{C'}}  \Omega_{A,B}(U(t)) &= \frac{d}{dt} \Omega_{A,B}(U_{C'}(t)).
\label{idcprime1}
\end{align}
Comparing (\ref{idcc1}) with (\ref{idc1}) and (\ref{idcprime1}), we deduce that
\begin{align}\label{domega}
\frac{d}{dt}\Omega_{A,B}(U(t)) = \frac{\partial}{ \partial t_C}\Omega_{A,B}(U(t)) + \frac{\partial }{\partial t_{C'}}\Omega_{A,B}(U(t)).
\end{align}
Integrating both sides from time $t = 0$ to $t = T$, we obtain
\begin{align}
\Omega_{A,B}^{C}(\{U(t)\})+\Omega_{A,B}^{C'}(\{U(t)\}) &= \int_0^Tdt\frac{d}{dt}\Omega_{A,B}(U(t)) \nonumber \\
&=0,
\end{align}
where the last equality follows from the fact that $U(T) = U(0) = \mathbbm{1}$. This proves (\ref{totflow2}). 

To prove (\ref{flowC}), we integrate (\ref{idc1}) from $t = 0$ to $t = T$ to obtain
\begin{align}
\Omega_{A,B}^{C}(\{U(t)\}) = \int_0^T dt \frac{d}{dt} \Omega_{A, B}(U_C(t)).
\end{align}
We then note that $\Omega_{A, B}(U_C(t)) = \Omega_{A \cap C, B \cap C}(U_C(t))$ for any flow: this again follows from property 3 in the definition of the flow, since $U_C(t)$ acts trivially outside of $C$. Eq.~(\ref{flowC}) follows immediately. 


Before concluding this section, it is worth noting that the bulk invariant (\ref{mag}) is odd under spatial reflections, just like the edge invariant. That is,
\begin{equation}\label{parityperm}
\Omega_{\sigma(A),\sigma(B)}^{\sigma(C)}(\{U(t)\}) =\mathrm{sgn}(\sigma) \Omega_{A,B}^C(\{U(t)\})
\end{equation}
where $A, B, C$ are three overlapping disk-like regions with the geometry of Fig.~\ref{fig:ABC}, and where $\sigma$ is a permutation of $A,B,C$ and $\mathrm{sgn}(\sigma)$ is the parity of $\sigma$. Eq.~(\ref{parityperm}) follows the corresponding property of the edge invariant (\ref{antisymmetryab}) together with the bulk-boundary correspondindence that we prove in the next section.

\subsection{Bulk-boundary correspondence}\label{sbulkboundarycorr}
We will now prove the bulk-boundary correspondence that we claimed earlier:
\begin{align}
F(U_{\mathrm{edge}})=M(\{U(t)\}).
\end{align}
Here $F(U_{\mathrm{edge}})$ is the edge invariant defined in (\ref{flow}) and $M(\{U(t)\})$ in the bulk invariant defined in (\ref{mag}), and $U_{\mathrm{edge}}$ is related to $U(t)$ via (\ref{Uedgedef}).

To this end, we note that (\ref{flowC}) implies that
\begin{align}
M(\{U(t)\}) &= \Omega_{A,B}^C(\{U(t)\}) \nonumber \\
&=  \int_0^T \frac{d}{dt}  \Omega_{A \cap C,B \cap C}(U_C(t))  dt \nonumber \\
&= \Omega_{A \cap C,B \cap C}(U_C(T)).
\label{Momegaabc}
\end{align}

Next, note that $U_C(T)=U_{\mathrm{edge}}$ is supported in the 1D circle $\partial_{\xi} C$, and the subsets of $A$ and $B$ that $U_{\mathrm{edge}}$ acts on are the intersections of $A$ and $B$ with $\partial_{\xi} C$, which form two overlapping intervals, like our setup for $F(U_{\mathrm{edge}})$ (Fig.~\ref{fig:ABintervals}). Therefore,
\begin{equation}\label{boundarybulkboundary}
\Omega_{A \cap C,B \cap C}(U_C(T)) = F(U_{\mathrm{edge}}).
\end{equation}

Putting together (\ref{Momegaabc}), (\ref{boundarybulkboundary}), we obtain the desired result $M(\{U(t)\})=F(U_{\mathrm{edge}})$.

\section{Spatially additive flows}\label{sadditive}

We say that a flow $\Omega_{A,B}(U)$ is ``spatially additive'' if it obeys
\begin{align}\label{additive}
\Omega_{A\cup B,C}(U)&= \Omega_{A,C}(U)+ \Omega_{B,C}(U) \nonumber \\
\Omega_{A,B\cup C}(U)&=\Omega_{A,B}(U)+ \Omega_{A,C}(U).
\end{align}
where in the first line, $A$ and $B$ are two \emph{disjoint} sets of lattice sites and in the second line, $B$ and $C$ are similarly two disjoint sets of lattice sites. Equivalently, a flow is spatially additive if it can be written as a sum of the form:
\begin{equation}\label{sumsites}
\Omega_{A,B}(U)=\sum_{a\in A}\sum_{b\in B}\Omega_{a,b}(U).
\end{equation}
Note that $\Omega_{a,b}(\{U(t)\})$ must vanish when the indices $a,b$ are far apart, in order to be consistent with Theorem 1.2.

A nice property of spatially additive flows is that we can write down alternative expressions for the edge invariant $F(U_{\mathrm{edge}})$ and the bulk invariant $M(\{U(t)\})$ that are based on a non-overlapping geometry. In particular, the formula for $F(U_{\mathrm{edge}})$ is
\begin{equation}\label{flownonoverlap}
F(U_{\mathrm{edge}})=\Omega_{I,J}(U_{\mathrm{edge}})-\Omega_{J,I}(U_{\mathrm{edge}})
\end{equation}
where $I,J$ are two adjacent, \emph{non-overlapping} intervals. Likewise, the formula for $M(\{U(t)\})$ is
\begin{align}\label{nonoverlap}
M(\{U(t)\})&=\Omega_{J,K}^{I}(\{U(t)\}) - \Omega_{K,J}^{I}(\{U(t)\}) \nonumber \\
&+ \Omega_{K,I}^{J}(\{U(t)\}) - \Omega_{I,K}^{J}(\{U(t)\}) \nonumber \\
&+ \Omega_{I,J}^{K}(\{U(t)\}) - \Omega_{J,I}^{K}(\{U(t)\}) 
\end{align}
where $I,J,K$ are three disjoint regions, meeting at a single point, of the form shown in Fig.~\ref{fig:abcnonoverlap}. We derive these formulas and discuss some technical advantages of additive flows in Appendix \ref{snonoverlapderivation}. Note that (\ref{nonoverlap}) is reminiscent of the real space Chern number formula in Ref.~\onlinecite{kitaev}; we make this connection more explicit in Appendix~\ref{sstationary}.

\begin{figure}[tb]
   \centering
   \includegraphics[width=.45\columnwidth]{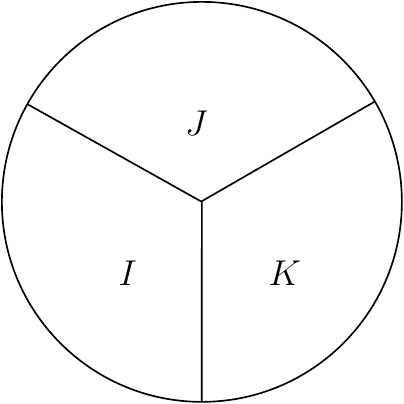} 
   \caption{For spatially additive flows, our bulk invariant can be computed using three non-overlapping adjacent regions, $I, J,$ and $K$ (see Eq.~(\ref{nonoverlap})).}
   \label{fig:abcnonoverlap}
\end{figure}

\section{Single-particle systems}\label{snoninteracting}

We begin by applying our construction to single-particle systems, expanding on the example that we introduced at the beginning of Sec.~\ref{sflows}. 

\subsection{Definition of $F(U_{\mathrm{edge}})$ and $M(\{U(t)\})$}\label{sflownonint}
Our starting point is the single-particle flow $\omega_{A,B}(U)$ given in (\ref{nonintflowU}). We can write this flow in a more convenient way in terms of projection matrices $P_A$ and $P_B$ into the sets $A$ and $B$ (Fig.~\ref{fig:ABintervals}):
\begin{equation}\label{WABnonint}
\omega_{A,B}(U)=\mathrm{Tr}(U^{\dagger}P_AU P_B)-\mathrm{Tr}(P_AP_B).
\end{equation}
Here, $P_A$ is a $|\Lambda|\times|\Lambda|$ diagonal matrix with matrix elements equal to $1$ for the sites in $A$ and $0$ elsewhere, and $P_B$ is defined similarly. As we mentioned earlier, it is easy to see that $\omega_{A,B}(U)$ satisfies the definition of flow (in the single-particle sense). 

Using Eq.~(\ref{flow}), we can construct an edge invariant:
\begin{equation}\label{flownonintedge}
F(U_{\mathrm{edge}})=\mathrm{Tr}(U_{\mathrm{edge}}^{\dagger}P_AU_{\mathrm{edge}}P_B)-\mathrm{Tr}(P_AP_B).
\end{equation}
To get some intuition for this edge invariant, consider the case where $U_{\mathrm{edge}}$ is a translation by $x$: i.e. $U_{\mathrm{edge}}^{\dagger}P_rU_{\mathrm{edge}} = P_{r+x}$ where $P_r$ is the projector $P_r = |r\>\<r|$. Then, $U_{\mathrm{edge}}$ shifts the overlap of $P_A$ and $P_B$ by $x$ so that $F(U_{\mathrm{edge}})=x$.

Moving on to the bulk invariant, Eq.~(\ref{mag}) gives
\begin{align}\label{magnonint0}
M(\{U(t)\})&= \Omega_{A,B}^C(\{U(t)\}) \nonumber \\
&= i \int_0^Tdt \ \mathrm{Tr}\left(U(t)^{\dagger}[H_C(t),P_A]U(t)P_B\right)
\end{align}
To make sense of this invariant, we have to define what we mean by $H_C(t)$ for single-particle Hamiltonians. As in the many-body case, we define $H_C(t) = \sum_{r \in C} H_r(t)$, where $H(t) = \sum_{r} H_r(t)$ is a decomposition of $H$ into local terms supported near $r$. In the single-particle case, there is a natural way to define the local terms $H_r(t)$, namely
\begin{align}
H_r(t) = \frac{1}{2} \{H(t), P_r\}
\end{align}
Again, $P_r$ denotes the projection onto site $r$, and $\{\cdot,\cdot\}$ denotes the anticommutator. By construction $H_r(t)$ is supported in a finite neighborhood around $r$ (assuming $H(t)$ is a finite range Hamiltonian). Substituting this into the definition of $H_C(t)$, we obtain
\begin{align}
H_C(t) = \frac{1}{2} \{H(t), P_C\}
\end{align}
so that our bulk invariant takes the form
\begin{align}\label{magnonint}
M(\{U(t)\}) =\frac{i}{2}\int_0^Tdt \ \mathrm{Tr}\left(U(t)^\dagger[\{H(t), P_C\},P_A]U(t)P_B\right)
\end{align}

\subsection{Relation to previously known invariants}\label{sclassnonint}
We now relate our invariants (\ref{flownonintedge}), (\ref{magnonint}) to previously known edge and bulk invariants for 2D unitary loops, discussed in Refs.~\onlinecite{kitagawa2010,anomalousedge}. 
We start with the edge invariant of Refs.~\onlinecite{kitagawa2010,anomalousedge}, which applies to translationally invariant systems. It is given by the momentum space formula
\begin{equation}\label{nedge}
n(U_{\mathrm{edge}})=-\frac{i}{2\pi}\int dk \mathrm{Tr}\left(U_{\mathrm{edge}}^\dagger\frac{\partial}{\partial k}U_{\mathrm{edge}}\right).
\end{equation}

We claim that our edge invariant $F(U_{\mathrm{edge}})$ (\ref{flownonintedge}) is equivalent to $n(U_{\mathrm{edge}})$ in the translationally invariant case, i.e.
\begin{align}
F(U_{\mathrm{edge}}) = n(U_{\mathrm{edge}})
\end{align}
To show this, we make a particular choice for the two overlapping intervals $A, B$ in the definition of $F(U_{\mathrm{edge}})$ (\ref{flownonintedge}). Specifically, we choose $A=(-\infty,0]$ and $B = [-L, \infty)$ where $L$ is a large positive number which we will send to $\infty$. For this choice of $A$ and $B$, Eq.~(\ref{flownonintedge}) reduces to
\begin{align}
F(U_{\mathrm{edge}})&= \lim_{L\rightarrow \infty} \mathrm{Tr}(U_{\mathrm{edge}}^\dagger P_{(-\infty,0]}U_{\mathrm{edge}}P_{[-L, \infty)}- P_{[-L,0]}) \nonumber \\
 &= \mathrm{Tr}(U_{\mathrm{edge}}^{\dagger} P_{(-\infty,0]} U_{\mathrm{edge}}- P_{(-\infty,0]}) \nonumber \\
 &= \mathrm{Tr}\left(U_{\mathrm{edge}}^{\dagger}\left[P_{(-\infty,0]}, U_{\mathrm{edge}}\right]\right)
\end{align}

We note that the above formula is exactly the expression for the flow $\mathcal{F}(U_{\mathrm{edge}})$ given in Eq. (112) of Ref.~\onlinecite{kitaev}, except with a projector onto $(-\infty,0]$ rather than $[0,\infty)$. To proceed further, one can use the argument given in Appendix C.1.3 of Ref.~\onlinecite{kitaev} to rewrite this expression in $k$-space. As explained in Ref.~\onlinecite{kitaev}, when we go to $k$-space, the real space trace is replaced by an integral over a $k$-space trace:
\begin{align}\label{ktrace}
\mathrm{Tr}(\cdot) \rightarrow \frac{1}{2\pi} \int d k \mathrm{Tr}(\cdot)
\end{align}
while the commutator is replaced by a derivative
\begin{align}\label{kderiv}
\left[P_{(-\infty,0]}, U_{\mathrm{edge}}\right] \rightarrow -i\frac{\partial U_{\mathrm{edge}}}{\partial k}
\end{align}
Note that there is an extra minus sign because we use the projector onto sites $(-\infty,0]$ rather than $[0,\infty)$. Making these replacements, we recover the previously known formula (\ref{nedge}).

Next, consider the bulk invariant $\mathcal{W}(\{U(t)\})$ of Ref.~\onlinecite{anomalousedge}, which is given by the momentum space formula
\begin{align}
\begin{split}\label{wkspace}
\mathcal{W}(\{U(t)\})&=\frac{1}{8\pi^2}\int dtdk_xdk_y\\
&\times\mathrm{Tr}\left(U^\dagger \frac{\partial}{\partial t} U\left[U^\dagger \frac{\partial}{\partial k_x} U,U^\dagger \frac{\partial}{\partial k_y} U\right]\right),
\end{split}
\end{align}
where we've dropped the $t$ dependence from $U(t)$ for brevity. 

We claim that our bulk invariant $M(\{U(t)\})$ (\ref{magnonint}) is equivalent to $\mathcal{W}(\{U(t)\})$:
\begin{align}
M(\{U(t)\}) = \mathcal{W}(\{U(t)\}).
\end{align}
To see this, we make a particular choice for the three regions $A, B, C$ in the definition of $M(\{U(t)\})$ (\ref{magnonint}). Specifically, we choose $A$ to be the left half plane $X_-$, $B$ to be the upper half plane $Y_+$, and $C$ to be a disk $D_L$ centered at the origin with a radius $L$, where $L$ is a large number which we will send to infinity. With these choices, (\ref{magnonint}) reduces to
\begin{align}
M(\{U(t)\})&= \lim_{L \rightarrow \infty} \frac{i}{2}\int_0^Tdt \nonumber \\
& \times \mathrm{Tr}\left( U^\dagger [\{H(t), P_{D_L}\}, P_{X_-} ]UP_{Y_+}\right) \nonumber \\
&= \lim_{L \rightarrow \infty} \frac{i}{2}\int_0^Tdt \nonumber \\
&\times \mathrm{Tr}\left( U^\dagger \{H(t), P_{D_L}\} U [U^\dagger P_{X_-}U, P_{Y_+}]\right) \nonumber \\
&=  i \int_0^Tdt \mathrm{Tr}\left( U^\dagger H(t) U  [U^\dagger P_{X_-}U,P_{Y_+} ]\right) 
\end{align}
where the second equality follows from the cyclicity of the trace. To proceed further, we replace $U^\dagger P_{X_-}U \rightarrow U^\dagger [P_{X_-}, U]$ in the above expression. Once can then use the same method as in Appendix C.1.3 of Ref.~\onlinecite{kitaev} to rewrite this expression in $k$-space, replacing the commutators with derivatives as in Eq.~(\ref{kderiv}). The result is
\begin{align}
\begin{split}
M(\{U(t)\})&=\frac{1}{4\pi^2}\int dtdk_xdk_y\\
&\times\mathrm{Tr}\left(U^\dagger\frac{\partial}{\partial t}U \frac{\partial}{\partial k_y} \left[U^\dagger\frac{\partial}{\partial k_x}U\right]\right),
\end{split}
\end{align}
One can then massage this expression into the form (\ref{wkspace}) by adding the following derivative term to the
integrand:
\begin{align}
-\frac{1}{2} \partial_{k_y} \mathrm{Tr}( U^\dagger \partial_t U U^\dagger \partial_{k_x} U ) 
&+ \frac{1}{2} \partial_{t} \mathrm{Tr}(U^\dagger \partial_{k_y} \partial_{k_x} U ) \nonumber \\
&-\frac{1}{2} \partial_{k_x} \mathrm{Tr}(U^\dagger \partial_{k_y} \partial_{t} U  )
\end{align}

\subsection{Relation to current}\label{scurrent}
In this section, we relate the bulk invariant $M(\{U(t)\})$ to the current -- a more familiar physical quantity. We do this in two different ways. First, we express $M(\{U(t)\})$ in terms of circulating bulk currents, which are related to the quantized orbital magnetization density described in Ref.~\onlinecite{magnetization}. Second, we relate $M(\{U(t)\})$ to the quantized current that flows between a fully filled region and an empty region in a non-interacting fermion system, which was also described in Ref.~\onlinecite{magnetization}.

We begin by deriving the circulating current formula. Our derivation starts with the non-overlapping formula for $M(\{U(t)\})$, given in Eq.~(\ref{nonoverlap}). This formula consists of a sum of six terms of the form $\Omega^{K}_{I,J}(\{U(t)\})$. Using the explicit formula for $\Omega^{K}_{I,J}(\{U(t)\})$ (\ref{magnonint}), together with the fact that $I, J, K$ are non-overlapping, we can expand each of these terms as
\begin{align}
&\Omega^K_{I,J}(\{U(t)\}) = \nonumber \\
&\frac{i}{2} \int_0^Tdt \ \mathrm{Tr}\left( U(t)^\dagger(P_K H(t)P_I - P_I H(t) P_K) U(t)P_J \right).
\label{omegaijksp}
\end{align}

Recall that, according to the standard definition of the current operator, the (Heisenberg-evolved) current operator from site $k$ to $i$ is given by
\begin{align}
\mathcal{I}_{ki}(t) = i U^\dagger(t) (P_k H(t) P_{i} - P_{i} H(t) P_k) U(t).
\end{align}

Comparing this definition with (\ref{omegaijksp}) we see that
\begin{align}\label{omegacurrent}
\Omega^K_{I,J}(\{U(t)\}) = \frac{1}{2} \int_0^T dt \ \mathcal{J}_{K,I}^J(t),
\end{align}
where
\begin{align}
\mathcal{J}_{K,I}^{J}(t) = \sum_{i \in I} \sum_{j \in J} \sum_{k \in K} \<j|\mathcal{I}_{ki}(t)|j\>.
\end{align}
and where $|j\>$ denotes the single-particle state where the particle is on site $j$. Substituting this expression into the nonoverlapping formula for $M(\{U(t)\})$ (\ref{nonoverlap}) and using the the observation that $\Omega_{I,J}^K=-\Omega_{K,J}^I$ (or equivalently, that the current is antisymmetric), we derive
\begin{align} \label{magcurrent}
M(\{U(t)\}) =  \int_0^Tdt \ [\mathcal{J}_{K,I}^{J}(t)+\mathcal{J}_{J,K}^{I}(t)+\mathcal{J}_{I,J}^{K}(t)].
\end{align}
 
The above formula for $M(\{U(t)\})$ has a nice intuitive picture: $M(\{U(t)\})$ is given by the time integral of the expectation value of current across the $K,I$ boundary in states initially in region $J$ (together the cyclic permutations). Therefore, it measures the cyclic micromotion of localized bulk states. We show in Appendix \ref{sstationary} that if $U(t)=\mathrm{exp}(-i2\pi Pt/T)$ for a time-independent Chern band projector $P$, we can perform the time integral explicitly to obtain the real space formula for the Chern number given in Ref.~\onlinecite{kitaev}.
\begin{figure}[tb]
   \centering
   \includegraphics[width=0.8\columnwidth]{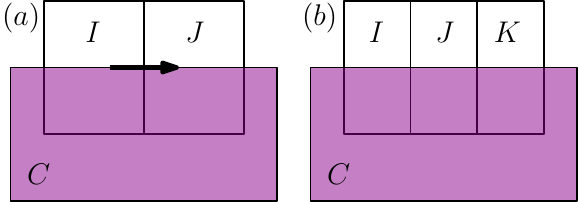} 
   \caption{(a) A quantized current flows from $I$ to $J$ at the boundary of a fully filled region $C$. (b) To show that this quantized current is equal to $\frac{M(\{U(t)\})}{T}$, we use a topologically equivalent setup to our overlapping geometry (Fig.~\ref{fig:ABC}) with $A = I \cup J$ and $B = J \cup K$.}
   \label{fig:currentbdry}
\end{figure}

Next, we relate $M(\{U(t)\})$ to the quantized current that flows at the boundary of a fully filled region and an empty region, as explained in Ref.~\onlinecite{magnetization}. This current is defined as
\begin{equation}\label{currentfilling}
\mathcal{I}(\{U(t)\})=\frac{1}{T}\int_0^Tdt \mathcal{J}_{I,J}^C(t)
\end{equation}
where $I$ and $J$ are finite, adjacent regions as illustrated in Fig.~\ref{fig:currentbdry}a and $C$ is a large region that overlaps with both $I$ and $J$. 

We begin by defining three adjacent regions $I,J,$ and $K$ as illustrated in Fig.~\ref{fig:currentbdry}b. We can connect this setup to our overlapping geometry (Fig.~\ref{fig:ABC}) by defining $A=I\cup J$ and $B=J\cup K$. According to (\ref{parityperm}), $\Omega_{A,B}^C(\{U(t)\}) = \Omega_{B,C}^A(\{U(t)\})=-\Omega_{A,C}^B(\{U(t)\})$, so 
\begin{equation}
M(\{U(t)\})=\frac{1}{2}\left[\Omega_{B,C}^A(\{U(t)\})-\Omega_{A,C}^B(\{U(t)\})\right]
\end{equation}

Because the flow is spatially additive in this case, we have (dropping the argument $(\{U(t)\})$ for clarity of notation)
\begin{align}
\begin{split}
M(\{U(t)\})&=\frac{1}{2}\left(\Omega_{J,C}^{I}+\Omega_{K,C}^{I}+\Omega_{J,C}^{J}+\Omega_{K,C}^{J}\right)\\
&-\frac{1}{2}\left(\Omega_{I,C}^{J}+\Omega_{J,C}^{J}+\Omega_{I,C}^{K}+\Omega_{J,C}^{K}\right)
\end{split}
\end{align}

Simplifying by canceling the $\Omega^J_{J,C}$ terms and using $\Omega^I_{K,C}=\Omega^K_{I,C}=0$ (because $I$ is far separated from $K$), we have
\begin{equation}\label{mcurrent}
M(\{U(t)\})=\frac{1}{2}\left(\Omega_{J,C}^{I}+\Omega_{K,C}^{J}-\Omega_{I,C}^{J}-\Omega_{J,C}^{K}\right)
\end{equation}
We can now write (\ref{mcurrent}) in terms of currents:
\begin{equation}\label{mcurrent2}
M(\{U(t)\})=\frac{1}{2}\int_0^Tdt\mathcal{J}^C_{I,J}(t)+\mathcal{J}^C_{J,K}(t)
\end{equation}

Finally, we claim that $\mathcal{J}^C_{J,K}(t)=\mathcal{J}^C_{I,J}(t)$. Intuitively, this is true because all the quantities above are topological and do not depend on the choice of location in the lattice. More rigorously, $\mathcal{J}^C_{J,K}(t)+\mathcal{J}^C_{J,I}+\mathcal{J}^C_{J,J}(t)+\mathcal{J}^C_{J,\Lambda\setminus(I\cup J\cup K)}(t)=0$ by current conservation. Also, $\mathcal{J}^C_{J,J}(t)=0$, and $\mathcal{J}^C_{J,\Lambda\setminus(I\cup J\cup K)}(t) = 0$ where the second current vanishes because there is no current flowing through the top and bottom edges of $J$. This means that $\mathcal{J}^C_{J,K}(t)+\mathcal{J}^C_{J,I}(t)=0$ so $\mathcal{J}^C_{J,K}(t)=-\mathcal{J}^C_{J,I}(t)=\mathcal{J}^C_{I,J}(t)$. In conclusion,
\begin{equation}
M(\{U(t)\})=\int_0^Tdt\mathcal{J}^C_{I,J}(t)
\end{equation}
Putting this together with Eq.~(\ref{currentfilling}), we obtain
\begin{equation}
\mathcal{I}(\{U(t)\})=\frac{M(\{U(t)\})}{T}
\end{equation}
This is the desired formula relating $M$ to the quantized current $\mathcal{I}(\{U(t)\})$.
\section{Interacting systems with $U(1)$ symmetry}\label{su1}
We now apply our methods to interacting systems with $U(1)$ symmetry, expanding on the example from Sec.~\ref{sexamples}. Many of our results closely parallel the single-particle case discussed above.
%

\subsection{Definition of $F(U_{\mathrm{edge}})$ and $M(\{U(t)\})$}\label{sflowu1}
Our basic setup is the same as the example discussed in Sec.~\ref{sexamples}: we consider a 2D lattice with a finite dimensional local Hilbert space on each site, each with an identical on-site charge operator $Q_r$ that has nonnegative integer eigenvalues. We assume that the Hamiltonian $H(t)$ conserves the total $U(1)$ charge $Q=\sum_{r}Q_r$. Our task is to construct bulk and edge invariants for unitary loops of this kind. 

Our starting point is the flow given in (\ref{chargeomega}):
\begin{equation}\label{flowu10}
\Omega_{A,B}(U)=\langle U^\dagger Q_AUQ_B\rangle_\rho-\langle Q_AQ_B\rangle_\rho,
\end{equation}
Here the expectation value $\<\cdot\>_\rho$ is taken in the mixed state
\begin{equation}\label{rho}
\rho=\frac{e^{\mu Q}}{Z}\qquad Z=\mathrm{Tr}e^{\mu Q},
\end{equation}
where $\mu$ is a real-valued ``chemical potential'' and $Q_A = \sum_{r \in A} Q_r$ and $Q_B = \sum_{r \in B} Q_r$ denote the total charge in regions $A, B$.

To construct an edge invariant, we subsitute this flow into (\ref{flow}), which gives
\begin{equation}\label{flowu1}
F(U_{\mathrm{edge}})=\langle U_{\mathrm{edge}}^\dagger Q_AU_{\mathrm{edge}}Q_B\rangle_\rho-\langle Q_AQ_B\rangle_\rho,
\end{equation}
where $A$ and $B$ are overlapping intervals (Fig.~\ref{fig:ABintervals}). 

Likewise, we can obtain a bulk invariant by substituting this flow into Eq.~(\ref{mag}):
\begin{align}
M(\{U(t)\})&= \Omega_{A,B}^C(\{U(t)\}) \nonumber \\
&=i\int_0^Tdt\langle U(t)^\dagger[H_{C}(t),Q_{A}] U(t)Q_{B}\rangle_\rho.
\label{mformu1int}
\end{align}

%
%
%
%
\subsection{Relation to previously known invariants}\label{sclassu1}
We begin by discussing the connection between (\ref{flowu1}) and the edge invariant of Ref.~\onlinecite{u1floquet}. The latter invariant takes values in the set of rational functions of a formal parameter $z$, and is denoted by $\tilde{\pi}(z)$. To define $\tilde{\pi}(z)$, let $A$ be a large interval and let $Q_A = \sum_{r \in A} Q_r$. Consider the action of the edge unitary $U_{\mathrm{edge}}$ on $Q_A$. Since $U_{\mathrm{edge}}$ is a $U(1)$ symmetric LPU, we know that
\begin{equation}\label{UQAU}
U_{\mathrm{edge}}^{\dagger} Q_A U_{\mathrm{edge}}= Q_A +O_L+O_R,
\end{equation}
where $O_L$ and $O_R$ are local operators acting near the left and right endpoints of $A$. Next, we write $Q_A = Q_L + Q_R$, where $Q_L, Q_R$ are the total charges within the left and right half of the interval, for some partition of the interval into two subintervals. The invariant $\tilde{\pi}(z)$ is then defined as
\begin{equation}
\tilde{\pi}(z)=\frac{\mathrm{Tr}\left(z^{Q_R+O_R}\right)}{\mathrm{Tr}\left(z^{Q_R}\right)},
\end{equation}

What is the relationship between $F(U_{\mathrm{edge}})$ and $\tilde{\pi}(z)$? Below we will show that
\begin{equation}\label{flowpiz}
F(U_{\mathrm{edge}})=\frac{d^2}{d\mu^2}\log\tilde{\pi}(e^{\mu})
\end{equation}
Note that this formula makes sense on general grounds, since $F(U_{\mathrm{edge}})$ and $\frac{d^2}{d\mu^2}\log\tilde{\pi}(e^{\mu})$ are both additive under tensor product, and they both have the same units (namely the units of $Q^2$). In fact, one might have been able to guess this formula based on these considerations.

We now derive the above relation (\ref{flowpiz}). Substituting (\ref{UQAU}) into the expression for the edge invariant (\ref{flowu1}) gives
\begin{equation}
F(U_{\mathrm{edge}})=\< O_RQ_B\>_\rho+\< O_LQ_B\>_\rho.
\end{equation}
To simplify this further, we note that the correlation function $\<O_L Q_B\>_\rho$  can be factored as
\begin{align}
\<O_L Q_B\>_\rho = \<O_L\>_\rho\<Q_B\>_\rho
\end{align}
since $O_L$ and $Q_B$ are supported in nonoverlapping regions and $\rho$ has vanishing correlation length. At the same time, we can see that $\<O_L\>_\rho = -\<O_R\>_\rho$
by taking expectation values off both sides of (\ref{UQAU}) above. Putting this together, we derive
\begin{align}\label{ORQB}
F(U_{\mathrm{edge}})=\< O_RQ_B\>_\rho-\<O_R\>_\rho\<Q_B\>_\rho
\end{align}
The next step is to use the factorization property again to deduce that $\<O_R Q_{\overline{B}}\>_\rho = \<O_R\>_\rho \<Q_{\overline{B}}\>_\rho$ where $\overline{B}$ denotes the complement of $B$. Therefore we are free to add $\<O_R Q_{\overline{B}}\>_\rho - \<O_R\>_\rho \<Q_{\overline{B}}\>_\rho$ to the right hand side of (\ref{ORQB}), which gives
\begin{align}
F(U_{\mathrm{edge}})=\< O_RQ\>_\rho-\<O_R\>_\rho\<Q\>_\rho
\end{align}
where $Q$ is the total charge. To complete the derivation, we rewrite the right hand side as
\begin{align}
F(U_{\mathrm{edge}}) &=\frac{d}{d\mu}\< O_R\>_\rho \nonumber \\
&=\frac{d^2}{d\mu^2}\log\tilde{\pi}(e^\mu).
\end{align}
where the second line follows from the identity $\< O_R\>=\frac{d}{d\mu}\log\tilde{\pi}(e^{\mu})$ derived in Ref.~\onlinecite{u1floquet}.


As for the bulk invariant (\ref{mformu1int}), there is nothing to compare it to: we are not aware of any other bulk invariants for strongly interacting Floquet systems with $U(1)$ symmetry.

\subsection{Relation to current}\label{scurrentu1}
In this section we discuss how to express the bulk invariant in terms of $U(1)$ currents. This discussion parallels the single-particle case (Sec.~\ref{scurrent}). As in that section, we derive two different expressions for $M(\{U(t)\})$: one in terms of circulating currents and one in terms of a $U(1)$ current that flows at the boundary between two regions at different chemical potentials\cite{u1floquet}. 

We begin by deriving a formula for $M(\{U(t)\})$ in terms of circulating currents. The first step is to define the Heisenberg evolved $U(1)$ current operator $\mathcal{I}_{ki}(t)$. We use the following definition:
\begin{align}\label{currentmb}
\mathcal{I}_{ki}(t) =i (U^\dagger(t) [H_{k}(t),Q_{i}]U(t) - U^\dagger(t)[H_{i}(t),Q_{k}]U(t)).
\end{align}
Note that this is a reasonable definition since $\mathcal{I}_{ki} = - \mathcal{I}_{ik}$ and $\sum_{i} \mathcal{I}_{ki}(t) = -\frac{d Q_k}{dt}$. 

Next, consider the expression (\ref{mformu1int}) for $\Omega^C_{A,B}$, and set $A = I$, $B = J$ and $C = K$ where $I, J, K$ are non-overlapping regions with the geometry shown in Fig.~\ref{fig:abcnonoverlap}. Comparing this expression with (\ref{currentmb}), we see that 
\begin{align}
\Omega^K_{I,J}(\{U(t)\}) - \Omega^I_{K,J}(\{U(t)\})  = \int_0^T dt \ \mathcal{J}_{K,I}^J(t)
\label{omegakijnonoverlap}
\end{align}
where $\mathcal{J}_{K,I}^{J}(t)$ is now given by 
\begin{align}
\mathcal{J}_{K,I}^{J}(t) = \sum_{i \in I} \sum_{j \in J} \sum_{k \in K} \<\mathcal{I}_{ki}(t) Q_j\>_\rho
\label{jkijinter}
\end{align}
Substituting (\ref{omegakijnonoverlap}) into the non-overlapping formula for $M(\{U(t)\})$ (\ref{nonoverlap}), we arrive at an expression for $M(\{U(t)\})$ which looks just like the single-particle case (\ref{magcurrent}):
\begin{align*}
M(\{U(t)\}) =  \int_0^Tdt \ [\mathcal{J}_{K,I}^{J}(t)+\mathcal{J}_{J,K}^{I}(t)+\mathcal{J}_{I,J}^{K}(t)].
\end{align*}
The only difference from (\ref{magcurrent}) is that $\mathcal{J}_{K,I}^{J}(t)$ now given by Eq.~(\ref{jkijinter}). This is our desired formula for $M(\{U(t)\})$ in terms of circulating currents.

We now move on to our second formula for $M(\{U(t)\})$. Again, this formula looks identical to the single-particle case:
\begin{equation}
M(\{U(t)\})=\int_0^Tdt \ \mathcal{J}^C_{I,J}(t)
\label{bulkcurrbd}
\end{equation}
where $I, J, C$ are three regions with the geometry shown in Fig.~\ref{fig:currentbdry}. The derivation of this formula is also the same as the single-particle case (see Sec.~\ref{scurrent}), but the physical interpretation of this formula is different. To understand this interpretation, let $\mu$ and $\mu'$ be two real numbers and consider a mixed state $\sigma(\mu, \mu')$ of the form
\begin{equation}
\sigma(\mu, \mu') =\frac{e^{\sum_r\mu_rQ_r}}{Z}\qquad Z=\mathrm{Tr}\left(e^{\sum_r\mu_rQ_r}\right),
\end{equation}
where 
\begin{align}
\mu_r = \begin{cases} \mu & \ \text{if } r\in C \\
			\mu'& \ \text{if } r \not \in C
			\end{cases}
\end{align}
We can think of $\sigma(\mu, \mu')$ as describing a state in which $C$ is held at chemical potential $\mu$, while the complement of $C$ is at chemical potential $\mu'$. Previously, Ref.~\onlinecite{u1floquet} argued that if we initialize a Floquet system with Hamiltonian $H(t)$ in such a state, then there will be a time-averaged current $\mathcal{I}$ that flows along the boundary of $C$, and that the size of this current depends only on $\mu$ and $\mu'$. By definition, this current is given by
\begin{align}
\mathcal{I}(\mu, \mu' ) = \frac{1}{T} \int_0^Tdt \sum_{i \in I} \sum_{j \in J} \<\mathcal{I}_{ij}(t)\>_{\sigma(\mu, \mu')}
\label{currmuinout}
\end{align}
where $\mathcal{I}_{ij}(t)$ is defined as in (\ref{currentmb}). We will now show that there is a close connection between this current $\mathcal{I}$ and the right hand side of (\ref{bulkcurrbd}), namely
\begin{align}
\frac{\partial}{\partial \mu} \mathcal{I}(\mu, \mu' )|_{\mu' = \mu} = \frac{1}{T}\int_0^Tdt \ \mathcal{J}_{I,J}^{C}(t)
\label{dIdmuJ}
\end{align}
To see this, note that
\begin{align}
\frac{\partial}{\partial \mu} \sigma(\mu, \mu')|_{\mu' = \mu} = (Q_C - \<Q_C\>_{\sigma(\mu, \mu)}) \sigma(\mu, \mu)
\end{align}
Substituting this into (\ref{currmuinout}), and using the fact that $\<\mathcal{I}_{ij}(t)\>_{\sigma(\mu, \mu)} = 0$, gives the desired identity (\ref{dIdmuJ}).

Eq.~(\ref{dIdmuJ}) is interesting because it provides a simple physical interpretation to our bulk invariant $M(\{U(t)\})$: comparing with (\ref{bulkcurrbd}), we see that the bulk invariant $M(\{U(t)\})$ is equal to the derivative $T \frac{\partial}{\partial \mu} \mathcal{I}(\mu, \mu' )|_{\mu' = \mu}$. In other words, $M(\{U(t)\})$ describes the linear response of the current $\mathcal{I}$ to changing the chemical potential $\mu$ (with $\mu'$ fixed).

\subsection{Bulk invariant from flux threading}\label{sflux}
\begin{figure}[tb]
   \centering
   \includegraphics[width=0.8\columnwidth]{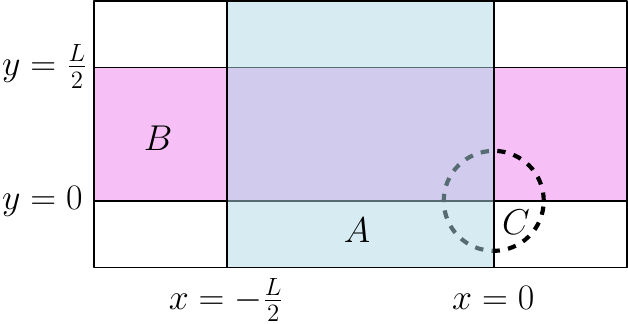} 
   \caption{In the case of $U(1)$ symmetric systems, we can compute our bulk invariant using the above torus geometry (opposite sides of the rectangle are identified). Our construction involves choosing vertical and horizontal strips $A$ and $B$, and then ``twisting'' the Hamiltonian $H(t)$ using the corresponding charge operators $Q_A$ and $Q_B$ (\ref{fluxham}).}
   \label{fig:flux}
\end{figure}
We now derive an expression for $M(\{U(t)\})$ which is based on flux threading through an $L\times L$ torus and which is analogous to the single-particle $k$-space formula (\ref{wkspace}). Specifically, our flux threading formula for $M(\{U(t)\})$ is: 
\begin{align}
M&(\{U(t)\}) \nonumber \\
&= -\frac{1}{2}\int_0^Tdt\left\langle U_f^\dagger \frac{\partial}{\partial t} U_f\left[U_f^\dagger\frac{\partial}{\partial\theta_x}U_f,U_f^\dagger\frac{\partial}{\partial\theta_y}U_f\right]\right\rangle_\rho,
\label{wintkspace}
\end{align}
where $U_f \equiv U_f(t, \theta_x, \theta_y)$ describes the unitary time evolution in the presence of flux $\theta_x$ and $\theta_y$ through the two holes of the torus. That is, $U_f$ is defined by
\begin{align}
U_f(t, \theta_x, \theta_y) &= \mathcal{T}e^{-i\int_0^t dt' H_f(t',\theta_x, \theta_y)}, 
\end{align}
where $H_f(t, \theta_x, \theta_y)$ is given by ``twisting' $H(t)$ by $\theta_x$ and $\theta_y$ across two branch cuts running along $x=0$ and $y=0$. More precisely, to define $H_f(t, \theta_x, \theta_y)$ let $A$ to be the vertical strip $-L/2 \leq x \leq 0$ and $B$ to be the horizontal strip $0 \leq y \leq L/2$ as shown in Fig.~\ref{fig:flux}. Then $H_f(t, \theta_x, \theta_y)$ is defined by
\begin{align}\label{fluxham}
 &H_f(t,\theta_x, \theta_y) = e^{i(\theta_xQ_A + \theta_y Q_B)}H(t)e^{-i(\theta_xQ_A + \theta_y Q_B)} \nonumber \\
&-\left(e^{i(\theta_xQ_A + \theta_y Q_B)}H_{L/2}(t)e^{-i(\theta_xQ_A+ \theta_y Q_B)}-H_{L/2}(t)\right),
\end{align}
where $H_{L/2}(t)$ is the sum of the terms in $H(t)$ with support near the lines at $x = -L/2$ or $y=L/2$. 

We emphasize that Eq.~(\ref{wintkspace}) holds for \emph{any} choice of fluxes $\theta_x, \theta_y$, so in particular, the right hand side is independent of $\theta_x, \theta_y$ (at least in the limit $L \rightarrow \infty$).

We now derive the above formula (\ref{wintkspace}) for $M(\{U(t)\})$. Our derivation proceeds in four steps. First, we claim that 
\begin{align}
M(\{U(t)\})&= \Omega_{A,B}^C(\{U(t)\}),
\label{Mformtorus0}
\end{align}
where $A$ and $B$ are the two regions defined above, $C$ is a disk around $x=y=0$, and $\Omega_{A,B}^C(\{U(t)\})$ is defined as in Eq.~(\ref{mformu1int}). 

This statement is not as obvious as it sounds since $A, B, C$ do not have the usual topology of three overlapping disklike regions. To prove (\ref{Mformtorus0}), it suffices to show that 
$\Omega_{A,B}^C(\{U(t)\}) = \Omega_{\mathcal{A},\mathcal{B}}^C(\{U(t)\})$, where $\mathcal{A}, \mathcal{B}$
are the disklike regions shown in Fig.~\ref{fig:ABC}. Once we show this, then the claim follows immediately, since $\mathcal{A}, \mathcal{B}, C$ have the usual topology and therefore 
$\Omega_{\mathcal{A},\mathcal{B}}^C(\{U(t)\}) = M(\{U(t)\})$. To see why $\Omega_{A,B}^C = \Omega_{\mathcal{A},\mathcal{B}}^C$, note that replacing $Q_B \rightarrow Q_{\mathcal{B}}$ in the integrand of Eq.~(\ref{mformu1int}) amounts to removing a collection of terms of the form $\text{Tr}( [U^\dagger H_C U, U^\dagger Q_A U] Q_i\rho)$, with $i \in B \setminus \mathcal{B}$. One can then check that each of these terms vanishes since (1) each $Q_i$ commutes with $U^\dagger H_C U$ (they have nonoverlapping support); (2) all three of \{$Q_i$, $U^\dagger H_C U$, $U^\dagger Q_A U\}$ commute with $\rho$; and (3) the trace is invariant under cyclic permutations of operators. The same argument explains why we can replace $Q_A \rightarrow Q_{\mathcal{A}}$.

Having establishing (\ref{Mformtorus0}), our next claim is that
\begin{align}
\begin{split}\label{mtheta}
&M(\{U(t)\}\\
&=\frac{-i}{2}\int_0^Tdt\left\langle U^\dagger H_C U\left[U^\dagger[Q_A,U],U^\dagger[Q_B,U]\right]\right\rangle_\rho.
\end{split}
\end{align}
To derive this claim, recall the antisymmetry property of $\Omega$ (\ref{parityperm}) which implies that
\begin{align}
\Omega_{A,B}^C = -\Omega_{B,A}^C.
\end{align}
Given this antisymmetry relation, (\ref{mtheta}) follows directly from (\ref{Mformtorus0}) since the right hand side of (\ref{mtheta}) is exactly the antisymmetrized combination $\frac{1}{2}(\Omega_{A,B}^C - \Omega_{B,A}^C)$. 
%
 
To state our next claim, define 
\begin{equation}
\tilde{U}(t,\theta_x,\theta_y)=e^{i(\theta_xQ_A+\theta_yQ_B)}U(t)e^{-i(\theta_xQ_A + \theta_yQ_B)}.
\end{equation}

We claim that we can replace $U \rightarrow \tilde{U}$ in the right hand side of (\ref{mtheta}): that is,
\begin{align}
\begin{split}\label{mtheta2}
&M(\{U(t)\})\\
&=\frac{-i}{2}\int_0^Tdt\left\langle \tilde{U}^\dagger \tilde{H}_C \tilde{U}\left[\tilde{U}^\dagger[Q_A,\tilde{U}],\tilde{U}^\dagger[Q_B,\tilde{U}]\right]\right\rangle_\rho.
\end{split}
\end{align}
where $\tilde{H}_C = e^{i(\theta_xQ_A + \theta_yQ_B)} H_C e^{-i(\theta_xQ_A + \theta_yQ_B)}$
This identity follows from (\ref{mtheta}) using the fact that the extra factors of $e^{\pm i(\theta_xQ_A + \theta_yQ_B)}$ commute with $Q_A, Q_B$ together with the cyclicity of the trace. In particular, using these two facts, one can commute through the $e^{\pm i(\theta_xQ_A + \theta_yQ_B)}$ terms so that they cancel with one another. 

To complete the derivation we need to show that
\begin{align}
&\frac{-i}{2}\int_0^Tdt\left\langle \tilde{U}^\dagger \tilde{H}_C \tilde{U}\left[\tilde{U}^\dagger[Q_A,\tilde{U}],\tilde{U}^\dagger[Q_B,\tilde{U}]\right]\right\rangle_\rho \nonumber\\
&=-\frac{1}{2}\int_0^Tdt\left\langle U_f^\dagger \frac{\partial}{\partial t} U_f\left[U_f^\dagger\frac{\partial}{\partial\theta_x}U_f,U_f^\dagger\frac{\partial}{\partial\theta_y}U_f\right]\right\rangle_\rho.
\end{align}

To this end, notice that $U_f(t,\theta_x,\theta_y)$ can be written as a product of the form
\begin{equation}
U_f(t,\theta_x,\theta_y)=U_{L/2}(t,\theta_x,\theta_y)\tilde{U}(t,\theta_x,\theta_y),
\end{equation}
where $U_{L/2}$ is a unitary operator supported along the two lines $x=-L/2$ and $y=L/2$ on the torus. 
We then have
\begin{align}
\begin{split}
U_f^\dagger\frac{\partial}{\partial\theta_x}U_f&=\tilde{U}^\dagger\frac{\partial}{\partial\theta_x}\tilde{U}+\tilde{U}^\dagger U_{L/2}^\dagger\left(\frac{\partial}{\partial\theta_x}U_{L/2}\right)\tilde{U}\\
&=i\tilde{U}^\dagger\left[Q_A,\tilde{U}\right]+U_{\theta_x,-L/2},
\end{split}
\end{align}
where $U_{\theta_x,-L/2}$ is an operator that varies with $\theta_x$ and is supported near $x = -L/2$, and where we have suppressed the $(t,\theta_x,\theta_y)$ arguments for brevity, 
Similarly, we have
\begin{align}
U_f^\dagger\frac{\partial}{\partial\theta_y}U_f= i\tilde{U}^\dagger\left[Q_B,\tilde{U}\right]+U_{\theta_y,L/2},
\end{align}
where $U_{\theta_y,L/2}$ is an operator that varies with $\theta_y$ and is supported near $y = L/2$.
Putting these together, we get
\begin{align}
\begin{split}\label{fluxxy}
&-\left[\tilde{U}^\dagger[Q_A,\tilde{U}],\tilde{U}^\dagger[Q_B,\tilde{U}]\right]\\
&=\left[U_f^\dagger\frac{\partial}{\partial\theta_x}U_f-U_{\theta_x,-L/2},U_f^\dagger\frac{\partial}{\partial\theta_y}U_f-U_{\theta_y,L/2}\right]\\
&=\left[U_f^\dagger\frac{\partial}{\partial\theta_x}U_f,U_f^\dagger\frac{\partial}{\partial\theta_y}U_f\right]+O_{L/2},
\end{split}
\end{align}
where $O_{L/2}$ is defined by
\begin{align}
O_{L/2} &= [U_f^\dagger\frac{\partial}{\partial\theta_y}U_f, U_{\theta_x,-L/2}] + [ U_{\theta_y,L/2}, U_f^\dagger\frac{\partial}{\partial\theta_x}U_f] \nonumber \\
&+ [U_{\theta_x,-L/2}, U_{\theta_y,L/2}].
\end{align}

Notice that $O_{L/2}$ is supported along the lines $x=-L/2$ and $y=L/2$. 

Substituting (\ref{fluxxy}) into (\ref{mtheta2}), and using the fact that
\begin{equation}\label{fluxt}
U_f^\dagger \tilde{H}_{C} U_f=\tilde{U}^\dagger \tilde{H}_C \tilde{U},
\end{equation}
we get
\begin{align}
\begin{split}
M&(\{U(t)\})\\
&=\frac{i}{2}\int_0^Tdt\left\langle U_f^\dagger \tilde{H}_{C} U_f\left[U_f^\dagger\frac{\partial}{\partial\theta_x}U_f,U_f^\dagger\frac{\partial}{\partial\theta_y}U_f\right]\right\rangle_\rho\\
& +\frac{i}{2}\int_0^T dt \left\langle U_f^\dagger\tilde{H}_{C} U_f\right\rangle_\rho\left\langle O_{L/2}\right\rangle_\rho .
\end{split}
\label{mtheta3}
\end{align}

We now claim that 
\begin{align}
\langle O_{L/2}\rangle_\rho = 0,
\end{align}
so that the second term vanishes. To see this, notice that $O_{L/2}$ is a sum of commutators of operators ($U_f^\dagger\frac{\partial}{\partial\theta_x}U_f$, $U_{\theta_x,-L/2}$, and $U_{\theta_y,L/2}$) all of which commute with $\rho$. Hence $\langle O_{L/2}\rangle = \mathrm{Tr}(O_{L/2} \rho)$ vanishes by the cyclicity of the trace.

All that remains is to show that we can replace $\tilde{H}_{C} \rightarrow H_f$ in the first term of (\ref{mtheta3}) above. To see this, note that $H_f - \tilde{H}_{C} = \sum_{r} H_{fr}$ is a sum of local terms $H_{fr}$ supported far away from $x=0$, $y = 0$. Thus, replacing $\tilde{H}_{C} \rightarrow H_f$ amounts to adding a collection of terms of the form $\mathrm{Tr}(U_f^\dagger H_{fr} U_f [U_f^\dagger\frac{\partial}{\partial\theta_x}U_f,U_f^\dagger\frac{\partial}{\partial\theta_y}U_f] \rho)$. But each of these terms vanishes by the cyclicity of the trace since $U_f^\dagger H_{fr} U_f$ commutes with either $U_f^\dagger\frac{\partial}{\partial\theta_x}U_f$, which is supported near $x = 0$, or $U_f^\dagger\frac{\partial}{\partial\theta_y}U_f$, which is supported near $ y = 0$, and all three of the operators $\{U_f^\dagger H_{fr} U_f, U_f^\dagger\frac{\partial}{\partial\theta_x}U_f,U_f^\dagger\frac{\partial}{\partial\theta_y}U_f\}$ commute with $\rho$.

\section{Interacting systems without symmetry}\label{snosymm}
We now discuss the case of interacting systems without any symmetry, expanding on the example discussed in Sec.~\ref{sexamples}. In this case, because the flow is not spatially additive, we can only obtain a bulk invariant in the overlapping geometry, and there is no obvious analogue of ``current" and ``magnetization" in these systems.

\subsection{Definition of $F(U_{\mathrm{edge}})$ and $M(\{U(t)\})$}\label{sflownosymm}
Our starting point is the flow given in (\ref{flownosymm_ex}):
\begin{equation}\label{flownosymm0}
\Omega_{A,B}(U)=\log\left[\frac{\eta(U^\dagger\mathcal{A}U,\mathcal{B})}{\eta(\mathcal{A},\mathcal{B})}\right].
\end{equation}

Recall that $\mathcal{A}, \mathcal{B}$ are operator algebras consisting of all operators supported on the two subsets of lattice sites, $A, B$, while $\eta$ is an overlap for operator algebras defined by
\begin{equation}\label{etaoperators}
\eta(\mathcal{A},\mathcal{B})=\sqrt{\sum_{\substack{O_a\in\mathcal{A} \\ O_b\in\mathcal{B}}}|\mathrm{tr}(O_a^\dagger O_b)|^2},
\end{equation}
where the sum runs over an orthonormal basis of operators in $\mathcal{A}, \mathcal{B}$ satisfying $\mathrm{tr}(O_a^\dagger O_{a'}) = \delta_{aa'}$ and $\mathrm{tr}(O_b^\dagger O_{b'}) = \delta_{bb'}$. Here, the lowercase symbol ``$\mathrm{tr}$'' denotes a normalized trace defined by $\mathrm{tr}(\mathbbm{1}) = 1$.

We can construct an edge invariant by substituting this flow into (\ref{flow}):
\begin{equation}\label{flownosymm1}
F(U_{\mathrm{edge}})=\Omega_{A,B}(U_{\mathrm{edge}}),
\end{equation}
where $A$ and $B$ are intervals illustrated in Fig.~\ref{fig:ABintervals}. 

Likewise, we can construct a bulk invariant by substituting this flow into Eq.~(\ref{mag}):
\begin{widetext}
\begin{equation}\label{magswap}
M(\{U(t)\}=\frac{i}{2}\int_0^Tdt\frac{\sum_{O_a,O_b}\mathrm{tr}\left(U^\dagger(t)[H_C(t),O_a^\dagger]U(t)O_b\right)\mathrm{tr}\left(U^\dagger(t)O_a^\dagger U(t)O_b\right)^* + \mathrm{c.c}}{\sum_{O_a,O_b}\left|\mathrm{tr}\left(U^\dagger(t)O_a^\dagger U(t)O_b\right)\right|^2.}
\end{equation}
\end{widetext}
where ``c.c" denotes the complex conjugate of the first term in the numerator. 

\subsection{Relation to previously known invariants}\label{sclassnosymm}
We now discuss the relationship between our edge invariant and the edge invariant $\mathrm{ind}(U_{\mathrm{edge}})$ presented in Ref.~\onlinecite{chiralbosons,GNVW}. The latter invariant (also known as the GNVW index) takes \emph{rational} values, $p/q \in \mathbb{Q}$, and is defined as follows. Let $A,B$ be two large \emph{adjacent} intervals, and let $\mathcal{A}, \mathcal{B}$ be the corresponding operator algebras consisting of all operators supported on $A,B$. Then the edge invariant $\mathrm{ind}(U_{\mathrm{edge}})$ is defined by
\begin{equation}\label{indexeq}
\mathrm{ind}(U_{\mathrm{edge}})=\frac{\eta(U_{\mathrm{edge}}^{\dagger}\mathcal{A} U_{\mathrm{edge}},\mathcal{B})}{\eta(\mathcal{A},U_{\mathrm{edge}}^{\dagger}\mathcal{B}U_{\mathrm{edge}})}.
\end{equation}
We prove in Appendix \ref{soverlapindex} that
\begin{equation}\label{edgeGNVWid}
F(U_{\mathrm{edge}})=\log\left[\mathrm{ind}(U_{\mathrm{edge}})\right].
\end{equation}
Thus, our edge invariant $F(U_{\mathrm{edge}})$ is closely related to the previously known invariant for classifying 1D locality preserving unitaries without any symmetries. Notice that while $F(U_{\mathrm{edge}})$ uses overlapping intervals $A$ and $B$, $\log[\mathrm{ind}(U_{\mathrm{edge}})]$ is defined in (\ref{indexeq}) with adjacent intervals, so the proof of (\ref{edgeGNVWid}) is nontrivial. 

Once again, there is no previously known bulk invariant that we can compare with $M(\{U(t)\})$. 

\section{General MBL Floquet circuits}\label{smblgen}
In this section, we will show how to generalize our edge and bulk invariants from unitary loops to general MBL Floquet systems. 

We begin with the edge invariants. To describe these, we first have to explain how to define edge unitaries for general MBL Floquet systems. This definition is similar to the unitary loop case: given a (2D) MBL Floquet system with Hamiltonian $H(t)$, we restrict the Hamiltonian to a finite disk $C$ by discarding all terms that have support outside of $C$. Denoting the restricted Hamiltonian by $H_C(t)$, we then define an edge unitary by\cite{chiralbosons} 
\begin{align}
U_{\mathrm{edge}} =  \mathcal{T}e^{-i\int_0^T dt H_C(t)} \cdot \prod_{r\in C}U_r^\dagger
\label{Uedgedef2}
\end{align}
where the $U_r$ operators are those that appear in the decomposition $U_F = \prod_r U_r$ (\ref{MBLcond1}). Just like the unitary loop case, $U_{\mathrm{edge}}$ is a 1D LPU supported near the boundary of $C$.

Having defined $U_{\mathrm{edge}}$, we can now describe the edge invariant. As before, our invariant $F(U_{\mathrm{edge}})$ is defined on 1D LPUs $U_{\mathrm{edge}}$. Given such an LPU, we choose two large overlapping intervals $A, B$ and then we define our edge invariant $F(U_{\mathrm{edge}})$ in exactly the same way as in the unitary loop case: 
\begin{equation}
F(U_{\mathrm{edge}})= \Omega_{A,B}(U_{\mathrm{edge}})
\label{edgembl}
\end{equation}

We now move on to the bulk invariant $M(\{U(t)\})$. Let $A, B, C$ be three overlapping disklike regions as in Fig.~\ref{fig:ABC}. We define $M(\{U(t)\})$ similarly to the unitary loop case, except that we time-average over many periods:
\begin{align}
M(\{U(t)\}) = \lim_{A,B,C,n\to\infty} \frac{1}{n}\int_0^{nT}dt\frac{\partial}{\partial t_C}\Omega_{A,B}(U(t))
\label{bulkmbl}
\end{align}

Here, the notation ``$\lim_{A,B, C, n \to \infty}$'' means that we should take the size of the regions $A, B, C$ to infinity, in addition to taking $n$ to infinity. More specifically, it is important that this limit is taken in such a way that the linear size of the regions $A, B, C$ grows faster than $n$. This will ensure that $A, B, C$ are much larger than the relevant Lieb-Robinson length $\ell = v_{LR} n T$ -- the length scale at which our invariant converges.

To complete our discussion, we now show that the above invariants (\ref{edgembl}), (\ref{bulkmbl}) obey the same bulk-boundary correspondence as in the unitary loop case:
\begin{equation}\label{avgcorr}
F(U_{\mathrm{edge}})=M(\{U(t)\})
\end{equation}

Our derivation proceeds in two steps. First, we show that 
\begin{equation}\label{avgbulk}
M(\{U(t)\}) = \lim_{A,B,C, n\to\infty} \frac{1}{n} \Omega_{A,B}(U_C(T)^n).
\end{equation}
where $U_C(t)$ is the unitary generated by $H_C(t)$:
\begin{equation}
U_C(t) =  \mathcal{T}e^{-i\int_0^t dt' H_C(t')}
\end{equation}
Then we show that
\begin{align}
\lim_{A,B,C, n\to\infty} \frac{1}{n} \Omega_{A,B}(U_C(T)^n) = F(U_{\mathrm{edge}})
\label{avgedge}
\end{align}
Together, Eqs. (\ref{avgbulk}) and (\ref{avgedge}) imply (\ref{avgcorr}).

To show (\ref{avgbulk}), we use the identity (\ref{idc1}), that is,
\begin{align*}
\frac{\partial}{\partial t_C}\Omega_{A,B}(U(t))=\frac{d}{dt}\Omega_{A,B}(U_C(t))
\end{align*}
This gives
\begin{align}
\begin{split}
M(\{U(t)\}) &= \lim_{A,B,C, n\to\infty} \frac{1}{n}\int_0^{nT}\frac{d}{dt}\Omega_{A,B}(U_C(t))\\
&=\lim_{A,B,C, n\to\infty}  \frac{1}{n} \Omega_{A,B}(U_C(T)^n)
\end{split}
\end{align}
where in the second line we used $U_C(nT)=\left[U_C(T)\right]^n$. 

To show (\ref{avgedge}), we use
\begin{align}
U_C(T) = U_{\mathrm{edge}} \cdot \prod_{r\in C}U_r
\end{align}
which follows from the definition of $U_{\mathrm{edge}}$ (\ref{Uedgedef2}). We will assume that all of the $U_r$ terms in this expression commute with $U_{\mathrm{edge}}$: we can make this assumption without loss of generality since we can always incorporate any $U_r$ terms that do not commute into the definition of $U_{\mathrm{edge}}$ without affecting the value of the edge invariant $F(U_{\mathrm{edge}})$.

Substituting this expression into $\Omega_{A,B}(U_C(T)^n)$, we obtain 
\begin{equation}
\Omega_{A,B}(U_C(T)^n) =
\Omega_{A,B}\left( \left(\prod_{r\in C}U_r^n\right) U_{\mathrm{edge}}^n \right).
\end{equation}
To proceed further we note that we can remove all the $U_r$ terms that are supported entirely in $A$ or $\overline{A}$ using Definition~\ref{flowdef}.1, since we can freely move these operators to the beginning of the product using the fact that all the operators commute. Likewise, we can remove all the $U_r$ terms that are supported entirely in $B$ or $\overline{B}$ by moving them to the end of the product and using Definition~\ref{flowdef}.2. After removing these terms, we are left with only the terms that have support in all four regions, $A, \overline{A}, B, \overline{B}$ -- i.e. terms that lie at the intersection of $\partial A$ and $\partial B$: 
\begin{equation}
\Omega_{A,B}(U_C(T)^n) =\Omega_{A,B}\left(\left(\prod_{r\in\partial A\cap\partial B}U_r^n\right) U_{\mathrm{edge}}^n  \right),
\end{equation}

Given that ultimately we will be interested in the limit of large $A, B, C$, we can assume in particular that $A, B, C$ are large enough that $\left(\prod_{r\in\partial A\cap\partial B}U_r^n\right)$ and $U_{\mathrm{edge}}$ are supported on disjoint regions. Then, we can apply Lemma~\ref{stack} from Appendix~\ref{slemmas} to write the flow as a sum of two flows:
\begin{align}
\begin{split}\label{omegabulkedge}
\Omega_{A,B}(U_C(T)^n) &= \Omega_{A,B}\left(U_{\mathrm{edge}}^n\right) \\
&+\Omega_{A,B}\left(\prod_{r\in\partial A\cap\partial B}U_r^n\right).
\end{split}
\end{align}

To evaluate the first term, $\Omega_{A,B}\left(U_{\mathrm{edge}}^n\right)$, we note that $U_{\mathrm{edge}}$ is supported near the boundary of $C$, so we can truncate $A, B$ to two intervals supported near the boundary of $C$. After this truncation, $\Omega_{A,B}\left(U_{\mathrm{edge}}^n\right)$ reduces to the edge invariant
\begin{align}
\Omega_{A,B}\left(U_{\mathrm{edge}}^n\right) &= F\left(U_{\mathrm{edge}}^n\right) \nonumber \\
&= n  F\left(U_{\mathrm{edge}}\right),
\end{align}
where the second equality follows from the additivity of the edge invariants under composition (see Corollary~\ref{composition} in Appendix~\ref{slemmas}). Notice that in this setup we've taken $A,B\to\infty$ faster than $n$, so $A$ and $B$ are sufficiently large (according to the definition of $F(U_{\mathrm{edge}})$ in Sec.~\ref{sflowedge}) compared to the operator spreading length of $U_{\mathrm{edge}}^n$.

Next consider the second term, $\Omega_{A,B}\left(\prod_{r\in\partial A\cap\partial B}U_r^n\right)$. This term involves a unitary that is supported in a disk of radius $\xi$ (the length scale associated with the quasilocal unitaries, $U_r$). Since $\xi$ is independent of the size of $A, B, C$ or $n$, it follows that $\Omega_{A,B}\left(\prod_{r\in\partial A\cap\partial B}U_r^n\right)$  is bounded by a constant that is independent of the size of $A, B, C$ or $n$.\footnote{Here we are assuming that $\Omega_{A,B}(U)$ is a continuous function of $U$ and hence it is bounded on a compact set.}

Therefore, the second term vanishes in the limit of interest, and we obtain
\begin{align}
\lim_{A,B,C, n\to\infty}  \frac{1}{n} \Omega_{A,B}(U_C(T)^n) = F\left(U_{\mathrm{edge}}\right)
\end{align}
This completes our derivation of the bulk-boundary correspondence for general MBL Floquet systems (\ref{avgcorr}).

\section{Discussion}\label{sdiscussion}
In this work, we have shown how to derive bulk and edge invariants for 2D MBL Floquet systems using a special mathematical object which we call a flow. Using this approach, we have obtained bulk and edge invariants for single-particle Floquet systems, interacting many-body Floquet systems with $U(1)$ symmetry, and interacting Floquet systems without any symmetry.

Throughout this paper, we have focused on two symmetry groups: the $U(1)$ symmetry group and the trivial group (i.e. no symmetry at all). More generally, we expect that our approach should give topological invariants that at least partially classify systems with other continuous symmetry groups.\footnote{Note that there may be additional difficulties to many-body localizing systems with non-Abelian symmetries\cite{nonabelian}.} On the other hand, finite symmetry groups may be problematic. The issue is that Floquet phases with finite symmetry group $G$ are believed to be classified by both the GNVW index and an additional index that takes values in the (finite) cohomology group $H^2(G,U(1))$. The latter, cohomology-valued index is probably out of reach of our flow-based approach. One way to see the obstruction is to note that our bulk invariant (\ref{mag}-\ref{omegaabc}) is expressed in terms of an \emph{integral}, which seems incompatible with the finite group structure of $H^2(G,U(1))$. Therefore we probably need other methods to construct invariants in this case. (As an aside, we note that the main problem here involves \emph{bulk} invariants; by contrast, it is possible to construct \emph{edge} invariants using similar ideas to the ones presented here, using a different kind of flow which is multiplicative and complex valued, rather than additive and real valued\cite{sptentanglers}).

While we have focused on bosonic systems in this paper, our results can be straightforwardly generalized to fermionic systems. In particular, the flows that we constructed for bosonic systems with $U(1)$ symmetry (\ref{flowu10}) and without symmetry (\ref{flownosymm0}) apply equally well to the fermionic case. The corresponding edge and bulk invariants are also valid in the fermionic case. The only new element is that these invariants can take values that are not possible in purely bosonic systems. For example, in the case of fermionic systems without symmetry, the edge invariant $F(U_{\mathrm{edge}})$ can take the value $\log(\sqrt{2})$ when $U_{\mathrm{edge}}$ is a ``Majorana translation''\cite{fermionic}.

One question raised by this work is whether there is any connection between our invariants for Floquet systems and previously known invariants for stationary topological phases. In the single-particle case, there is indeed a close relationship between these two types of invariants. For example, the single-particle invariant $M(\{U(t)\})$ (\ref{magnonint0}) is closely related to the Chern number, as shown in Appendix \ref{sstationary}. By analogy, one might wonder if our many-body Floquet invariants, with and without $U(1)$ symmetry, are related to many-body stationary invariants like the electric or thermal Hall conductance (see e.g.~the modular commutator formula for the thermal Hall conductance\cite{kim2021,kim2022,fan2022}). If such a connection exists, it would be very interesting since the two types of invariants describe different objects: the stationary invariants describe properties of (gapped) ground states, while our Floquet invariants describe properties of unitary operators.

Another question is to understand the physical interpretation of (\ref{magswap}), i.e.~the bulk counterpart of the GNVW index. Unlike the invariants for $U(1)$-symmetric systems, we do not know how to relate this invariant to current operators. On the other hand, previous work has shown that the edge invariant (\ref{indexeq}) can be interpreted in terms of transport of quantum information\cite{tracking,gong2021,ranard2022} so it is possible that the bulk invariant could also have an interpretation of this kind.

One possible direction for future work would be to consider the generalization of MBL Floquet systems discussed in Refs.\onlinecite{dynamically, radical}. In this generalization, one requires that $U_F^N$ is many-body localized for some finite integer $N$, but $U_F$ itself need not be many-body localized. (An illustrative example of such a system is the dynamical Kitaev honeycomb model studied in detail in Ref.~\onlinecite{radical}, which becomes many-body localized after two periods). In these systems, we cannot use Eq.~(\ref{Uedgedef2}) to define an effective edge unitary, so it is not possible to write down a meaningful edge invariant. However, it may be possible to find bulk invariants for these systems.

It would also be interesting to consider the \emph{partially} many-body localized Floquet systems discussed in Ref.~\onlinecite{hierarchy}. These systems are built out of fermionic degrees of freedom and are localized up to $n$-body terms. Ref.~\onlinecite{hierarchy} showed that multi-particle correlations in these systems produce a family of integer valued topological invariants that generalize the winding number $W(\{U(t)\})$. It would be interesting to try to study flows and the bulk-boundary correspondence for these systems.

\acknowledgments

We thank Ari Turner for explaining the SWAP formulation of the GNVW index. C.Z. and M.L. acknowledge the support of the Kadanoff Center for Theoretical Physics at the University of Chicago.
This work was supported by the Simons Collaboration on Ultra-Quantum Matter, which is a grant from the 
Simons Foundation (651440, M.L.), and the National Science Foundation Graduate Research Fellowship under Grant No. 1746045.
\appendix

\section{Equivalence of classification of unitary loops and edge unitaries}\label{shomotopy}
In this appendix we show that if two unitary loops, $\{U(t)\}$ and $\{U'(t)\}$ are equivalent in the sense of Sec.~\ref{sloopdef}, then the corresponding edge unitaries $U_{\mathrm{edge}}$ and $U_{\mathrm{edge}}'$ are equivalent in the sense of Eq.~(\ref{simW}).

Let $\{U(t)\}$ and $\{U'(t)\}$ be two $d$-dimensional unitary loops that are equivalent in the sense that there exists a one-parameter family of unitary loops $\{U_s(t)\}$, depending smoothly on $s$, with $U_0(t) = U(t)$ and $U_1(t) = U'(t)$. Let $U_{\mathrm{edge}}, U_{\mathrm{edge}}'$ be the corresponding $(d-1)$-dimensional edge unitaries, defined as in Eq.~(\ref{Uedgedef}). We wish to show that $U_{\mathrm{edge}}' = W U_{\mathrm{edge}}$ for some $(d-1)$-dimensional locally generated unitary $W$. To see this, consider the edge unitary corresponding to $\{U_s(t)\}$, which we denote by $U_{\mathrm{edge}}(s)$, and then define a Hermitian operator $H_{\mathrm{edge}}(s)$ by
\begin{equation}
H_{\mathrm{edge}}(s)=i\left(\frac{d}{ds}U_{\mathrm{edge}}(s)\right)U_{\mathrm{edge}}^\dagger(s).
\end{equation}

By construction,
\begin{equation}
\frac{d}{ds}U_{\mathrm{edge}}(s) = - i H_{\mathrm{edge}}(s) U_{\mathrm{edge}}(s)
\end{equation}
so that
\begin{align}
U_{\mathrm{edge}}(1) = \mathcal{T} \exp \left( - i \int_0^1 H_{\mathrm{edge}}(s) ds \right) \cdot U_{\mathrm{edge}}(0)
\end{align}
Using $U_{\mathrm{edge}}(1) = U_{\mathrm{edge}}'$ and $U_{\mathrm{edge}}(0) = U_{\mathrm{edge}}$, we deduce that
\begin{align}
U_{\mathrm{edge}}' = \mathcal{T} \exp \left( - i \int_0^1 H_{\mathrm{edge}}(s) ds \right) \cdot U_{\mathrm{edge}}
\end{align}
To complete the proof, we need to show $H_{\mathrm{edge}}(s)$ is a \emph{local} $(d-1)$ dimensional Hamiltonian. To this end, let $O_r, O_{r'}$ be local operators supported on sites $r, r'$, and consider the double commutator $[[H_{\mathrm{edge}}(s),O_r], O_{r'}]$. We now argue that the operator norm of this double commutator is exponentially small in the distance $|r - r'|$, which establishes the locality of $H_{\mathrm{edge}}(s)$. First, we rewrite the commutator $[H_{\mathrm{edge}}(s),O_r]$ as
\begin{align}
\begin{split}
&[H_{\mathrm{edge}}(s),O_r]\\
&=-i U_{\mathrm{edge}}(s)\frac{d}{ds}\left(U_{\mathrm{edge}}^\dagger(s)O_rU_{\mathrm{edge}}(s)\right)U_{\mathrm{edge}}^\dagger(s)
\end{split}
\end{align}
It follows that
\begin{align}
& \| [[H_{\mathrm{edge}}(s),O_r], O_{r'}] \|  \nonumber \\
& = \left \| \left[\frac{d}{ds}\left(U_{\mathrm{edge}}^\dagger(s)O_rU_{\mathrm{edge}}(s)\right), U_{\mathrm{edge}}^\dagger(s) O_{r'} U_{\mathrm{edge}}(s) \right] \right \|
\end{align}
Now, by Lieb-Robinson bounds, the operator $U_{\mathrm{edge}}^\dagger(s) O_{r'} U_{\mathrm{edge}}(s)$ is supported within a finite distance of site $r'$ with exponential tails. Similarly, the operator $\frac{d}{ds}\left(U_{\mathrm{edge}}^\dagger(s)O_rU_{\mathrm{edge}}(s)\right)$ is supported within a finite distance of site $r$, again with exponential tails. It follows that the commutator between these operators is exponentially small in the distance $|r-r'|$, as we wished to show.

\section{Proof of Theorem~\ref{theorem1}}\label{slemmas}
In this appendix, we prove Theorem~\ref{theorem1}. 

\subsection{Two lemmas}
Our proof uses two lemmas which apply to any flow $\Omega_{A,B}(U)$. The first lemma says that flows are additive under composition of unitaries supported in disjoint regions:
\begin{lemma}\label{stack}
\emph{(generalized stacking)} Let $U_1, U_2$ be ($G$-symmetric) unitaries supported on disjoint subsets $\Lambda_1, \Lambda_2 \subset \Lambda$. For any $A_1, B_1 \subset \Lambda_1$, and
$A_2, B_2 \subset \Lambda_2$,
\begin{equation}
\Omega_{A_1\cup A_2,B_1\cup B_2}(U_1U_2)=\Omega_{A_1,B_1}(U_{1}) + \Omega_{A_2,B_2}(U_{2})
\end{equation}
\end{lemma}

\emph{Proof.} The claim follows straightforwardly from Definition \ref{flowdef}.3 and \ref{flowdef}.4 by thinking of $U_1$ as a tensor product $U_1 \otimes \mathbbm{1}$ and $U_2$ as $\mathbbm{1} \otimes U_2$, and using $(U_1\otimes\mathbbm{1})(\mathbbm{1}\otimes U_2)=U_1\otimes U_2$ where $U_1$ is defined on $\Lambda_1$ and $U_2$ is defined on $\Lambda_2$. \\

The second lemma says that for any LPU $U$, the tensor product $U \otimes U^\dagger$ is always an FDLU:
\begin{lemma}\label{UUdag}
Let $U$ be a ($G$-symmetric) strict LPU with an operator spreading length $\xi$, defined on a lattice $\Lambda$. For any such $U$, the tensor product $U\otimes U^\dagger$, acting on the bilayer system $\Lambda \times \{1,2\}$, can be realized as a ($G$-symmetric) FDLU of depth $2$ built out of gates of radius $\xi$. 
\end{lemma}

\emph{Proof.} We rewrite $U\otimes U^\dagger$ as
\begin{equation}
U\otimes U^\dagger=(\mathrm{SWAP}')(\mathrm{SWAP}),
\end{equation}
where $\mathrm{SWAP}$ is the unitary transformation that swaps the two layers and
\begin{equation}
\mathrm{SWAP}'=(\mathbbm{1}\otimes U^\dagger)\mathrm{SWAP}(\mathbbm{1}\otimes U).
\end{equation}
It is easy to see that $\mathrm{SWAP}$ is an FDLU of depth $1$ built out of gates of radius $1$ while $\mathrm{SWAP}'$ is an FDLU of depth $1$ built out of gates of radius $\xi$.  
Since $U\otimes U^\dagger$ is a composition of these two FDLUs, the claim follows immediately.

\subsection{Main argument}\label{smainargument}
We are now ready to prove Theorem~\ref{theorem1}.


\emph{Proof. } Item (1): 
Let $U$ be a strict LPU with an operator spreading length $\xi$, and let $W$ be an FDLU of depth $n$:
\begin{align}
W=W_nW_{n-1}\cdots W_1
\end{align}
Let 
\begin{align}
W' = W_n' W_{n-1}' \cdots W_1'
\end{align}
where each $W_i'$ is obtained by removing all unitary gates from $W_i$ except for those fully supported in $\partial_{2n \lambda} A \cap \partial_{2n \lambda + \xi} B$. We wish to show that $\Omega_{A,B}(WU) = \Omega_{A,B}(W'U)$. To this end, we decompose each $W_i$ as a product, $W_i = W_i' V_i^A V_i^B$, where $V_i^A$ consists of all the gates in $W_i$ whose region of support contains sites deeper than $2n \lambda$ within $A$ or $\overline{A}$, and where $V_i^B$ consists of all the remaining gates in $W_i$ whose region of support contains sites deeper than $2 n \lambda + \xi$ within $B$ or $\overline{B}$. We now show that we can remove each $V_i^A$ and $V_i^B$ without affecting $\Omega_{A,B}(W U)$. First consider $V_1^A$ and $V_1^B$. Note that
\begin{align}
\Omega_{A,B}(W U) &= \Omega_{A,B}(W_n \cdots W_1' V_1^A V_1^B U) \nonumber \\
&= \Omega_{A,B}( \tilde{V}_1^A W_n \cdots W_1' U \tilde{V}_1^B) 
\label{wu1}
\end{align}
where $\tilde{V}_1^A = (W_n \cdots W_2) V_1^A (W_n \cdots W_2)^\dagger$ and $\tilde{V}_1^B = U^\dagger V_1^B U$. Next notice that $\tilde{V}_1^A$ can be written as a product of unitaries, each of which is supported either entirely in $A$ or entirely in $\overline{A}$, since $W_n \cdots W_2$ has an operator spreading length of at most $2(n-1) \lambda$. Therefore, by Definition \ref{flowdef}.1, we can remove $\tilde{V}_1^A$ without affecting the value of $\Omega_{A,B}$. Similarly, $\tilde{V}_1^B$ is a product of unitaries, each of which is supported entirely in $B$ or entirely in $\overline{B}$, since $U$ has an operator spreading length $\xi$. Therefore, we can also remove $\tilde{V}_1^B$ according to Definition \ref{flowdef}.2. Removing these two operators from (\ref{wu1}), we obtain 
\begin{align}
\Omega_{A,B}(W U) = \Omega_{A,B}(W_n \cdots W_1' U)
\end{align}
In exactly the same way, we can remove $V_2^A, V_2^B$ by moving $V_2^A$ to the left and $V_2^B$ to the right and then applying Definition \ref{flowdef}.1 and \ref{flowdef}.2 to remove the conjugated operators $\tilde{V}_2^A$ and $\tilde{V}_2^B$. Continuing in this way, we can remove all the $V_i^A, V_i^B$ operators until we are left with
\begin{align}
\Omega_{A,B}(W U) &= \Omega_{A,B}(W_n' \cdots W_1' U) \nonumber \\
&= \Omega_{A,B}(W' U)
\end{align}
This completes the proof of the first part of the theorem.
\begin{figure}[tb]
   \centering
   \includegraphics[width=0.8\columnwidth]{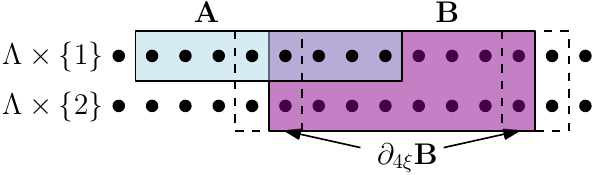} 
   \caption{The bilayer system used in the proof of Theorem~\ref{theorem1}.2. Here, $\mathbf{A} = A \times \{1\}$ is a subset of $\Lambda\times\{1\}$ while $\mathbf{B} = B \times \{1, 2\}$ is a subset of $\Lambda\times\{1,2\}$. The thickened boundary $\partial_{4\xi}\mathbf{B}$ consists of sites within $4\xi$ of the left and right edges of $\mathbf{B}$.}
   \label{fig:bilayer}
\end{figure}

Item (2): Let $U$ be a strict LPU with operator spreading length $\xi$, defined on a lattice $\Lambda$, and let $A, B \subset \Lambda$. We wish to show that $\Omega_{A, B}(U) = \Omega_{A \setminus a,B}(U)$ for any site $a \in A$ such that $a \not \in \partial_{4\xi} B$. To prove this, consider the bilayer system $\Lambda \times \{1,2\}$, and
define two subsets $\v{A}, \v{B} \subset \Lambda \times \{1,2\}$ by 
\begin{align}
\v{A} = A \times \{1\}, \quad \v{B} = B \times \{1,2\}
\end{align}\
This setup is illustrated in Fig.~\ref{fig:bilayer}. Consider the unitary $W = U \otimes U^\dagger$, acting on $\Lambda \times \{1,2\}$. From Definition~\ref{flowdef}.3, it is easy to see that
\begin{align}
\Omega_{\v{A}, \v{B}} (W) = \Omega_{A,B}(U)
\label{bilayer}
\end{align}
e.g. by setting 
\begin{align*}
U_1 &= U, \quad U_2 = U^\dagger, \nonumber \\
A_{1} &= A, \quad A_2 = \emptyset, \nonumber \\
B_{1} &= B, \quad B_2 = B
\end{align*}

At the same time, using Lemma~\ref{UUdag}, we know that $W$ is an FDLU of depth $2$ built out of gates of radius $\xi$. Therefore, using Theorem \ref{theorem1}.1, 
\begin{align}
\Omega_{\v{A}, \v{B}} (W) = \Omega_{\v{A}, \v{B}} (W')
\label{Wtrun}
\end{align}
where $W'$ is obtained from $W$ by removing all the unitary gates in $W$ except for those fully supported in $\partial_{4 \xi}\v{B}$ (In fact, Theorem \ref{theorem1}.1 tells us that we can remove all the gates except for those supported in $\partial_{4 \xi}{\v{A}} \cap \partial_{4 \xi} \v{B}$ so it is actually a stronger statement than what we need here -- where we remove \emph{fewer} gates). Note that here, by $\partial_{4\xi}\mathbf{B}$, we mean sites that are within $4\xi$ of both $B$ and $\overline{B}$ in the direction parallel to the two layers. 

To proceed further, note that the support of $W'$ does not contain the point $a \times 1$; therefore, by Lemma~\ref{stack}, 
\begin{align}
\Omega_{\v{A}, \v{B}} (W') = \Omega_{\v{A} \setminus \{a \times 1\}, \v{B}} (W')
\label{Wstack}
\end{align}

Also, by the same reasoning as in (\ref{Wtrun}), 
\begin{align}
\Omega_{\v{A} \setminus \{a \times 1\}, \v{B}} (W') = \Omega_{\v{A} \setminus \{a \times 1\}, \v{B}} (W)
\label{Wtrun2}
\end{align}
while by the same reasoning as in (\ref{bilayer}), 
\begin{align}
\Omega_{\v{A} \setminus \{a \times 1\}, \v{B}} (W) = \Omega_{A \setminus a ,B}(U)
\label{bilayer2}
\end{align}

Combining {\color{blue} (\ref{bilayer} - \ref{bilayer2})}, we deduce that
\begin{align}
\Omega_{A,B}(U) = \Omega_{A \setminus a,B}(U),
\end{align}
proving the claim. In exactly the same way, we can show that $\Omega_{A, B}(U) = \Omega_{A,B \setminus b}(U)$ for any site $b \in B$ such that $b \not \in \partial_{4\xi} A$. This completes our proof of item (2).

\subsection{Three more corollaries}
In Sec.~\ref{sproperties}, we listed two corollaries of Theorem \ref{theorem1}. We now discuss three additional corollaries:

\begin{corollary}\label{conservation}
\emph{(conservation law)} Let $U$ be a ($G$-symmetric) strict LPU with operator spreading length $\xi$. Then $\Omega_{A,B}(U)=0$ if $\partial_{4\xi}A\cap B=\emptyset$ or $A\cap \partial_{4\xi}B=\emptyset$ . 
\end{corollary}
\emph{Proof.} This is an immediate consequence of Theorem \ref{theorem1}.2.

\begin{corollary}\label{composition}
\emph{(additivity under composition)} Let $U_1$ and $U_2$ be ($G$-symmetric) strict LPUs defined on a lattice $\Lambda$, with operator spreading length $\xi$. Then $\Omega_{A,B}(U_1U_2)=\Omega_{A,B}(U_1)+\Omega_{A,B}(U_2)$ if $\partial_{4\xi} A\cap\partial_{5\xi} B=\emptyset$.
\end{corollary}

\emph{Proof.} The basic idea is to relate the composition of two unitaries to a tensor product. Consider a bilayer system $\Lambda \times \{1,2\}$, and define subsets
$\v{A} = A \times \{1,2\}$, and $\v{B} = B \times \{1,2\}$. Consider the unitary $U_1 U_2 \otimes \mathbbm{1}$ acting on this bilayer system. By Definition \ref{flowdef}.3 and \ref{flowdef}.4,
\begin{align}
\Omega_{\v{A}, \v{B}}(U_1 U_2 \otimes \mathbbm{1}) = \Omega_{A,B}(U_1 U_2)
\end{align}
At the same time,
\begin{equation}\label{u2u2d}
\Omega_{\v{A},\v{B}}(U_1U_2 \otimes\mathbbm{1}) =\Omega_{\v{A},\v{B}}((U_1\otimes U_1^\dagger)(U_2\otimes U_1)) 
\end{equation}
Notice that $U_1\otimes U_1^\dagger$ is an FDLU of depth two with gates of radius $\xi$, according to Lemma~\ref{UUdag}. Therefore by Corollary~\ref{corollary1}, $U_1 \otimes U_1^\dagger$ can be dropped -- that is,
\begin{equation}
\Omega_{\v{A},\v{B}}((U_1\otimes U_1^\dagger)(U_2\otimes U_1)) = \Omega_{\v{A},\v{B}}(U_2\otimes U_1)
\end{equation}
Putting this all together, we deduce that
\begin{align}
\Omega_{A,B}(U_1 U_2) &= \Omega_{\v{A},\v{B}}(U_2\otimes U_1) \nonumber \\
&=\Omega_{A,B}(U_1)+\Omega_{A,B}(U_2)
\end{align}
where the second equality follows from Definition~\ref{flowdef}.3.

\begin{figure}[tb]
   \centering
   \includegraphics[width=0.9\columnwidth]{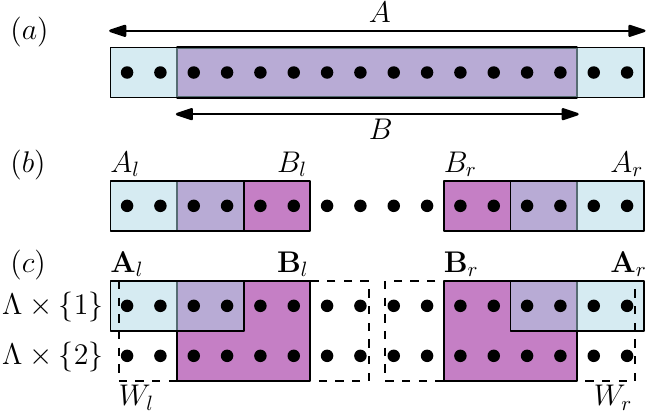} 
   \caption{The bilayer system used in the proof of Corollary~\ref{antisymmetric}. (a) We consider two intervals $A=[0,40\xi]$ and $B=[5,35\xi]$ on a spin chain. (b) Using Theorem~\ref{theorem1}.2, we remove sites in $A$ and $B$ to get $A'=A_l\cup A_r$ and $B'=B_l\cup B_r$ respectively. (c) To complete the proof, we again consider a bilayer system, and we use Theorem~\ref{theorem1}.1 to truncate $W=U\otimes U^\dagger$ to $W'=W_lW_r$.}
   \label{fig:bilayer2}
\end{figure}

\begin{corollary}\label{antisymmetric}
\emph{(antisymmetry)} Let $U$ be a ($G$-symmetric) strict LPU with operator spreading length $\xi$ defined on a 1D lattice $\Lambda$. Then $\Omega_{A,B}(U)=-\Omega_{B,A}(U)$ for any two overlapping intervals $A = [a_1, a_2]$ and $B = [b_1, b_2]$ such that $b_1 - a_1$, $a_2 - b_1$ and $b_2 - a_2$ are larger than $4\xi$.
\end{corollary}

\emph{Proof.} We begin with two intervals $A', B'$ defined by $A' = [0, 40\xi]$ and $B' = [5\xi, 35\xi]$, as shown in Fig.~\ref{fig:bilayer2}a. By Corollary~\ref{conservation}, we know that
\begin{equation}\label{sum0}
\Omega_{A', B'}(U) = 0
\end{equation}
Also, using Theorem~\ref{theorem1}.2, we can remove any sites in $A'$ that are outside of $\partial_{4\xi} B'$, without affecting the value of $\Omega_{A', B'}(U)$. In particular, we have
\begin{equation}
\Omega_{A', B'}(U)=\Omega_{A_l \cup A_r, B'}(U)
\label{trunA}
\end{equation}
where $A_l = [0, 10\xi]$ and $A_r = [30 \xi, 40\xi]$. Applying Theorem~\ref{theorem1}.2 again, but this time to the sites in $B'$, we have
\begin{equation}
\Omega_{A_l \cup A_r, B'}(U) = \Omega_{A_l \cup A_r, B_l \cup B_r}(U)
\label{trunB}
\end{equation}
where $B_l = [5 \xi, 15 \xi]$ and $B_r = [25 \xi, 35 \xi]$. The resulting system is shown in Fig.~\ref{fig:bilayer2}b.
Combining (\ref{sum0}-\ref{trunB}), we derive
\begin{align}
\Omega_{A_l \cup A_r, B_l \cup B_r}(U) = 0
\label{aabb0}
\end{align}
Below, we will argue that
\begin{align}
\Omega_{A_l \cup A_r, B_l \cup B_r}(U) = \Omega_{A_l, B_l}(U) + \Omega_{A_r, B_r}(U) 
\label{sumaabb}
\end{align}
Once we establish (\ref{sumaabb}), the Corollary follows easily. Indeed, let $A = [a_1, a_2]$ and $B = [b_1, b_2]$ be any two overlapping intervals such that $b_1 - a_1$, $a_2 - b_1$ and $b_2 - a_2$ are larger than $4\xi$. Then, since $\Omega_{A,B}(U)$ is independent of the choice of $A, B$ for large enough intervals (see Sec.~\ref{sflowedge}), we know that 
\begin{align}
\Omega_{A,B}(U) = \Omega_{A_l, B_l}(U)
\end{align}
(since $A_l$ is located to the left of $B_l$) and
\begin{align}
\Omega_{B,A}(U) = \Omega_{A_r, B_r}(U)
\end{align}
(since $A_r$ is located to the \emph{right} of $B_r$). The Corollary now follows from these equalities together with (\ref{aabb0}-\ref{sumaabb}).

All that remains is to show (\ref{sumaabb}). To do this, we will use the same trick as in the main proof in Sec.~\ref{smainargument}: we consider a bilayer system $\Lambda\times\{1,2\}$ and define subsets
\begin{equation}
\mathbf{A_l} = A_l \times\{1\}\quad \mathbf{B_l}= B_l \times\{1,2\}.
\end{equation}
and similarly for $\mathbf{A_r}, \mathbf{B_r}$. Again, we consider the unitary $W=U\otimes U^\dagger$ acting on $\Lambda\times\{1,2\}$, and we note that Definition~\ref{flowdef}.3 implies that
\begin{align}
\Omega_{A_l \cup A_r, B_l \cup B_r}(U) = \Omega_{\mathbf{A_l} \cup \mathbf{A_r}, \mathbf{B_l} \cup \mathbf{B_r}}(W)
\label{WUab}
\end{align}
Also, using Theorem~\ref{theorem1}.1, we know that
\begin{equation}
\Omega_{\mathbf{A_l} \cup \mathbf{A_r}, \mathbf{B_l} \cup \mathbf{B_r}}(W) =\Omega_{\mathbf{A_l} \cup \mathbf{A_r}, \mathbf{B_l} \cup \mathbf{B_r}}(W') 
\label{Wabtrun}
\end{equation}
where $W'$ is obtained from $W$ by removing all unitary gates except for those contained in $\partial_{4\xi}  (\mathbf{B_l} \cup \mathbf{B_r})$. 

To proceed further, we decompose $W'$ into a product of two unitaries supported in disjoint regions, shown in Fig.~\ref{fig:bilayer2}c. Specifically, we use $W'=W_l W_r$, where $W_l$ is supported in $ [\xi, 19\xi]$  and $W_r$ is supported in $[21 \xi, 39\xi]$. Then, by Lemma~\ref{stack}, we have
\begin{align} 
\Omega_{\mathbf{A_l} \cup \mathbf{A_r}, \mathbf{B_l} \cup \mathbf{B_r}}(W') = \Omega_{\mathbf{A_l}, \mathbf{B_l} }(W_l) + \Omega_{\mathbf{A_r}, \mathbf{B_r}}(W_r) 
\end{align}
Also, by the same reasoning as in (\ref{Wabtrun}), we know that
\begin{align}
\Omega_{\mathbf{A_l}, \mathbf{B_l} }(W_l) = \Omega_{\mathbf{A_l}, \mathbf{B_l} }(W), \quad \Omega_{\mathbf{A_r}, \mathbf{B_r}}(W_r) = \Omega_{\mathbf{A_r}, \mathbf{B_r}}(W)
\end{align}
and by the same reasoning as (\ref{WUab}),
\begin{align}
\Omega_{\mathbf{A_l}, \mathbf{B_l} }(W) = \Omega_{A_l, B_l}(U), \quad \Omega_{\mathbf{A_r}, \mathbf{B_r} }(W) = \Omega_{A_r, B_r}(U)
\label{WUab2}
\end{align}
Combining (\ref{WUab}-\ref{WUab2}) proves the claim (\ref{sumaabb}).

\section{Derivation of non-overlapping formulas}\label{snonoverlapderivation}
\begin{figure}[tb]
   \centering
   \includegraphics[width=0.67\columnwidth]{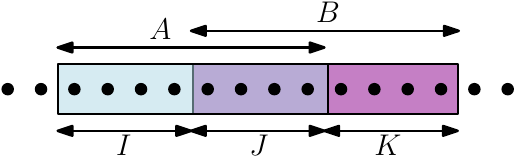} 
   \caption{To derive the non-overlapping formula for the edge invariant, we partition $A\cup B$ into three non-overlapping intervals $I,J,$ and $K$.}
   \label{fig:edgenonoverlap}
\end{figure}
\begin{figure}[tb]
   \centering
   \includegraphics[width=0.8\columnwidth]{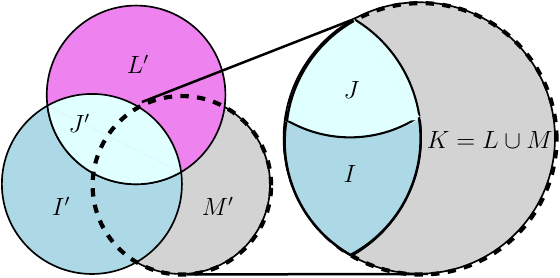} 
   \caption{To derive the non-overlapping formula for the bulk invariant, we partition $A\cup B\cup C$ into four non-overlapping sets $I',J',L',$ and $M'$. We denote their intersection with $C$ by $I,J,L,$ and $M$, with $K=L\cup M$.}
   \label{fig:nonoverlap}
\end{figure}

In this appendix, we consider spatially additive flows, i.e. flows obeying
\begin{align}\label{additive2}
\Omega_{AB,C}(U)&= \Omega_{A,C}(U)+ \Omega_{B,C}(U) \nonumber \\
\Omega_{A,BC}(U)&=\Omega_{A,B}(U)+ \Omega_{A,C}(U).
\end{align}
and we derive the ``non-overlapping'' formulas (\ref{flownonoverlap}) and (\ref{nonoverlap}) for their edge and bulk invariants.

We begin with the edge invariant $F(U_{\mathrm{edge}})$. Recall that this invariant is defined by $F(U_{\mathrm{edge}}) = \Omega_{A,B}(U_{\mathrm{edge}})$ where $A, B$ are two overlapping intervals. To derive the non-overlapping formula (\ref{flownonoverlap}), we decompose $A, B$ into three non-overlapping intervals $I$, $J$, and $K$,
with $A = I \cup J$ and $B = J \cup K$. This is illustrated in Fig.~\ref{fig:edgenonoverlap}. Using (\ref{additive2}) and omitting the argument $U_{\mathrm{edge}}$ in $\Omega_{A,B}(U_{\mathrm{edge}})$ for brevity, we have
\begin{align}
\begin{split}
F(U_{\mathrm{edge}})&=\Omega_{A,B}\\
&=\Omega_{I,J}+\Omega_{I,K}+\Omega_{J,J}+\Omega_{J,K}.
\end{split}
\end{align}
Next we simplify the above expression using Corollary~\ref{conservation}, which says that $\Omega_{A,B}=0$ if the boundaries of $A$ and $B$ are much further apart than the operator spreading length of  $U_{\mathrm{edge}}$. This means that $\Omega_{I,K}=0$ and 
\begin{align}
\Omega_{J,J}+\Omega_{J,K}+\Omega_{J,I}=\Omega_{J,I\cup J\cup K}=0.
\end{align}
Substituting $-\Omega_{J,I}$ for $\Omega_{J,J}+\Omega_{J,K}$, we obtain the desired non-overlapping formula for $F(U_{\mathrm{edge}})$:
\begin{equation}
F(U_{\mathrm{edge}})=\Omega_{I,J} - \Omega_{J,I}
\end{equation}

Next we consider the bulk invariant $M(\{U(t)\})$, which is defined by $M(\{U(t)\}) = \Omega_{A,B}^C(\{U(t)\})$. First, we define four non-overlapping regions $I',J',L',$ and $M'$, as shown in Fig.~\ref{fig:nonoverlap}. In particular, $A=I'\cup J'$ and $B=J'\cup L'$.  Using (\ref{additive2}) and omitting $U(t)$ in $\Omega_{A,B}(U(t))$ for brevity, we have
\begin{align}
\begin{split}\label{mnonoverlap}
M&(\{U(t)\})=\int_0^Tdt\frac{\partial}{ \partial t_C}\Omega_{A,B}(U(t))\\
&=\int_0^Tdt\frac{\partial}{\partial t_{C}}\left(\Omega_{I',J'}+\Omega_{I',L'}+\Omega_{J',J'}+\Omega_{J',L'}\right).
\end{split}
\end{align}
We claim that the second term, $\int_0^Tdt\frac{\partial}{\partial t_{C}}\Omega_{I',L'}$, vanishes. To see this, note that Eq.~(\ref{Momegaabc}) implies that
\begin{align*}
 \int_0^Tdt\frac{\partial}{\partial t_{C}}\Omega_{I',L'}=\Omega_{I' \cap C,L' \cap C}(U_{C}(T)), 
\end{align*} 
One can see that the right hand side vanishes using the fact that $U_C(T)$ is supported near the boundary of $C$, and the fact that $I' \cap \partial C$ and $L' \cap \partial C$ are far apart and then applying Corollary~\ref{corollary2}. By the same reasoning, $\int_0^Tdt\frac{\partial}{\partial t_{C}}\Omega_{J',M'}=0$. Subtracting $\int_0^Tdt\frac{\partial}{\partial t_{C}}\Omega_{I',L'}$ and adding $\int_0^Tdt\frac{\partial}{\partial t_{C}}\Omega_{J',M'}$ to  (\ref{mnonoverlap}), and defining $K'=L'\cup M'$, we get
\begin{equation}
M(\{U(t)\})=\int_0^Tdt\frac{\partial}{\partial t_{C}}\left(\Omega_{I',J'}+\Omega_{J',J'}+\Omega_{J',K'}\right).
\label{mnonoverlap2})
\end{equation}

Next we define $I=I'\cap C$, $J=J'\cap C$, and $K=(L'\cup M')\cap C$ as in Fig.~\ref{fig:nonoverlap} and we split $\frac{\partial}{\partial t_C}$ into three pieces:
\begin{align}
\frac{\partial}{\partial t_C} = \frac{\partial}{\partial t_I}+\frac{\partial}{\partial t_J}+\frac{\partial}{\partial t_K}.
\end{align}
Substituting this expression into (\ref{mnonoverlap2}) gives three terms involving $\frac{\partial}{\partial t_I}$, $\frac{\partial}{\partial t_J}$, and $\frac{\partial}{\partial t_K}$. We start with the $\frac{\partial}{\partial t_I}$ term. To simplify this term, we note that 
\begin{align}
\frac{\partial}{\partial t_I}\Omega_{I'\cup J'\cup K',J'}=0
\end{align}
by Corollary~\ref{corollary2} together with the observation that $I$ is far away from the point where the boundaries of $I'\cup J'\cup K'$ and $J'$ intersect. It follows that
\begin{align}
\frac{\partial}{\partial t_I} \Omega_{I',J'} +  \frac{\partial}{\partial t_I} \Omega_{J',J'} = - \frac{\partial}{\partial t_I} \Omega_{K',J'},
\end{align}
so that the $\frac{\partial}{\partial t_I}$ term can be rewritten as
\begin{align}
\begin{split}\label{tI}
\int_0^Tdt&\frac{\partial}{\partial t_I}\left(\Omega_{I',J'}+\Omega_{J',J'}+\Omega_{J',K'}\right)\\
&=\int_0^Tdt\frac{\partial}{\partial t_I}\left(\Omega_{J',K'}-\Omega_{K',J'}\right).
\end{split}
\end{align}
Similarly, using $\frac{\partial}{\partial t_K}\Omega_{J',I'\cup J'\cup K'}=0$, we can rewrite the $\frac{\partial}{\partial t_K}$ term as
\begin{align}
\begin{split}\label{tK}
\int_0^Tdt&\frac{\partial}{\partial t_K}\left(\Omega_{I',J'}+\Omega_{J',J'}+\Omega_{J',K'}\right)\\
&=\int_0^Tdt\frac{\partial}{\partial t_K}\left(\Omega_{I',J'}-\Omega_{J',I'}\right)
\end{split}
\end{align}
Finally, using 
\begin{align*}
\frac{\partial}{\partial t_J}\Omega_{I'\cup J'\cup K',J'}=\frac{\partial}{\partial t_J}\Omega_{I'\cup J'\cup K',K'}&=\frac{\partial}{\partial t_J}\Omega_{K',I'\cup J'\cup K'} \nonumber \\
&=0, 
\end{align*}
we can rewrite the $\frac{\partial}{\partial t_J}$ term as
\begin{align}
\begin{split}\label{tJ}
\int_0^Tdt&\frac{\partial}{\partial t_J}\left(\Omega_{I',J'}+\Omega_{J',J'}+\Omega_{J',K'}\right)\\
&=\int_0^Tdt\frac{\partial}{\partial t_J}\left(\Omega_{K',I'}-\Omega_{I',K'}\right)
\end{split}
\end{align}
Putting together (\ref{tI})-(\ref{tJ}), we get
\begin{align}
M(\{U(t)\}) &=\int_0^Tdt \frac{\partial}{\partial t_I}(\Omega_{J',K'} - \Omega_{K',J'}) \nonumber \\
&+ \int_0^Tdt \frac{\partial}{\partial t_J}(\Omega_{K',I'} - \Omega_{I',K'}) \nonumber \\
&+ \int_0^Tdt \frac{\partial}{\partial t_K}(\Omega_{I',J'} - \Omega_{J',I'}) 
\end{align}

To simplify further, we note that we can truncate $I'$, $J'$, and $K'$ to $I$, $J$, and $K$ using spatial additivity. For example, by spatial additivity, 
\begin{align}
\frac{\partial}{\partial t_I}\Omega_{J',K'}= \frac{\partial}{\partial t_I}\Omega_{J,K} + \frac{\partial}{\partial t_I}\Omega_{J,K_o} &+ \frac{\partial}{\partial t_I}\Omega_{J_o,K} \nonumber \\
& +\frac{\partial}{\partial t_I}\Omega_{J_o,K_o},
\end{align}
where $J_o=J'\setminus J$ and $K_o=K'\setminus K$. The latter three terms all vanish using Corollary~\ref{corollary2} since $I$ is far from the intersection of the boundaries of $J, K_o$, and $J_o, K$ and $J_o, K_o$ respectively. Hence, we deduce that $\frac{\partial}{\partial t_I}\Omega_{J',K'}= \frac{\partial}{\partial t_I}\Omega_{J,K}$. Applying the same truncation argument to the other terms, we obtain the desired nonoverlapping formula:
\begin{align}
M(\{U(t)\}) &=\int_0^Tdt \frac{\partial}{\partial t_I}(\Omega_{J, K} - \Omega_{K, J}) \nonumber \\
&+ \int_0^Tdt \frac{\partial}{\partial t_J}(\Omega_{K, I} - \Omega_{I, K}) \nonumber \\
&+ \int_0^Tdt \frac{\partial}{\partial t_K}(\Omega_{I, J} - \Omega_{J, I}) 
\end{align}

\section{$M(\{U(t)\})$ for a stationary Hamiltonian}\label{sstationary}
Consider a single-particle system whose time-independent Hamiltonian $H$ is a projector. For such a system, the time evolution operator $U(t)=e^{-iHt}$ satisfies $U(2\pi)=\mathbbm{1}$, so it forms a unitary loop with $T=2\pi$. For this system, we evaluate $M(\{U(t)\})$ using Eq.~(\ref{magcurrent}), and show that it is equal to the Chern number of the band that $H$ projects onto. 

Recall that to use (\ref{magcurrent}), we partition the plane into three non-overlapping regions $I, J,$ and $K$ that meet at a point. Note that $\mathcal{J}_{K,I}^I(t)$ is given by
\begin{equation}
\mathcal{J}_{K,I}^J(t)=i\mathrm{Tr}(U^\dagger(t)(P_KHP_I-P_IHP_K)U(t)P_J)
\end{equation}

Integrating $\mathcal{J}^J_{K,I}(t)$ over a period, using 
\begin{align}
U(t) = 1+(e^{-it}-1)H,
\end{align}
gives
\begin{equation}
\int_0^{2\pi}dt \mathcal{J}_{K,I}^J(t)=4\pi i\mathrm{Tr}(HP_KHP_IHP_J-HP_IHP_KHP_J)
\end{equation}

Then from Eq.~(\ref{magcurrent}) and the fact that the trace is invariant under cyclic permutations, we have
\begin{align}
\begin{split}\label{projector}
M(\{U(t)\})&=12\pi i\mathrm{Tr}(HP_KHP_IHP_J-HP_IHP_KHP_J)
\end{split}
\end{align}
The projector onto the ground state of $H$ is $P_{GS}=1-H$. Substituting $1-P_{GS}$ for $H$ in (\ref{projector}), we get precisely the real space formula for the Chern number of the ground state of $H$\cite{kitaev}. 

\section{An identity relating $\eta$ for sets and their complement}\label{sindexcomplement}

In this appendix, we derive an identity for $\eta$ that we will need in Appendix~\ref{soverlapindex}. Consider a unitary transformation $U$ defined on a lattice spin system. Let $A, B$ be two subsets of spins, and let $\overline{A}, \overline{B}$ be their complements. Also, let $\mathcal{A}, \mathcal{B}$ be operator algebras consisting of all operators supported in $A,B$, and let $\overline{\mathcal{A}}, \overline{\mathcal{B}}$ be the corresponding operator algebras for $\overline{A}, \overline{B}$. The identity that we will prove is as follows:
\begin{equation}\label{complement}
\eta(U^\dagger\mathcal{A}U, \mathcal{B})=\frac{d^{N_A+N_B}}{d^N}\eta(U^\dagger\overline{\mathcal{A}}U, \overline{\mathcal{B}}).
\end{equation}
Here, $N_A, N_B$ denote the number of spins in regions $A, B$, while $N$ denotes the total number of spins in the lattice.

To begin, we rewrite the definition of $\eta$ (\ref{etaoperators}) using the unnormalized trace $\mathrm{Tr}$ (instead of the  normalized trace ``$\mathrm{tr}$''):
\begin{align}
\begin{split}\label{etaoperators2}
\eta&(U^\dagger\mathcal{A}U,\mathcal{B})\\
&=\frac{d^{(N_A+N_B)/2}}{d^N}\sqrt{\sum_{O_a,O_b}|\mathrm{Tr}(U^{\dagger}O_a^\dagger U O_b )|^2}.
\end{split}
\end{align}
Here the  $O_a$ operators are normalized so that $\mathrm{Tr}(O_a^\dagger O_{a'}) = \delta_{aa'}$. 
Note that in (\ref{etaoperators}), the prefactor $\frac{d^{(N_A+N_B)/2}}{d^N}$ is hidden in the normalized trace ``$\mathrm{tr}$.'' 

To proceed further, it is useful to introduce a second copy of our lattice spin system. We then use the fact that a product of traces can be written as a trace over a tensor product to rewrite $\eta$ as expression involving two copies of our lattice:
\begin{widetext}
\begin{align}
\label{SWAPindex}
\eta(U^\dagger \mathcal{A}U, \mathcal{B} )&=\frac{d^{(N_A+N_B)/2}}{d^N}\sqrt{\sum_{O_a,O_b}\mathrm{Tr}[(U^\dagger\otimes U^\dagger)(O_{a}^\dagger \otimes O_{a}) (U \otimes U)(O_{b}\otimes O_{b}^\dagger) ]} \nonumber \\
&=\frac{d^{(N_A+N_B)/2}}{d^N}\sqrt{\mathrm{Tr}\left[(U^\dagger\otimes U^\dagger)\left(\sum_{O_a}O_{a}^\dagger \otimes O_{a}\right) (U \otimes U)\left(\sum_{O_b}O_{b}\otimes O_{b}^\dagger \right)\right]} 
\end{align}
Next, we use the identity
\begin{equation}\label{swapform}
\sum_{O_a}O_{a}^\dagger \otimes O_{a}= \text{SWAP}_{A}, \quad \sum_{O_b}O_{b}^\dagger \otimes O_{b}= \text{SWAP}_{B}
\end{equation}
where $\text{SWAP}_{A}$ denotes the unitary operator that acts like a $\text{SWAP}$ within region $A$ and acts like the identity outside of $A$, and similarly for $\text{SWAP}_{B}$. With this identity, we can write
\begin{align}
\eta(U^\dagger \mathcal{A}U, \mathcal{B} )
&=\frac{d^{(N_A+N_B)/2}}{d^N}\sqrt{\mathrm{Tr}[(U^\dagger\otimes U^\dagger)(\text{SWAP}_{A}) (U\otimes U)(\text{SWAP}_{B})]}.
\end{align}
Next, we insert $\text{SWAP}^2=1$ in this equation, where $\text{SWAP}$ exchanges the entire chains 1 and 2:
\begin{align}
\eta(U^\dagger \mathcal{A}U, \mathcal{B} )
&=\frac{d^{(N_A+N_B)/2}}{d^N}\sqrt{\mathrm{Tr}[(U^\dagger\otimes U^\dagger)(\text{SWAP}_{A})(\text{SWAP}^2) (U\otimes U)(\text{SWAP}_{B})]}.
\end{align}
Using the fact that $[\text{SWAP},U\otimes U]=0$, we can commute the $\text{SWAP}$ through and rewrite this expression as
\begin{align}
\eta(U^\dagger \mathcal{A}U, \mathcal{B} )
&=\frac{d^{(N_A+N_B)/2}}{d^N}\sqrt{\mathrm{Tr}[(U^\dagger\otimes U^\dagger)(\text{SWAP}_{A} \cdot \text{SWAP})  (U\otimes U)( \text{SWAP} \cdot \text{SWAP}_{B})]}.
\end{align}
Notice that 
\begin{equation}
\text{SWAP}_{A}\cdot\text{SWAP}=\text{SWAP}_{\overline{A}}\qquad \text{SWAP} \cdot\text{SWAP}_{B}=\text{SWAP}_{\overline{B}}.
\end{equation}
This allows us to simplify the above expression as follows:
\begin{align}
\begin{split}\label{cproof}
\eta(U^\dagger \mathcal{A}U, \mathcal{B} )&=\frac{d^{(N_A+N_B)/2}}{d^N}\sqrt{\mathrm{Tr}[(U^\dagger\otimes U^\dagger)(\text{SWAP}_{\overline{A}}) (U\otimes U)(\text{SWAP}_{\overline{B}})]}\\
&=\frac{d^{N_A+N_B}}{d^N}\eta(U^{\dagger}\overline{\mathcal{A}}U,\overline{\mathcal{B}}),
\end{split}
\end{align}
\end{widetext}
where in the last line we used $N=N_A+N_{\overline{A}}=N_B+N_{\overline{B}}$. This concludes the proof of (\ref{complement}).

\section{Overlapping formula for the GNVW index}\label{soverlapindex}
\begin{figure}[tb]
   \centering
   \includegraphics[width=.5\columnwidth]{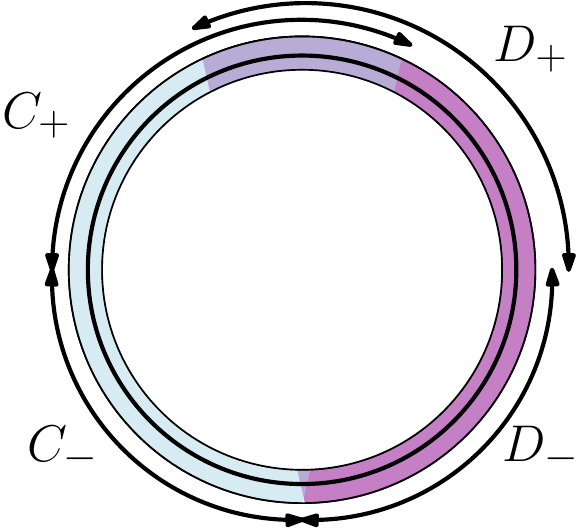} 
   \caption{To derive the overlapping formula for the GNVW index, we consider a circular spin chain and two intervals, $C=C_+\cup C_-$ and $D=D_+\cup D_-$, which are adjacent in the lower half of the spin chain and overlapping in the upper half of the spin chain.}
   \label{fig:ABcircle}
\end{figure}
In this appendix, we derive Eq.~(\ref{edgeGNVWid}), i.e. we show that our edge invariant $F(U)$ is related to the GNVW index $\text{ind}(U)$ by
\begin{align}
F(U) = \log \text{ind}(U)
\end{align}
This amounts to proving the following identity. Let $A, B$ be two large \emph{overlapping} intervals in some spin chain, and let $\mathcal{A}, \mathcal{B}$ be operator algebras consisting of all operators supported on $A, B$. Likewise, let $A', B'$ be two large \emph{non-overlapping} adjacent, intervals, and let $\mathcal{A}', \mathcal{B}'$ be the corresponding operator algebras. The identity we need to prove is: 
\begin{align}
\frac{\eta(U^{\dagger}\mathcal{A}U,\mathcal{B})}{\eta(\mathcal{A},\mathcal{B})}=\frac{\eta(U^\dagger \mathcal{A}' U, \mathcal{B}')}{\eta(\mathcal{A}', U^\dagger\mathcal{B}' U)}.
\label{GNVWid}
\end{align} 
Here the left-hand side is the exponential of our edge invariant $\exp[F(U)]$ while the right hand side is the standard formula for the GNVW index, $\text{ind}(U)$.

To establish the identity (\ref{GNVWid}), we consider a 1D chain in a periodic ring geometry. We consider two intervals $C, D$ that are adjacent (but non-overlapping) at the bottom part of the chain and that overlap at the top part of the chain (see Fig.~\ref{fig:ABcircle}). We partition $C, D$ into two pieces, $C = C_-\cup C_+$ and $D = D_- \cup D_+$ where $C_+$ and $C_-$ are the parts of $C$ in the upper and lower half of the chain, and similarly for $D_+$ and $D_-$. 

Let $\mathcal{C}, \mathcal{D}$ and $\mathcal{C}_\pm$ and $\mathcal{D}_\pm$ be the corresponding operator algebras and consider the quantity $\eta(\mathcal{C}, U^\dagger \mathcal{D} U)$. Because operators that have support near the middle of $C$ or $D$, do not contribute to $\eta$, we can factor $\eta(U^\dagger\mathcal{C}U,\mathcal{D})$ into two terms: 
\begin{equation}\label{ABindex}
\eta(U^\dagger\mathcal{C} U, \mathcal{D})=\eta(U^\dagger\mathcal{C}_+ U, \mathcal{D}_+)\eta(U^\dagger\mathcal{C}_-U, \mathcal{D}_- ),
\end{equation}

Next, let $\overline{C}, \overline{D}$ be the complements of $C, D$ and let $\overline{\mathcal{C}}, \overline{\mathcal{D}}$ be the corresponding algebras. By the identity (\ref{complement}),
\begin{equation}
\eta(U^\dagger \mathcal{C}U, \mathcal{D})=\frac{d^{N_C+N_D}}{d^N}\eta(U^\dagger\overline{\mathcal{C}}U,  \overline{\mathcal{D}}).
\end{equation}
At the same time, if we compare the interval $\overline{C}$ to $D_-$, and likewise we compare $\overline{D}$ to $C_-$, we can see that 
\begin{equation}
\eta(U^\dagger\overline{\mathcal{C}}U,  \overline{\mathcal{D}})= \eta(U^\dagger\mathcal{D}_- U,\mathcal{C}_- )
\end{equation}
since these two pairs of intervals are identical in the region where they touch, i.e. the region that contributes to $\eta$. Hence, we have
\begin{equation}
\eta(U^{\dagger}\mathcal{C}U,\mathcal{D} )=\frac{d^{N_C+N_D}}{d^N}\eta(U^\dagger\mathcal{D}_- U,\mathcal{C}_- ).
\label{etaCD}
\end{equation}

To proceed further we note that the prefactor $\frac{d^{N_C+N_D}}{d^N}$ can be rewritten as
\begin{align}
\frac{d^{N_C+N_D}}{d^N} &= d^{N_{C \cap D}} \nonumber \\
&= d^{N_{C_+ \cap D_+}} \nonumber \\
&= \eta(\mathcal{C}_+,\mathcal{D}_+)
\end{align}
so (\ref{etaCD}) can be written as
\begin{equation}
\eta (U^\dagger\mathcal{C}U, \mathcal{D})= \eta(\mathcal{C}_+,\mathcal{D}_+) \eta(U^\dagger\mathcal{D}_- U,\mathcal{C}_- ).
\end{equation}
Substituting this identity into (\ref{ABindex}) gives
\begin{align}
\begin{split}
\frac{\eta(U^{\dagger}\mathcal{C}_+U, \mathcal{D}_+ )}{\eta(\mathcal{C}_+,\mathcal{D}_+)}&=\frac{\eta(U^\dagger\mathcal{D}_- U,\mathcal{C}_- )}{\eta(U^\dagger \mathcal{C}_- U,\mathcal{D}_- )}\\
&=\frac{\eta(U^\dagger\mathcal{D}_- U,\mathcal{C}_- )}{\eta(\mathcal{D}_- ,U^\dagger \mathcal{C}_- U)}
\end{split}
\end{align}
Finally, identifying $C_+$ and $D_+$ with the overlapping intervals $A, B$, and identifying $D_-$ and $C_-$ with the non-overlapping intervals $A', B'$ we recover the
desired identity (\ref{GNVWid}).

\bibliography{flowsfloquetbib}

\begin{thebibliography}{37}%
\makeatletter
\providecommand \@ifxundefined [1]{%
 \@ifx{#1\undefined}
}%
\providecommand \@ifnum [1]{%
 \ifnum #1\expandafter \@firstoftwo
 \else \expandafter \@secondoftwo
 \fi
}%
\providecommand \@ifx [1]{%
 \ifx #1\expandafter \@firstoftwo
 \else \expandafter \@secondoftwo
 \fi
}%
\providecommand \natexlab [1]{#1}%
\providecommand \enquote  [1]{``#1''}%
\providecommand \bibnamefont  [1]{#1}%
\providecommand \bibfnamefont [1]{#1}%
\providecommand \citenamefont [1]{#1}%
\providecommand \href@noop [0]{\@secondoftwo}%
\providecommand \href [0]{\begingroup \@sanitize@url \@href}%
\providecommand \@href[1]{\@@startlink{#1}\@@href}%
\providecommand \@@href[1]{\endgroup#1\@@endlink}%
\providecommand \@sanitize@url [0]{\catcode `\\12\catcode `\$12\catcode
  `\&12\catcode `\#12\catcode `\^12\catcode `\_12\catcode `\%12\relax}%
\providecommand \@@startlink[1]{}%
\providecommand \@@endlink[0]{}%
\providecommand \url  [0]{\begingroup\@sanitize@url \@url }%
\providecommand \@url [1]{\endgroup\@href {#1}{\urlprefix }}%
\providecommand \urlprefix  [0]{URL }%
\providecommand \Eprint [0]{\href }%
\providecommand \doibase [0]{http://dx.doi.org/}%
\providecommand \selectlanguage [0]{\@gobble}%
\providecommand \bibinfo  [0]{\@secondoftwo}%
\providecommand \bibfield  [0]{\@secondoftwo}%
\providecommand \translation [1]{[#1]}%
\providecommand \BibitemOpen [0]{}%
\providecommand \bibitemStop [0]{}%
\providecommand \bibitemNoStop [0]{.\EOS\space}%
\providecommand \EOS [0]{\spacefactor3000\relax}%
\providecommand \BibitemShut  [1]{\csname bibitem#1\endcsname}%
\let\auto@bib@innerbib\@empty
\bibitem [{\citenamefont {Harper}\ \emph {et~al.}(2020)\citenamefont {Harper},
  \citenamefont {Roy}, \citenamefont {Rudner},\ and\ \citenamefont
  {Sondhi}}]{floquetreview}%
  \BibitemOpen
  \bibfield  {author} {\bibinfo {author} {\bibfnamefont {F.}~\bibnamefont
  {Harper}}, \bibinfo {author} {\bibfnamefont {R.}~\bibnamefont {Roy}},
  \bibinfo {author} {\bibfnamefont {M.~S.}\ \bibnamefont {Rudner}}, \ and\
  \bibinfo {author} {\bibfnamefont {S.}~\bibnamefont {Sondhi}},\ }\href
  {\doibase 10.1146/annurev-conmatphys-031218-013721} {\bibfield  {journal}
  {\bibinfo  {journal} {Annual Review of Condensed Matter Physics}\ }\textbf
  {\bibinfo {volume} {11}},\ \bibinfo {pages} {345} (\bibinfo {year}
  {2020})}\BibitemShut {NoStop}%
\bibitem [{\citenamefont {Rudner}\ and\ \citenamefont
  {Lindner}(2020)}]{rudnerband}%
  \BibitemOpen
  \bibfield  {author} {\bibinfo {author} {\bibfnamefont {M.~S.}\ \bibnamefont
  {Rudner}}\ and\ \bibinfo {author} {\bibfnamefont {N.~H.}\ \bibnamefont
  {Lindner}},\ }\href {\doibase 10.1038/s42254-020-0170-z} {\bibfield
  {journal} {\bibinfo  {journal} {Nature Reviews Physics}\ ,\ \bibinfo {pages}
  {1}} (\bibinfo {year} {2020})}\BibitemShut {NoStop}%
\bibitem [{\citenamefont {Kitagawa}\ \emph {et~al.}(2010)\citenamefont
  {Kitagawa}, \citenamefont {Berg}, \citenamefont {Rudner},\ and\ \citenamefont
  {Demler}}]{kitagawa2010}%
  \BibitemOpen
  \bibfield  {author} {\bibinfo {author} {\bibfnamefont {T.}~\bibnamefont
  {Kitagawa}}, \bibinfo {author} {\bibfnamefont {E.}~\bibnamefont {Berg}},
  \bibinfo {author} {\bibfnamefont {M.}~\bibnamefont {Rudner}}, \ and\ \bibinfo
  {author} {\bibfnamefont {E.}~\bibnamefont {Demler}},\ }\href {\doibase
  10.1103/PhysRevB.82.235114} {\bibfield  {journal} {\bibinfo  {journal} {Phys.
  Rev. B}\ }\textbf {\bibinfo {volume} {82}},\ \bibinfo {pages} {235114}
  (\bibinfo {year} {2010})}\BibitemShut {NoStop}%
\bibitem [{\citenamefont {Rudner}\ \emph {et~al.}(2013)\citenamefont {Rudner},
  \citenamefont {Lindner}, \citenamefont {Berg},\ and\ \citenamefont
  {Levin}}]{anomalousedge}%
  \BibitemOpen
  \bibfield  {author} {\bibinfo {author} {\bibfnamefont {M.~S.}\ \bibnamefont
  {Rudner}}, \bibinfo {author} {\bibfnamefont {N.~H.}\ \bibnamefont {Lindner}},
  \bibinfo {author} {\bibfnamefont {E.}~\bibnamefont {Berg}}, \ and\ \bibinfo
  {author} {\bibfnamefont {M.}~\bibnamefont {Levin}},\ }\href {\doibase
  10.1103/PhysRevX.3.031005} {\bibfield  {journal} {\bibinfo  {journal} {Phys.
  Rev. X}\ }\textbf {\bibinfo {volume} {3}},\ \bibinfo {pages} {031005}
  (\bibinfo {year} {2013})}\BibitemShut {NoStop}%
\bibitem [{\citenamefont {Nathan}\ \emph {et~al.}(2017)\citenamefont {Nathan},
  \citenamefont {Rudner}, \citenamefont {Lindner}, \citenamefont {Berg},\ and\
  \citenamefont {Refael}}]{magnetization}%
  \BibitemOpen
  \bibfield  {author} {\bibinfo {author} {\bibfnamefont {F.}~\bibnamefont
  {Nathan}}, \bibinfo {author} {\bibfnamefont {M.~S.}\ \bibnamefont {Rudner}},
  \bibinfo {author} {\bibfnamefont {N.~H.}\ \bibnamefont {Lindner}}, \bibinfo
  {author} {\bibfnamefont {E.}~\bibnamefont {Berg}}, \ and\ \bibinfo {author}
  {\bibfnamefont {G.}~\bibnamefont {Refael}},\ }\href {\doibase
  10.1103/PhysRevLett.119.186801} {\bibfield  {journal} {\bibinfo  {journal}
  {Phys. Rev. Lett.}\ }\textbf {\bibinfo {volume} {119}},\ \bibinfo {pages}
  {186801} (\bibinfo {year} {2017})}\BibitemShut {NoStop}%
\bibitem [{\citenamefont {Nathan}\ and\ \citenamefont
  {Rudner}(2015)}]{nathan2015}%
  \BibitemOpen
  \bibfield  {author} {\bibinfo {author} {\bibfnamefont {F.}~\bibnamefont
  {Nathan}}\ and\ \bibinfo {author} {\bibfnamefont {M.~S.}\ \bibnamefont
  {Rudner}},\ }\href
  {https://iopscience.iop.org/article/10.1088/1367-2630/17/12/125014/meta}
  {\bibfield  {journal} {\bibinfo  {journal} {New Journal of Physics}\ }\textbf
  {\bibinfo {volume} {17}},\ \bibinfo {pages} {125014} (\bibinfo {year}
  {2015})}\BibitemShut {NoStop}%
\bibitem [{\citenamefont {Titum}\ \emph {et~al.}(2016)\citenamefont {Titum},
  \citenamefont {Berg}, \citenamefont {Rudner}, \citenamefont {Refael},\ and\
  \citenamefont {Lindner}}]{afai}%
  \BibitemOpen
  \bibfield  {author} {\bibinfo {author} {\bibfnamefont {P.}~\bibnamefont
  {Titum}}, \bibinfo {author} {\bibfnamefont {E.}~\bibnamefont {Berg}},
  \bibinfo {author} {\bibfnamefont {M.~S.}\ \bibnamefont {Rudner}}, \bibinfo
  {author} {\bibfnamefont {G.}~\bibnamefont {Refael}}, \ and\ \bibinfo {author}
  {\bibfnamefont {N.~H.}\ \bibnamefont {Lindner}},\ }\href {\doibase
  10.1103/PhysRevX.6.021013} {\bibfield  {journal} {\bibinfo  {journal} {Phys.
  Rev. X}\ }\textbf {\bibinfo {volume} {6}},\ \bibinfo {pages} {021013}
  (\bibinfo {year} {2016})}\BibitemShut {NoStop}%
\bibitem [{\citenamefont {Graf}\ and\ \citenamefont {Tauber}(2018)}]{graf2018}%
  \BibitemOpen
  \bibfield  {author} {\bibinfo {author} {\bibfnamefont {G.~M.}\ \bibnamefont
  {Graf}}\ and\ \bibinfo {author} {\bibfnamefont {C.}~\bibnamefont {Tauber}},\
  }in\ \href {https://link.springer.com/article/10.1007/s00023-018-0657-7}
  {\emph {\bibinfo {booktitle} {Annales Henri Poincar{\'e}}}},\ Vol.~\bibinfo
  {volume} {19}\ (\bibinfo {organization} {Springer},\ \bibinfo {year} {2018})\
  pp.\ \bibinfo {pages} {709--741}\BibitemShut {NoStop}%
\bibitem [{\citenamefont {Shapiro}\ and\ \citenamefont
  {Tauber}(2019)}]{shapiro2019}%
  \BibitemOpen
  \bibfield  {author} {\bibinfo {author} {\bibfnamefont {J.}~\bibnamefont
  {Shapiro}}\ and\ \bibinfo {author} {\bibfnamefont {C.}~\bibnamefont
  {Tauber}},\ }in\ \href
  {https://link.springer.com/article/10.1007/s00023-019-00794-3} {\emph
  {\bibinfo {booktitle} {Annales Henri Poincar{\'e}}}},\ Vol.~\bibinfo {volume}
  {20}\ (\bibinfo {organization} {Springer},\ \bibinfo {year} {2019})\ pp.\
  \bibinfo {pages} {1837--1875}\BibitemShut {NoStop}%
\bibitem [{\citenamefont {Vu}(2022)}]{vu2022}%
  \BibitemOpen
  \bibfield  {author} {\bibinfo {author} {\bibfnamefont {D.}~\bibnamefont
  {Vu}},\ }\href {\doibase 10.1103/PhysRevB.105.064304} {\bibfield  {journal}
  {\bibinfo  {journal} {Phys. Rev. B}\ }\textbf {\bibinfo {volume} {105}},\
  \bibinfo {pages} {064304} (\bibinfo {year} {2022})}\BibitemShut {NoStop}%
\bibitem [{\citenamefont {Po}\ \emph {et~al.}(2016)\citenamefont {Po},
  \citenamefont {Fidkowski}, \citenamefont {Morimoto}, \citenamefont {Potter},\
  and\ \citenamefont {Vishwanath}}]{chiralbosons}%
  \BibitemOpen
  \bibfield  {author} {\bibinfo {author} {\bibfnamefont {H.~C.}\ \bibnamefont
  {Po}}, \bibinfo {author} {\bibfnamefont {L.}~\bibnamefont {Fidkowski}},
  \bibinfo {author} {\bibfnamefont {T.}~\bibnamefont {Morimoto}}, \bibinfo
  {author} {\bibfnamefont {A.~C.}\ \bibnamefont {Potter}}, \ and\ \bibinfo
  {author} {\bibfnamefont {A.}~\bibnamefont {Vishwanath}},\ }\href {\doibase
  10.1103/PhysRevX.6.041070} {\bibfield  {journal} {\bibinfo  {journal} {Phys.
  Rev. X}\ }\textbf {\bibinfo {volume} {6}},\ \bibinfo {pages} {041070}
  (\bibinfo {year} {2016})}\BibitemShut {NoStop}%
\bibitem [{\citenamefont {Harper}\ and\ \citenamefont
  {Roy}(2017)}]{harperorder}%
  \BibitemOpen
  \bibfield  {author} {\bibinfo {author} {\bibfnamefont {F.}~\bibnamefont
  {Harper}}\ and\ \bibinfo {author} {\bibfnamefont {R.}~\bibnamefont {Roy}},\
  }\href {\doibase 10.1103/PhysRevLett.118.115301} {\bibfield  {journal}
  {\bibinfo  {journal} {Phys. Rev. Lett.}\ }\textbf {\bibinfo {volume} {118}},\
  \bibinfo {pages} {115301} (\bibinfo {year} {2017})}\BibitemShut {NoStop}%
\bibitem [{\citenamefont {Gross}\ \emph {et~al.}(2012)\citenamefont {Gross},
  \citenamefont {Nesme}, \citenamefont {Vogts},\ and\ \citenamefont
  {Werner}}]{GNVW}%
  \BibitemOpen
  \bibfield  {author} {\bibinfo {author} {\bibfnamefont {D.}~\bibnamefont
  {Gross}}, \bibinfo {author} {\bibfnamefont {V.}~\bibnamefont {Nesme}},
  \bibinfo {author} {\bibfnamefont {H.}~\bibnamefont {Vogts}}, \ and\ \bibinfo
  {author} {\bibfnamefont {R.~F.}\ \bibnamefont {Werner}},\ }\href {\doibase
  10.1007/s00220-012-1423-1} {\bibfield  {journal} {\bibinfo  {journal}
  {Communications in Mathematical Physics}\ }\textbf {\bibinfo {volume}
  {310}},\ \bibinfo {pages} {419} (\bibinfo {year} {2012})}\BibitemShut
  {NoStop}%
\bibitem [{\citenamefont {Gong}\ \emph {et~al.}(2020)\citenamefont {Gong},
  \citenamefont {S\"underhauf}, \citenamefont {Schuch},\ and\ \citenamefont
  {Cirac}}]{mpu}%
  \BibitemOpen
  \bibfield  {author} {\bibinfo {author} {\bibfnamefont {Z.}~\bibnamefont
  {Gong}}, \bibinfo {author} {\bibfnamefont {C.}~\bibnamefont {S\"underhauf}},
  \bibinfo {author} {\bibfnamefont {N.}~\bibnamefont {Schuch}}, \ and\ \bibinfo
  {author} {\bibfnamefont {J.~I.}\ \bibnamefont {Cirac}},\ }\href {\doibase
  10.1103/PhysRevLett.124.100402} {\bibfield  {journal} {\bibinfo  {journal}
  {Phys. Rev. Lett.}\ }\textbf {\bibinfo {volume} {124}},\ \bibinfo {pages}
  {100402} (\bibinfo {year} {2020})}\BibitemShut {NoStop}%
\bibitem [{\citenamefont {Duschatko}\ \emph {et~al.}(2018)\citenamefont
  {Duschatko}, \citenamefont {Dumitrescu},\ and\ \citenamefont
  {Potter}}]{tracking}%
  \BibitemOpen
  \bibfield  {author} {\bibinfo {author} {\bibfnamefont {B.~R.}\ \bibnamefont
  {Duschatko}}, \bibinfo {author} {\bibfnamefont {P.~T.}\ \bibnamefont
  {Dumitrescu}}, \ and\ \bibinfo {author} {\bibfnamefont {A.~C.}\ \bibnamefont
  {Potter}},\ }\href {\doibase 10.1103/PhysRevB.98.054309} {\bibfield
  {journal} {\bibinfo  {journal} {Phys. Rev. B}\ }\textbf {\bibinfo {volume}
  {98}},\ \bibinfo {pages} {054309} (\bibinfo {year} {2018})}\BibitemShut
  {NoStop}%
\bibitem [{\citenamefont {Ranard}\ \emph {et~al.}(2022)\citenamefont {Ranard},
  \citenamefont {Walter},\ and\ \citenamefont {Witteveen}}]{ranard2022}%
  \BibitemOpen
  \bibfield  {author} {\bibinfo {author} {\bibfnamefont {D.}~\bibnamefont
  {Ranard}}, \bibinfo {author} {\bibfnamefont {M.}~\bibnamefont {Walter}}, \
  and\ \bibinfo {author} {\bibfnamefont {F.}~\bibnamefont {Witteveen}},\ }in\
  \href {https://link.springer.com/article/10.1007/s00023-022-01193-x} {\emph
  {\bibinfo {booktitle} {Annales Henri Poincar{\'e}}}}\ (\bibinfo
  {organization} {Springer},\ \bibinfo {year} {2022})\ pp.\ \bibinfo {pages}
  {1--75}\BibitemShut {NoStop}%
\bibitem [{\citenamefont {Zhang}\ and\ \citenamefont
  {Levin}(2021)}]{u1floquet}%
  \BibitemOpen
  \bibfield  {author} {\bibinfo {author} {\bibfnamefont {C.}~\bibnamefont
  {Zhang}}\ and\ \bibinfo {author} {\bibfnamefont {M.}~\bibnamefont {Levin}},\
  }\href {\doibase 10.1103/PhysRevB.103.064302} {\bibfield  {journal} {\bibinfo
   {journal} {Phys. Rev. B}\ }\textbf {\bibinfo {volume} {103}},\ \bibinfo
  {pages} {064302} (\bibinfo {year} {2021})}\BibitemShut {NoStop}%
\bibitem [{\citenamefont {Nathan}\ \emph
  {et~al.}(2019{\natexlab{a}})\citenamefont {Nathan}, \citenamefont {Abanin},
  \citenamefont {Berg}, \citenamefont {Lindner},\ and\ \citenamefont
  {Rudner}}]{nathanafi}%
  \BibitemOpen
  \bibfield  {author} {\bibinfo {author} {\bibfnamefont {F.}~\bibnamefont
  {Nathan}}, \bibinfo {author} {\bibfnamefont {D.}~\bibnamefont {Abanin}},
  \bibinfo {author} {\bibfnamefont {E.}~\bibnamefont {Berg}}, \bibinfo {author}
  {\bibfnamefont {N.~H.}\ \bibnamefont {Lindner}}, \ and\ \bibinfo {author}
  {\bibfnamefont {M.~S.}\ \bibnamefont {Rudner}},\ }\href {\doibase
  10.1103/PhysRevB.99.195133} {\bibfield  {journal} {\bibinfo  {journal} {Phys.
  Rev. B}\ }\textbf {\bibinfo {volume} {99}},\ \bibinfo {pages} {195133}
  (\bibinfo {year} {2019}{\natexlab{a}})}\BibitemShut {NoStop}%
\bibitem [{\citenamefont {Glorioso}\ \emph {et~al.}(2019)\citenamefont
  {Glorioso}, \citenamefont {Gromov},\ and\ \citenamefont {Ryu}}]{eft}%
  \BibitemOpen
  \bibfield  {author} {\bibinfo {author} {\bibfnamefont {P.}~\bibnamefont
  {Glorioso}}, \bibinfo {author} {\bibfnamefont {A.}~\bibnamefont {Gromov}}, \
  and\ \bibinfo {author} {\bibfnamefont {S.}~\bibnamefont {Ryu}},\ }\href
  {https://arxiv.org/abs/1908.03217} {\bibfield  {journal} {\bibinfo  {journal}
  {arXiv preprint arXiv:1908.03217}\ } (\bibinfo {year} {2019})}\BibitemShut
  {NoStop}%
\bibitem [{\citenamefont {D'Alessio}\ and\ \citenamefont
  {Rigol}(2014)}]{rigollongtime}%
  \BibitemOpen
  \bibfield  {author} {\bibinfo {author} {\bibfnamefont {L.}~\bibnamefont
  {D'Alessio}}\ and\ \bibinfo {author} {\bibfnamefont {M.}~\bibnamefont
  {Rigol}},\ }\href {\doibase 10.1103/PhysRevX.4.041048} {\bibfield  {journal}
  {\bibinfo  {journal} {Phys. Rev. X}\ }\textbf {\bibinfo {volume} {4}},\
  \bibinfo {pages} {041048} (\bibinfo {year} {2014})}\BibitemShut {NoStop}%
\bibitem [{\citenamefont {Lazarides}\ \emph {et~al.}(2014)\citenamefont
  {Lazarides}, \citenamefont {Das},\ and\ \citenamefont
  {Moessner}}]{lazaridesgeneric}%
  \BibitemOpen
  \bibfield  {author} {\bibinfo {author} {\bibfnamefont {A.}~\bibnamefont
  {Lazarides}}, \bibinfo {author} {\bibfnamefont {A.}~\bibnamefont {Das}}, \
  and\ \bibinfo {author} {\bibfnamefont {R.}~\bibnamefont {Moessner}},\ }\href
  {\doibase 10.1103/PhysRevE.90.012110} {\bibfield  {journal} {\bibinfo
  {journal} {Phys. Rev. E}\ }\textbf {\bibinfo {volume} {90}},\ \bibinfo
  {pages} {012110} (\bibinfo {year} {2014})}\BibitemShut {NoStop}%
\bibitem [{\citenamefont {Ponte}\ \emph
  {et~al.}(2015{\natexlab{a}})\citenamefont {Ponte}, \citenamefont {Chandran},
  \citenamefont {Papić},\ and\ \citenamefont {Abanin}}]{ponteergodic}%
  \BibitemOpen
  \bibfield  {author} {\bibinfo {author} {\bibfnamefont {P.}~\bibnamefont
  {Ponte}}, \bibinfo {author} {\bibfnamefont {A.}~\bibnamefont {Chandran}},
  \bibinfo {author} {\bibfnamefont {Z.}~\bibnamefont {Papić}}, \ and\ \bibinfo
  {author} {\bibfnamefont {D.~A.}\ \bibnamefont {Abanin}},\ }\href {\doibase
  https://doi.org/10.1016/j.aop.2014.11.008} {\bibfield  {journal} {\bibinfo
  {journal} {Annals of Physics}\ }\textbf {\bibinfo {volume} {353}},\ \bibinfo
  {pages} {196 } (\bibinfo {year} {2015}{\natexlab{a}})}\BibitemShut {NoStop}%
\bibitem [{\citenamefont {Lazarides}\ \emph {et~al.}(2015)\citenamefont
  {Lazarides}, \citenamefont {Das},\ and\ \citenamefont {Moessner}}]{fate}%
  \BibitemOpen
  \bibfield  {author} {\bibinfo {author} {\bibfnamefont {A.}~\bibnamefont
  {Lazarides}}, \bibinfo {author} {\bibfnamefont {A.}~\bibnamefont {Das}}, \
  and\ \bibinfo {author} {\bibfnamefont {R.}~\bibnamefont {Moessner}},\ }\href
  {\doibase 10.1103/PhysRevLett.115.030402} {\bibfield  {journal} {\bibinfo
  {journal} {Phys. Rev. Lett.}\ }\textbf {\bibinfo {volume} {115}},\ \bibinfo
  {pages} {030402} (\bibinfo {year} {2015})}\BibitemShut {NoStop}%
\bibitem [{\citenamefont {Ponte}\ \emph
  {et~al.}(2015{\natexlab{b}})\citenamefont {Ponte}, \citenamefont
  {Papi\ifmmode~\acute{c}\else \'{c}\fi{}}, \citenamefont {Huveneers},\ and\
  \citenamefont {Abanin}}]{ponte}%
  \BibitemOpen
  \bibfield  {author} {\bibinfo {author} {\bibfnamefont {P.}~\bibnamefont
  {Ponte}}, \bibinfo {author} {\bibfnamefont {Z.}~\bibnamefont
  {Papi\ifmmode~\acute{c}\else \'{c}\fi{}}}, \bibinfo {author} {\bibfnamefont
  {F.~m.~c.}\ \bibnamefont {Huveneers}}, \ and\ \bibinfo {author}
  {\bibfnamefont {D.~A.}\ \bibnamefont {Abanin}},\ }\href {\doibase
  10.1103/PhysRevLett.114.140401} {\bibfield  {journal} {\bibinfo  {journal}
  {Phys. Rev. Lett.}\ }\textbf {\bibinfo {volume} {114}},\ \bibinfo {pages}
  {140401} (\bibinfo {year} {2015}{\natexlab{b}})}\BibitemShut {NoStop}%
\bibitem [{\citenamefont {Abanin}\ \emph {et~al.}(2016)\citenamefont {Abanin},
  \citenamefont {Roeck},\ and\ \citenamefont {Huveneers}}]{abanintheory}%
  \BibitemOpen
  \bibfield  {author} {\bibinfo {author} {\bibfnamefont {D.~A.}\ \bibnamefont
  {Abanin}}, \bibinfo {author} {\bibfnamefont {W.~D.}\ \bibnamefont {Roeck}}, \
  and\ \bibinfo {author} {\bibfnamefont {F.}~\bibnamefont {Huveneers}},\ }\href
  {\doibase https://doi.org/10.1016/j.aop.2016.03.010} {\bibfield  {journal}
  {\bibinfo  {journal} {Annals of Physics}\ }\textbf {\bibinfo {volume}
  {372}},\ \bibinfo {pages} {1 } (\bibinfo {year} {2016})}\BibitemShut
  {NoStop}%
\bibitem [{\citenamefont {Roy}\ and\ \citenamefont
  {Harper}(2017)}]{alldimensions}%
  \BibitemOpen
  \bibfield  {author} {\bibinfo {author} {\bibfnamefont {R.}~\bibnamefont
  {Roy}}\ and\ \bibinfo {author} {\bibfnamefont {F.}~\bibnamefont {Harper}},\
  }\href {\doibase 10.1103/PhysRevB.95.195128} {\bibfield  {journal} {\bibinfo
  {journal} {Phys. Rev. B}\ }\textbf {\bibinfo {volume} {95}},\ \bibinfo
  {pages} {195128} (\bibinfo {year} {2017})}\BibitemShut {NoStop}%
\bibitem [{\citenamefont {Kitaev}(2006)}]{kitaev}%
  \BibitemOpen
  \bibfield  {author} {\bibinfo {author} {\bibfnamefont {A.}~\bibnamefont
  {Kitaev}},\ }\href {\doibase https://doi.org/10.1016/j.aop.2005.10.005}
  {\bibfield  {journal} {\bibinfo  {journal} {Annals of Physics}\ }\textbf
  {\bibinfo {volume} {321}},\ \bibinfo {pages} {2 } (\bibinfo {year} {2006})},\
  \bibinfo {note} {january Special Issue}\BibitemShut {NoStop}%
\bibitem [{\citenamefont {Potter}\ and\ \citenamefont
  {Vasseur}(2016)}]{nonabelian}%
  \BibitemOpen
  \bibfield  {author} {\bibinfo {author} {\bibfnamefont {A.~C.}\ \bibnamefont
  {Potter}}\ and\ \bibinfo {author} {\bibfnamefont {R.}~\bibnamefont
  {Vasseur}},\ }\href {\doibase 10.1103/PhysRevB.94.224206} {\bibfield
  {journal} {\bibinfo  {journal} {Phys. Rev. B}\ }\textbf {\bibinfo {volume}
  {94}},\ \bibinfo {pages} {224206} (\bibinfo {year} {2016})}\BibitemShut
  {NoStop}%
\bibitem [{\citenamefont {Zhang}()}]{sptentanglers}%
  \BibitemOpen
  \bibfield  {author} {\bibinfo {author} {\bibfnamefont {C.}~\bibnamefont
  {Zhang}},\ }\href@noop {} {\bibinfo  {journal} {to appear}\ }\BibitemShut
  {NoStop}%
\bibitem [{\citenamefont {Fidkowski}\ \emph {et~al.}(2019)\citenamefont
  {Fidkowski}, \citenamefont {Po}, \citenamefont {Potter},\ and\ \citenamefont
  {Vishwanath}}]{fermionic}%
  \BibitemOpen
\bibfield  {journal} {  }\bibfield  {author} {\bibinfo {author} {\bibfnamefont
  {L.}~\bibnamefont {Fidkowski}}, \bibinfo {author} {\bibfnamefont {H.~C.}\
  \bibnamefont {Po}}, \bibinfo {author} {\bibfnamefont {A.~C.}\ \bibnamefont
  {Potter}}, \ and\ \bibinfo {author} {\bibfnamefont {A.}~\bibnamefont
  {Vishwanath}},\ }\href {\doibase 10.1103/PhysRevB.99.085115} {\bibfield
  {journal} {\bibinfo  {journal} {Phys. Rev. B}\ }\textbf {\bibinfo {volume}
  {99}},\ \bibinfo {pages} {085115} (\bibinfo {year} {2019})}\BibitemShut
  {NoStop}%
\bibitem [{\citenamefont {Kim}\ \emph {et~al.}(2021)\citenamefont {Kim},
  \citenamefont {Shi}, \citenamefont {Kato},\ and\ \citenamefont
  {Albert}}]{kim2021}%
  \BibitemOpen
  \bibfield  {author} {\bibinfo {author} {\bibfnamefont {I.~H.}\ \bibnamefont
  {Kim}}, \bibinfo {author} {\bibfnamefont {B.}~\bibnamefont {Shi}}, \bibinfo
  {author} {\bibfnamefont {K.}~\bibnamefont {Kato}}, \ and\ \bibinfo {author}
  {\bibfnamefont {V.~V.}\ \bibnamefont {Albert}},\ }\href
  {https://arxiv.org/abs/2110.10400} {\bibfield  {journal} {\bibinfo  {journal}
  {arXiv preprint arXiv:2110.10400}\ } (\bibinfo {year} {2021})}\BibitemShut
  {NoStop}%
\bibitem [{\citenamefont {Kim}\ \emph {et~al.}(2022)\citenamefont {Kim},
  \citenamefont {Shi}, \citenamefont {Kato},\ and\ \citenamefont
  {Albert}}]{kim2022}%
  \BibitemOpen
  \bibfield  {author} {\bibinfo {author} {\bibfnamefont {I.~H.}\ \bibnamefont
  {Kim}}, \bibinfo {author} {\bibfnamefont {B.}~\bibnamefont {Shi}}, \bibinfo
  {author} {\bibfnamefont {K.}~\bibnamefont {Kato}}, \ and\ \bibinfo {author}
  {\bibfnamefont {V.~V.}\ \bibnamefont {Albert}},\ }\href {\doibase
  10.1103/PhysRevLett.128.176402} {\bibfield  {journal} {\bibinfo  {journal}
  {Phys. Rev. Lett.}\ }\textbf {\bibinfo {volume} {128}},\ \bibinfo {pages}
  {176402} (\bibinfo {year} {2022})}\BibitemShut {NoStop}%
\bibitem [{\citenamefont {Fan}(2022)}]{fan2022}%
  \BibitemOpen
  \bibfield  {author} {\bibinfo {author} {\bibfnamefont {R.}~\bibnamefont
  {Fan}},\ }\href {https://arxiv.org/abs/2206.02823} {\bibfield  {journal}
  {\bibinfo  {journal} {arXiv preprint arXiv:2206.02823}\ } (\bibinfo {year}
  {2022})}\BibitemShut {NoStop}%
\bibitem [{\citenamefont {Gong}\ \emph {et~al.}(2021)\citenamefont {Gong},
  \citenamefont {Piroli},\ and\ \citenamefont {Cirac}}]{gong2021}%
  \BibitemOpen
  \bibfield  {author} {\bibinfo {author} {\bibfnamefont {Z.}~\bibnamefont
  {Gong}}, \bibinfo {author} {\bibfnamefont {L.}~\bibnamefont {Piroli}}, \ and\
  \bibinfo {author} {\bibfnamefont {J.~I.}\ \bibnamefont {Cirac}},\ }\href
  {\doibase 10.1103/PhysRevLett.126.160601} {\bibfield  {journal} {\bibinfo
  {journal} {Phys. Rev. Lett.}\ }\textbf {\bibinfo {volume} {126}},\ \bibinfo
  {pages} {160601} (\bibinfo {year} {2021})}\BibitemShut {NoStop}%
\bibitem [{\citenamefont {Potter}\ and\ \citenamefont
  {Morimoto}(2017)}]{dynamically}%
  \BibitemOpen
  \bibfield  {author} {\bibinfo {author} {\bibfnamefont {A.~C.}\ \bibnamefont
  {Potter}}\ and\ \bibinfo {author} {\bibfnamefont {T.}~\bibnamefont
  {Morimoto}},\ }\href {\doibase 10.1103/PhysRevB.95.155126} {\bibfield
  {journal} {\bibinfo  {journal} {Phys. Rev. B}\ }\textbf {\bibinfo {volume}
  {95}},\ \bibinfo {pages} {155126} (\bibinfo {year} {2017})}\BibitemShut
  {NoStop}%
\bibitem [{\citenamefont {Po}\ \emph {et~al.}(2017)\citenamefont {Po},
  \citenamefont {Fidkowski}, \citenamefont {Vishwanath},\ and\ \citenamefont
  {Potter}}]{radical}%
  \BibitemOpen
  \bibfield  {author} {\bibinfo {author} {\bibfnamefont {H.~C.}\ \bibnamefont
  {Po}}, \bibinfo {author} {\bibfnamefont {L.}~\bibnamefont {Fidkowski}},
  \bibinfo {author} {\bibfnamefont {A.}~\bibnamefont {Vishwanath}}, \ and\
  \bibinfo {author} {\bibfnamefont {A.~C.}\ \bibnamefont {Potter}},\ }\href
  {\doibase 10.1103/PhysRevB.96.245116} {\bibfield  {journal} {\bibinfo
  {journal} {Phys. Rev. B}\ }\textbf {\bibinfo {volume} {96}},\ \bibinfo
  {pages} {245116} (\bibinfo {year} {2017})}\BibitemShut {NoStop}%
\bibitem [{\citenamefont {Nathan}\ \emph
  {et~al.}(2019{\natexlab{b}})\citenamefont {Nathan}, \citenamefont {Abanin},
  \citenamefont {Lindner}, \citenamefont {Berg},\ and\ \citenamefont
  {Rudner}}]{hierarchy}%
  \BibitemOpen
  \bibfield  {author} {\bibinfo {author} {\bibfnamefont {F.}~\bibnamefont
  {Nathan}}, \bibinfo {author} {\bibfnamefont {D.~A.}\ \bibnamefont {Abanin}},
  \bibinfo {author} {\bibfnamefont {N.~H.}\ \bibnamefont {Lindner}}, \bibinfo
  {author} {\bibfnamefont {E.}~\bibnamefont {Berg}}, \ and\ \bibinfo {author}
  {\bibfnamefont {M.~S.}\ \bibnamefont {Rudner}},\ }\href
  {https://arxiv.org/abs/1907.12228} {\bibfield  {journal} {\bibinfo  {journal}
  {arXiv:1907.12228}\ } (\bibinfo {year} {2019}{\natexlab{b}})}\BibitemShut
  {NoStop}%
\end{thebibliography}%

\end{document}